\documentclass[aps,prb,reprint,amsmath,amssymb,superscriptaddress,nofootinbib]{revtex4-2}
\usepackage{microtype}
\usepackage{amsmath,amssymb,amsthm,mathtools,bm}
\usepackage{mathrsfs}
\usepackage{booktabs}
\usepackage{enumitem}
\usepackage{comment}
\usepackage{braket}
\usepackage[dvipsnames]{xcolor}
\usepackage{tikz}
\usetikzlibrary{arrows.meta,positioning,shapes.geometric,calc}
\usepackage{hyperref}
\hypersetup{
    colorlinks=true,
    linkcolor=blue!60!black,
    citecolor=magenta!60!black,
    urlcolor=blue!60!black
}

\definecolor{TableBackground}{HTML}{F2F6F9}
\definecolor{TableAccent}{HTML}{315A75}
\definecolor{TFIBlue}{HTML}{1557C0}
\definecolor{TFILightBlue}{HTML}{F5F9FF}
\definecolor{TFIViolet}{HTML}{4C1D95}
\definecolor{TFIBlue}{HTML}{1557C0}
\definecolor{TFILightBlue}{HTML}{F5F9FF}
\definecolor{TFIViolet}{HTML}{4C1D95}

\theoremstyle{definition}

\newcommand{\ii}{\mathrm{i}}
\newcommand{\dd}{\mathrm{d}}

\newcommand{\Pf}{\mathrm{Pf}}
\newcommand{\CT}{\operatorname{CT}}
\newcommand{\abs}[1]{\left|#1\right|}

\newcommand{\cZ}{\mathcal{Z}}

\begin{document}


\title{Fugacity-Resolved Stabilizer Entropy in Critical Quantum Chains: Discrete Selberg Sums and Exactly Solvable R\'enyi Indices}

\author{Reyhaneh Khasseh}
\affiliation{Theoretical Physics III, Center for Electronic Correlations and Magnetism,
Institute of Physics, University of Augsburg, D-86135 Augsburg, Germany}

\author{M. A. Rajabpour}
\affiliation{Instituto de F\'isica, Universidade Federal Fluminense,
Av.~Gal.~Milton Tavares de Souza s/n, Gragoat\'a,
24210-346 Niter\'oi, RJ, Brazil}

\begin{abstract}
Stabilizer R\'enyi entropy quantifies the nonstabilizerness of a quantum
state through R\'enyi moments of its Pauli expectation-value distribution.
Its standard form sums over all contributing Pauli-string degrees and
therefore retains only the total R\'enyi weight. We introduce a fugacity-resolved partition function for the critical
transverse-field Ising chain that resolves this degree in the balanced Majorana representation and
generates its full counting statistics. The relevance of this Ising problem extends beyond a single model:
exact decimation identities and the correspondence with the half-filled
\(XX\) chain establish it as a common finite-size building block for
stabilizer and computational-basis Shannon--R\'enyi entropies. We map the resulting fugacity-resolved all-minors sum, for every positive
real R\'enyi index, to a checkerboard-weighted discrete Selberg ensemble
on a half-filled doubled root lattice. For positive integer indices, this
representation further reduces to a finite aliased Dyson constant term.
At \(\alpha=\tfrac12,1,2\), determinant and Pfaffian compressions yield
explicit product formulas. At \(\alpha=4\), the generic-fugacity problem
admits an exact inverse Jack--Kostka representation, while at unit
fugacity a complementary middle-minor identity relates it to the square
of the \(\alpha=2\) result. These exact generating functions also determine the full balanced-degree
statistics. Complement symmetry makes the distribution symmetric about
\(k=L/2\), while its width depends strongly on the R\'enyi index. It is
exactly binomial at \(\alpha=1\), has a variance proportional to \(L\) at
\(\alpha=\tfrac12\), and develops an \(L\log L\) enhancement at
\(\alpha=2\). Despite these different fluctuation scales, the centered
distributions, when rescaled by their standard deviations, converge to
Gaussian limits at all three indices. Beyond these solvable cases,
finite-size numerics reveal an evolution from a central peak to a
symmetric bimodal profile and eventually to endpoint dominance as the
R\'enyi index is increased.

\end{abstract}

\maketitle

 \tableofcontents 

\section{Introduction}
\label{sec:intro}

Information-theoretic observables have become central tools for describing
quantum many-body states beyond conventional correlation functions.
Entanglement entropies organize bipartite quantum correlations and reveal
universal data of critical theories, while Shannon--R\'enyi entropies of
wave-function amplitudes probe the probability distribution selected by a
measurement basis
\cite{CalabreseCardy2009,StephanEtAl2009,
StephanMisguichPasquier2011,AlcarazRajabpour2013,Stephan2014,
AlcarazRajabpour2014,LuitzAletLaflorencie2014,TarighiEtAl2022}.
A complementary perspective is provided by stabilizer states and Clifford
operations, which form an efficiently tractable sector of quantum theory and
underlie quantum error correction and classical simulation
\cite{Gottesman1997,BravyiKitaev2005}.  Departures from this sector are
quantified within the resource theory of nonstabilizerness, or magic,
through quasiprobability, contextuality, robustness, and operational
measures
\cite{VeitchFerrieGrossEmerson2012,HowardWallmanVeitchEmerson2014,
VeitchMousavianGottesmanEmerson2014,HowardCampbell2017}.

The stabilizer R\'enyi entropy recasts this resource in terms of R\'enyi
moments of the Pauli expectation-value distribution
\cite{OlivieroLeoneHamma2022}.  It avoids optimization over stabilizer
decompositions and admits direct measurement and algorithmic protocols,
while its precise resource-theoretic status, including its dependence on
the R\'enyi index and on the allowed stabilizer operations, has been
clarified in subsequent work
\cite{OlivieroLeoneHammaLloyd2022,HaugLeeKim2024,
HaugPiroliMonotones2023,LeoneBittel2024}.  This accessibility has made it
possible to study nonstabilizerness in many-body states using
tensor-network contractions, perfect Pauli sampling, Pauli-Markov chains,
and reduced-string sampling
\cite{HaugPiroli2023,LamiCollura2023,TarabungaEtAl2023,DingWangYan2025}.
These studies show that entanglement and nonstabilizerness need not track
one another, and that Clifford disentangling can reveal structures not
captured by entanglement alone
\cite{FuxEtAl2024,FrauEtAl2024,FanEtAl2025,FrauEtAl2025}.  For fermionic
Gaussian states, determinantal sampling now permits system sizes far beyond
direct Pauli enumeration \cite{ColluraEtAl2026}.

At criticality, nonstabilizerness has also emerged as a diagnostic of
scale-invariant many-body structure.  Studies based on robustness of magic,
mana, and related magic measures have found enhanced nonstabilizerness near
quantum critical points and argued that magic can be distributed across the
length scales of a conformal ground state
\cite{SarkarMukhopadhyayBayat2020,WhiteCaoSwingle2021,Tarabunga2024}.
A resource-theory analysis of the transverse-field Ising ground state
identified characteristic singular behavior of stabilizer-based measures
near the transition \cite{LeoneOlivieroHamma2022}.  More recently,
continuum descriptions have related stabilizer R\'enyi entropy to conformal
boundary conditions and topological defects, including universal subleading
terms and fusion data
\cite{HoshinoOshikawaAshida2026,HoshinoAshida2026}.  On the lattice, exact
correspondences have connected the Pauli distribution of paired Gaussian
states to the computational-basis distribution of doubled,
number-conserving Gaussian states, and have reduced broad families of
critical chains to a common transverse-field Ising building block
\cite{RamirezTrinoRajabpour2026}.  Related analyses have extended this
program to thermal boundary crossovers and massive finite-size regimes at
R\'enyi index one half
\cite{KhassehRajabpour2026,KhassehRamirezTrinoRajabpour2026}.  Together,
these results show that stabilizer entropy can retain universal information
while remaining sensitive to structures not directly visible in standard
thermodynamic observables.

A basic finite-size difficulty remains even for free fermions.  Wick's
theorem converts individual Pauli expectation values into determinants or
Pfaffians of submatrices, but a stabilizer R\'enyi moment requires powers of
all balanced minors.  Their number grows as a central binomial coefficient,
so knowing the Gaussian state, or even diagonalizing the underlying
quadratic Hamiltonian exactly, does not by itself compress the entropy
calculation.  Correlation-matrix methods, Gaussian-state amplitude
formulas, and matchgate techniques provide the appropriate local algebra
\cite{LiebSchultzMattis1961,Peschel2003,PeschelEisler2009,
TarighiKhassehRajabpour2024,RajabpourSeifiMirjafarlouKhasseh2025},
but the global sum over minors is a separate problem.  The main question of
this work is therefore not whether the underlying spin chain is integrable,
but which R\'enyi indices permit an exact finite-size compression of the
resulting all-minors sum, and which algebraic mechanism is responsible in
each case.

There is a substantial body of precedent for the probability ensembles that
emerge from this all-minors reformulation.  Vandermonde powers on a
discrete circular lattice were studied as one-dimensional Coulomb gases
\cite{Gaudin1973,MehtaMehta1975}, and closely related Jastrow structures
appear in Calogero--Sutherland and Haldane--Shastry settings
\cite{Sutherland1971,Haldane1988,Shastry1988,LamersSerban2024}.  In
quantum spin chains, discrete Dyson-gas partition functions arise in
Shannon and participation entropies, inverse participation ratios, and full
counting statistics
\cite{StephanEtAl2009,Stephan2014,StephanPollmann2017}.
Counting-field deformations of free fermions and spin chains are likewise
often governed by determinant, Toeplitz, or Pfaffian structures
\cite{IvanovAbanovCheianov2013,GrohaEsslerCalabrese2018,
AresRajabpourViti2021}.  These works provide essential benchmarks.  Many of
them, however, focus either on unrefined partition sums with varying
R\'enyi exponent, or on counting-field deformations at fixed classical
Vandermonde exponent.  The objective here is to keep both ingredients at
once: the R\'enyi-index dependence and the checkerboard fugacity
resolution.


Building on established stabilizer decimation identities and the
TFI--\(XX\) correspondence, we identify the critical TFI stabilizer
entropy as a common finite-size building block for several stabilizer and
computational-basis Shannon--R\'enyi observables. We extend these
connections by proving that, in compatible finite-size sectors, the full
computational-basis distribution of the range-\(m\) \(XX\) chain
factorizes over \(m\) squeezed sublattices. This yields an exact
Shannon--R\'enyi decimation for the entire family. The same \(XX\) route
also incorporates the Haldane--Shastry chain through the known order-two
escort relation between their ground-state distributions\cite{Haldane1988,Shastry1988}. These results
reduce a broad class of observables to the same TFI all-minors problem.
Accordingly, the remainder of this work focuses on the exact
finite-size analysis of this all-minors structure.

We resolve this all-minors sum by introducing a fugacity conjugate to the
balanced Majorana degree labeling the contributing Pauli-string sectors.
We prove a termwise bijection between balanced minors of the critical TFI
correlation matrix and half-filled configurations on a doubled
root-of-unity lattice. For every positive real R\'enyi index, this gives
an exact checkerboard-weighted discrete Selberg ensemble that preserves
the complete degree-generating polynomial. At unit fugacity, the
construction reduces, after normalization, to discrete Dyson-gas moments
previously encountered for the half-filled \(XX\) chain
\cite{Gaudin1973,Stephan2014}.

For positive integer indices, the finite Fourier bandwidth of the
Vandermonde weight reduces the discrete root measure to finitely many
visible aliases, yielding a neutral aliased Dyson--Morris constant-term
representation. At the classical indices
\(\alpha=\tfrac12,1,2\), determinant and Pfaffian compressions simplify
further to explicit fugacity-resolved product formulas. The case
\(\alpha=4\) is structurally different: at generic fugacity it admits an
exact rectangular inverse Jack--Kostka representation, whereas at unit
fugacity a complementary middle-minor identity relates it to the square
of the \(\alpha=2\) result. More generally, perfect-square indices form a
confluent hyperpfaffian hierarchy, while positive integer indices admit
complementary-minor and shifted-Dyson organizations. These constructions
also distinguish exact structural representations from polynomial-size
determinant and Pfaffian evaluations.

Because the fugacity-resolved partition function is also the generating
function of the balanced Majorana degree, the exact formulas determine
its full counting statistics. Complement symmetry makes the degree
distribution symmetric about half system size. The distribution is exactly binomial at
\(\alpha=1\), has a variance growing linearly with system size at
\(\alpha=\tfrac12\), and develops an \(L\log L\) enhancement at
\(\alpha=2\). Despite this anomalous broadening, the centered and
variance-rescaled distributions converge to Gaussian limits at all three
product-solvable indices. Beyond these indices, finite-size results show
that increasing the R\'enyi tilt reorganizes the degree-resolved Pauli
weight from a central peak toward bimodal and eventually
endpoint-dominated profiles.

The paper is organized as follows. Section~2 establishes the common TFI
building block and proves the range-\(m\) \(XX\) factorization.
Sections~3 and~4 derive the fugacity-resolved discrete Selberg mapping and
its aliased constant-term representation. Sections~5 and~6 present the
special-index formulas and organize their algebraic and computational
status. Section~7 derives the degree-resolved full counting statistics,
while the appendices contain the technical proofs and generic-weight
constructions.

\section{The TFI Stabilizer Entropy as a Common Building Block}
\label{sec:web}

\begin{figure*}[t]
    \centering
    \includegraphics[width=0.8\textwidth]
    {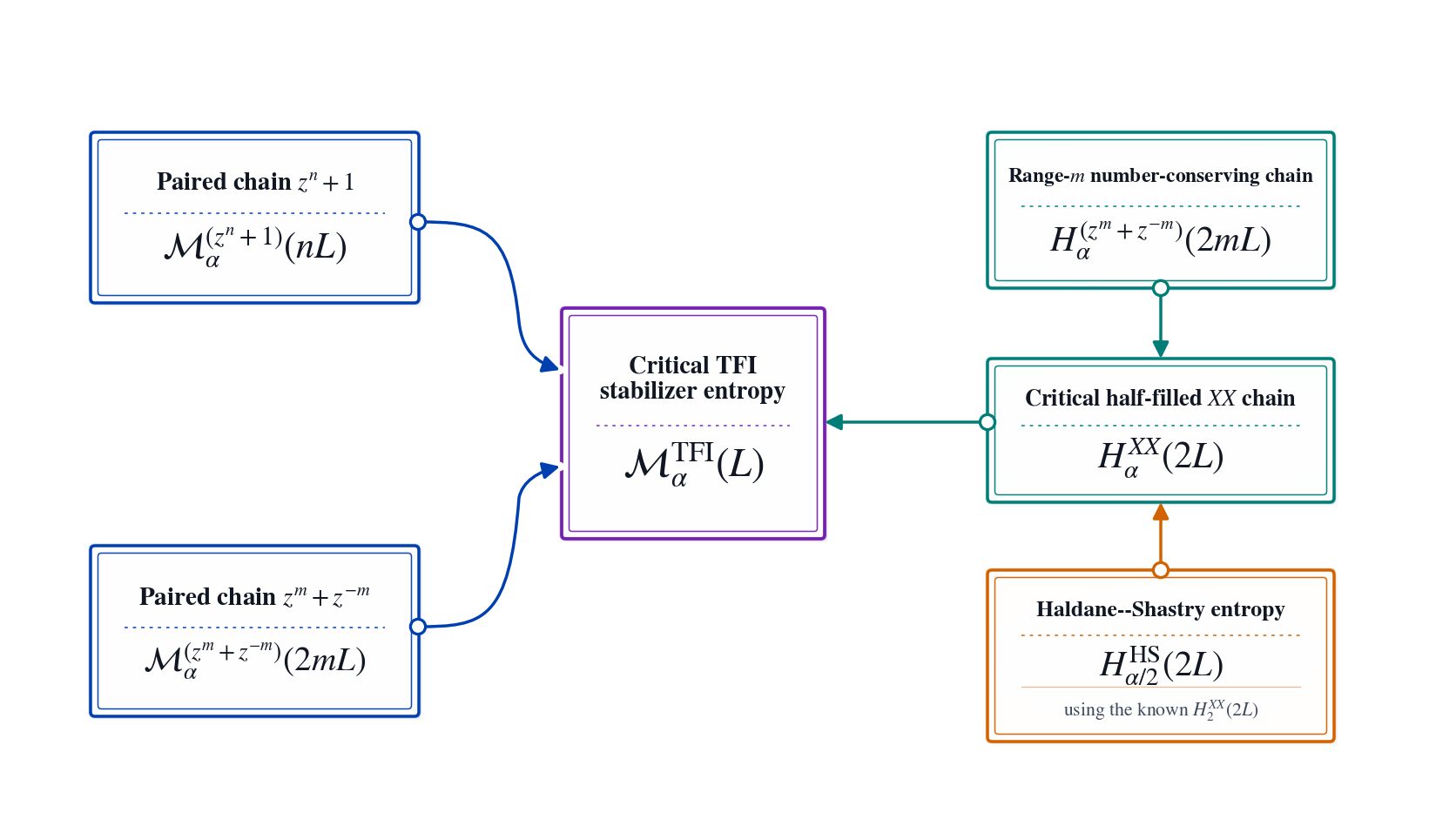}
    \caption{
    Exact entropy reductions to the critical TFI stabilizer R\'enyi
    entropy. Stabilizer decimation applies to paired free-fermion chains,
    probability factorization applies to the range-\(m\)
    number-conserving chains, and the Haldane--Shastry distribution is the
    order-two escort of the half-filled \(XX\) distribution. In every
    case, the required input is supplied by
    \(\mathcal M_\alpha^{\rm TFI}\).
    }
    \label{fig:tfi-building-block}
\end{figure*}

The critical transverse-field Ising chain provides the common entropy
building block used throughout this work. Several paired free-fermion
chains reduce to independent TFI copies through stabilizer decimation,
whereas number-conserving chains are connected to the same building block
through the TFI--\(XX\) correspondence and a probability-level
factorization. The computational-basis distribution of the interacting
Haldane--Shastry ground state supplies a further escort relation. The
resulting web of exact reductions is displayed in
Fig.~\ref{fig:tfi-building-block} and collected in
Table~\ref{tab:tfi-web}.

\subsection{Definitions and model conventions}
\label{subsec:tfi-building-block}

We consider a periodic chain of \(L\) spins, with site labels understood
modulo \(L\). Our basic model is the critical transverse-field Ising chain,
\begin{equation}
\label{eq:TFI-spin-Hamiltonian}
{\cal H}_{\rm TFI}
=
-\frac12
\sum_{j=0}^{L-1}
\left(
\sigma_j^x\sigma_{j+1}^x+\sigma_j^z
\right).
\end{equation}
For a pure state \(\ket{\psi}\) of these \(L\) spins, the stabilizer
R\'enyi entropy is
\begin{equation}
\label{eq:SRE-section2}
\mathcal M_\alpha[\psi]
=
\frac{1}{1-\alpha}
\log
\left[
\frac{1}{2^L}
\sum_{P\in\mathcal P_L}
\abs{\bra{\psi}P\ket{\psi}}^{2\alpha}
\right],
\qquad
\alpha\neq1,
\end{equation}
where \(\mathcal P_L\) is the \(L\)-spin Pauli group modulo phases. With
this normalization, every stabilizer state has
\(\mathcal M_\alpha=0\). We denote the ground-state value of
Eq.~\eqref{eq:TFI-spin-Hamiltonian} by
\(\mathcal M_\alpha^{\rm TFI}(L)\).

For a state with computational-basis probabilities
\(p_{\mathbf n}\), the corresponding Shannon--R\'enyi entropy is
\begin{equation}
\label{eq:Shannon-Renyi-section2}
H_\alpha
=
\frac{1}{1-\alpha}
\log
\sum_{\mathbf n}p_{\mathbf n}^{\alpha},
\qquad
\alpha\neq1.
\end{equation}

To formulate the fermionic families entering the decimation relations, we
use the Jordan--Wigner convention
\begin{equation}
\label{eq:JW-section2}
c_j
=
\left[
\prod_{\ell<j}(-\sigma_\ell^z)
\right]\sigma_j^-,
\qquad
\sigma_j^-=\frac{\sigma_j^x-i\sigma_j^y}{2},
\end{equation}
and introduce the Majorana operators
\[
a_j=c_j+c_j^\dagger,
\qquad
b_j=i(c_j-c_j^\dagger).
\]
A translation-invariant paired chain can then be parametrized by a Laurent
symbol \(f(z)\) as
\begin{equation}
\label{eq:paired-symbol-section2}
{\cal H}_f
=
-\frac{i}{2}\sum_{j,r}t_r\,b_j a_{j+r},
\qquad
f(z)=\sum_r t_r z^r.
\end{equation}
In this notation, the critical TFI chain corresponds to $f(z)=1+z.$
The two families that enter the stabilizer decimation identities are
\begin{equation}
\label{eq:decimation-symbols-section2}
f_n(z)=1+z^n,
\qquad
f_m(z)=z^m+z^{-m}.
\end{equation}

With the Jordan--Wigner convention in Eq.~\eqref{eq:JW-section2}, a
convenient bulk spin representative of the first family is
\begin{equation}
\label{eq:zn1-spin-section2}
{\cal H}_n^{(1+z^n)}
=
-\frac12\sum_j
\left[
\sigma_j^z
+\sigma_j^x
\left(
\prod_{a=1}^{n-1}\sigma_{j+a}^z
\right)
\sigma_{j+n}^x
\right].
\end{equation}
For \(n=1\), this reduces to the critical TFI Hamiltonian in
Eq.~\eqref{eq:TFI-spin-Hamiltonian}. The second family has the bulk spin
representation
\begin{equation}
\label{eq:XXm-spin-section2}
\begin{aligned}
{\cal H}_m^{(z^m+z^{-m})}
&=
\frac{(-1)^m}{2}\sum_j
\Biggl[
\sigma_j^x
\left(
\prod_{a=1}^{m-1}\sigma_{j+a}^z
\right)
\sigma_{j+m}^x
+{}
\\[-1mm]
&\qquad\qquad
\sigma_j^y
\left(
\prod_{a=1}^{m-1}\sigma_{j+a}^z
\right)
\sigma_{j+m}^y
\Biggr].
\end{aligned}
\end{equation}
Equivalently, it is the range-\(m\) hopping chain
\begin{equation}
\label{eq:XXm-fermion-section2}
{\cal H}_m^{(z^m+z^{-m})}
=
-\sum_{j=0}^{L-1}
\left(
c_j^\dagger c_{j+m}
+
c_{j+m}^\dagger c_j
\right).
\end{equation}
The intervening \(\sigma^z\) string in
Eq.~\eqref{eq:XXm-spin-section2} is therefore required by the
Jordan--Wigner transformation and is essential for \(m>1\).

For periodic systems, we write the fermionic boundary condition as
\begin{equation}
\label{eq:boundary-twist-section2}
c_{j+L}=e^{i\phi}c_j,
\qquad
\theta_q=\frac{2\pi q+\phi}{L},
\end{equation}
with \(\phi=\pi\) in the Neveu--Schwarz sector and \(\phi=0\) in the
Ramond sector. All finite-size reductions below assume matching boundary
sectors and exclude cases in which the corresponding one-particle symbol
contains an unresolved zero mode.

\subsection{Reduction mechanisms}
\label{subsec:stabilizer-route}

Three distinct mechanisms underlie the reductions summarized in
Table~\ref{tab:tfi-web}: stabilizer decimation, probability-level
factorization, and an escort relation between computational-basis
distributions.

First, the paired symbols in Eq.~\eqref{eq:decimation-symbols-section2}
split into independent momentum sectors. The stabilizer decimation
identities of Ref.~\cite{RamirezTrinoRajabpour2026} then reduce their
stabilizer R\'enyi entropies to copies of the critical TFI result. The
required divisibility conditions are \(n\mid L\) for the family
\(1+z^n\) and \(2m\mid L\) for the family \(z^m+z^{-m}\).

\begin{table*}[t]
\centering
\setlength{\fboxsep}{8pt}

\colorbox{TableBackground}{%
\begin{minipage}{0.94\textwidth}
\centering
\setlength{\tabcolsep}{7pt}
\renewcommand{\arraystretch}{1.35}

\begin{tabular}{
@{}
p{0.20\linewidth}
p{0.46\linewidth}
p{0.27\linewidth}
@{}
}
\toprule

\textcolor{TableAccent}{\textbf{Quantity}}
&
\textcolor{TableAccent}{\textbf{Exact reduction}}
&
\textcolor{TableAccent}{\textbf{Origin}}
\\
\midrule

$\mathcal M_\alpha^{\rm TFI}(L)$
&
$H_\alpha^{XX}(2L)-L\log 2$
&
TFI--\(XX\) stabilizer--Shannon relation
\\[1mm]

$\mathcal M_\alpha^{(z^n+1)}(L)$
&
$n\,\mathcal M_\alpha^{\rm TFI}(L/n)$
&
known stabilizer decimation
\\[1mm]

$\mathcal M_\alpha^{(z^m+z^{-m})}(L)$
&
$2m\,\mathcal M_\alpha^{\rm TFI}(L/(2m))$
&
known stabilizer decimation
\\[1mm]

$H_\alpha^{(z^m+z^{-m})}(L)$
&
$m\,H_\alpha^{XX}(L/m)$
&
Shannon probability factorization
\\[1mm]

$H_\alpha^{(z^m+z^{-m})}(2mL)$
&
$m\,\mathcal M_\alpha^{\rm TFI}(L)+mL\log 2$
&
Shannon factorization plus TFI--\(XX\) relation
\\[1mm]

$H_{\alpha}^{\rm HS}(2L)$
&
$\displaystyle
L\log 2+
\frac{
(1-2\alpha)\mathcal M_{2\alpha}^{\rm TFI}(L)
+
\alpha\,\mathcal M_2^{\rm TFI}(L)
}{
1-\alpha
}$
&
\(XX\) escort relation plus TFI--\(XX\) relation
\\

\bottomrule
\end{tabular}

\end{minipage}%
}

\caption{
Exact reductions showing that the critical TFI stabilizer R\'enyi entropy
is the common building block behind several stabilizer and
computational-basis Shannon--R\'enyi quantities. The
Haldane--Shastry formula holds for \(\alpha\neq1\), with its Shannon limit
obtained by continuity.
}
\label{tab:tfi-web}
\end{table*}

Second, the number-conserving range-\(m\) chain admits a stronger
probability-level decomposition. For \(L=m\ell\), reordering the sites
according to their residue classes modulo \(m\) separates the system into
\(m\) squeezed \(XX\) chains of length \(\ell\). More precisely, the
complete computational-basis probability distribution factorizes as
\begin{equation}
\label{eq:XXm-probability-factorization-section2}
p_{\mathbf n}^{(z^m+z^{-m}),\phi}(L)
=
\prod_{r=0}^{m-1}
p_{\mathbf n^{(r)}}^{XX,\phi}(\ell),
\end{equation}
where
$
\mathbf n^{(r)}
=
\left(
n_r,n_{r+m},\ldots,n_{r+(\ell-1)m}
\right)$
is the configuration on the \(r\)-th squeezed sublattice. The correlation
matrix proof of Eq.~\eqref{eq:XXm-probability-factorization-section2} is
given in Appendix~\ref{app:XXm-factorization}. Combined with the known
TFI--\(XX\) correspondence, this factorization gives the
number-conserving entries in Table~\ref{tab:tfi-web}.

Third, the Shannon branch extends to the interacting Haldane--Shastry
chain. On \(2L\) sites, its Hamiltonian can be written in a conventional
normalization as
\begin{equation}
\label{eq:HS-Hamiltonian-section2}
{\cal H}_{\rm HS}
=
J\left(\frac{\pi}{2L}\right)^2
\sum_{0\le j<k\le 2L-1}
\frac{\mathbf S_j\cdot\mathbf S_k}
{\sin^2\!\left[\frac{\pi(j-k)}{2L}\right]},
\end{equation}
up to an additive constant
\cite{Haldane1988,Shastry1988}. At zero magnetization, a
computational-basis configuration is specified by an index set
\[
U=\{u_1<\cdots<u_L\}
\subseteq
\{0,1,\ldots,2L-1\},
\]
whose elements give the positions of the \(L\) down spins. Associate with
each site the root of unity
\begin{equation}
\label{eq:HS-roots-section2}
z_j
=
e^{\pi i j/L},
\qquad
j=0,\ldots,2L-1,
\end{equation}
and define the Vandermonde factor of the occupied roots by
\begin{equation}
\label{eq:HS-Vandermonde-section2}
\Delta(z_U)
:=
\prod_{1\le a<b\le L}
\left(
z_{u_b}-z_{u_a}
\right).
\end{equation}
For brevity, we write \(\Delta(U)\equiv\Delta(z_U)\). The
Haldane--Shastry and half-filled \(XX\) probabilities on the same
configuration space then satisfy
\begin{equation}
\label{eq:HS-XX-probabilities-section2}
p_{\rm HS}(U)\propto\abs{\Delta(U)}^4,
\qquad
p_{\rm XX}(U)\propto\abs{\Delta(U)}^2.
\end{equation}
Since both distributions are normalized, it follows that
\begin{equation}
\label{eq:HS-escort-section2}
p_{\rm HS}(U)
=
\frac{p_{\rm XX}(U)^2}
{\displaystyle\sum_{\substack{V\subseteq\{0,\ldots,2L-1\}\\|V|=L}}
p_{\rm XX}(V)^2}.
\end{equation}
Thus, the Haldane--Shastry probability distribution is the order-two
escort of the half-filled \(XX\) distribution. The resulting reduction of
its computational-basis Shannon--R\'enyi entropy, derived in
Ref.~\cite{StephanPollmann2017}, is summarized in
Table~\ref{tab:tfi-web}. This relation applies specifically to the
Haldane--Shastry wave-function probabilities in the computational basis
and does not imply a corresponding identity for its stabilizer
R\'enyi entropy.

Table~\ref{tab:tfi-web} contains all entropy-level relations needed in the
remainder of the paper. We therefore focus henceforth on the fugacity-resolved all-minors problem
of the critical TFI ground state. The resulting formulas can then be
transferred to the other models through the reductions summarized above.
\section{From the half-shift matrix to a discrete Selberg gas}
\label{sec:cauchy}

The critical TFI stabilizer problem can be rewritten exactly as a
half-filled Vandermonde gas on a doubled root-of-unity lattice. The mapping
holds term by term and preserves the fugacity that resolves the balanced
Majorana degree. This representation provides the starting point for the
aliased constant-term formulation developed in the next section.

\subsection{Fugacity-resolved all-minors problem}
\label{subsec:resolved-minors}

\begin{figure*}[t]
    \centering
    \includegraphics[width=0.9\textwidth]
    {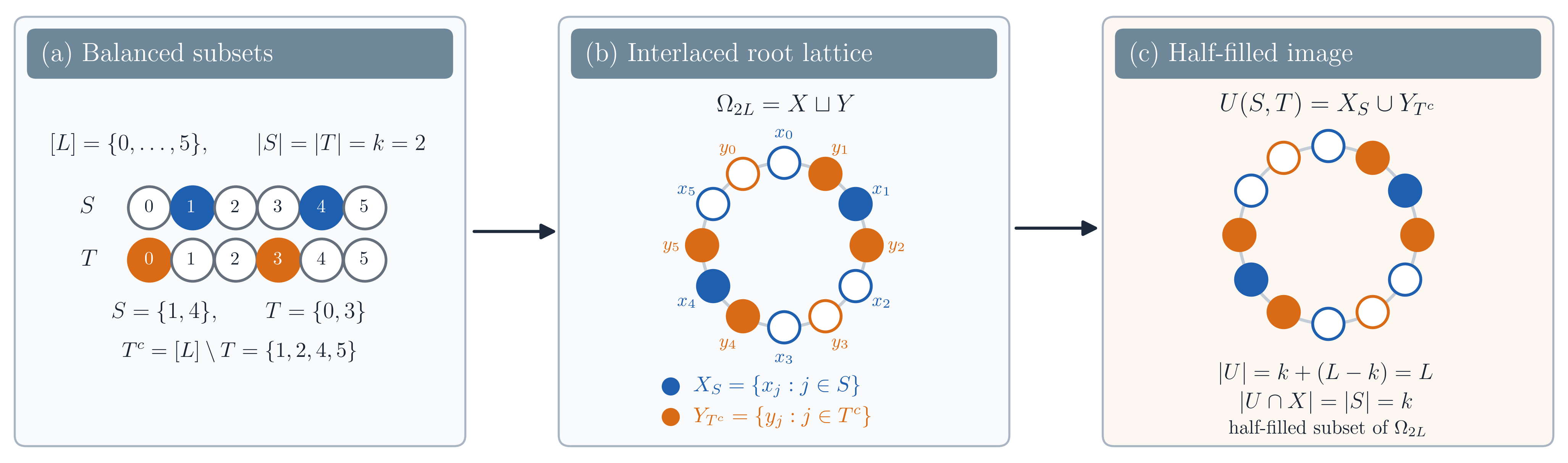}
   \caption{
Schematic of the bijection between a balanced minor label and a
half-filled configuration on the doubled root lattice, illustrated for
a six-site example with two selected indices. Panel (a) shows the two
balanced subsets that label the minor. Panel (b) maps the first subset
to selected roots on one sublattice and the complement of the second
subset to selected roots on the interlaced sublattice. Panel (c) combines
these selections into a half-filled configuration on the doubled lattice.
The construction is reversible, so the original balanced subsets can be
recovered uniquely from the final configuration.}
   \label{fig:doubled-root-map}
\end{figure*}

Let \(
[L]=\{0,1,\ldots,L-1\}.
\)
The critical TFI ground state is characterized by the real \(L\times L\)
matrix
\begin{equation}
\label{eq:Gdefinition}
G_{jk}
=
\frac{(-1)^{j-k}}{L}
\csc\!\left[
\frac{\pi}{L}
\left(j-k+\frac12\right)
\right],
\qquad
j,k\in[L].
\end{equation}
We refer to \(G\) as the half-shift matrix because its kernel depends on
the integer separation \(j-k\) shifted by one half. Equivalently, it
couples the two Majorana sublattices with a relative half-lattice
displacement.

A balanced Majorana monomial contains the same number \(k\) of
\(a\)- and \(b\)-Majorana operators. We denote their respective position
sets by
\[
S,T\subseteq[L],
\qquad
|S|=|T|=k,
\]
and write \(P_{S,T}\) for the corresponding Pauli string, up to an
irrelevant overall phase. In the Majorana representation,
\[
P_{S,T}
\ \longleftrightarrow\
\prod_{j\in S}a_j
\prod_{k\in T}b_k .
\]
Wick's theorem then reduces its expectation value to the corresponding
minor of \(G\),
\[
\abs{\langle P_{S,T}\rangle}
=
\abs{\det G[S,T]}.
\]
The convention for \(k=0\) is that the empty minor has determinant one,
corresponding to the identity Pauli string.

This motivates the fugacity-resolved all-minors polynomial
\begin{equation}
\label{eq:centralZ}
\cZ_{\alpha,L}(u)
=
\sum_{k=0}^{L}
u^k
\sum_{\substack{S,T\subseteq[L]\\|S|=|T|=k}}
\abs{\det G[S,T]}^{2\alpha}.
\end{equation}
The power of \(u\) records the balanced Majorana degree \(k\). Since each
sector contains \(k\) operators of type \(a\) and \(k\) operators of type
\(b\), the corresponding Majorana monomial has total degree \(2k\). This
balanced degree labels sectors of the Pauli spectrum, but it should not be
confused with the ordinary Pauli support, namely the number of lattice
sites on which the Pauli string is nontrivial.

At unit fugacity, Eq.~\eqref{eq:centralZ} reduces to the unrefined moment,
\begin{equation}
\label{eq:centralZ-unrefined}
\cZ_{\alpha,L}
:=
\cZ_{\alpha,L}(1),
\end{equation}
and the critical TFI stabilizer R\'enyi entropy is
\begin{equation}
\label{eq:SRE-Z-section}
\mathcal M_\alpha^{\rm TFI}(L)
=
\frac{1}{1-\alpha}
\log\!\left[
\frac{\cZ_{\alpha,L}(1)}{2^L}
\right],
\qquad
\alpha\neq1.
\end{equation}

More explicitly, define the normalized balanced-degree distribution
\begin{equation}
\label{eq:balanced-degree-distribution-section3}
\mathbb P_{\alpha,L}(k)
=
\frac{
\displaystyle
\sum_{\substack{S,T\subseteq[L]\\|S|=|T|=k}}
\abs{\det G[S,T]}^{2\alpha}
}{
\cZ_{\alpha,L}(1)
}.
\end{equation}
Then
\begin{equation}
\label{eq:balanced-degree-pgf-section3}
\frac{\cZ_{\alpha,L}(u)}{\cZ_{\alpha,L}(1)}
=
\sum_{k=0}^{L}
\mathbb P_{\alpha,L}(k)\,u^k
\end{equation}
is the probability-generating function of the balanced Majorana degree in
the \(\alpha\)-weighted Pauli ensemble. The fugacity refinement therefore
retains the sector-resolved information that is lost after setting
\(u=1\).

In the following subsection, the same polynomial is mapped term by term
to a checkerboard-weighted half-filled Vandermonde gas, equivalently a
discrete Selberg sum. The intermediate neutral two-component defect-gas
form and the terminology relating these equivalent representations are
summarized in Appendix~\ref{app:taxonomy}. For positive integer
\(\alpha\), the polynomial also admits the aliased Dyson constant-term
representation derived in the next section.

\subsection{Cauchy reduction and doubled-root representation}
\label{subsec:discrete-selberg-mapping}

The half-integer displacement in the kernel of
Eq.~\eqref{eq:Gdefinition} naturally leads to a doubled root-of-unity
lattice. Let
\begin{equation}
\label{eq:doubled-root-lattice}
\zeta=e^{\pi i/L},
\qquad
\Omega_{2L}=\{\zeta^a:0\leq a<2L\},
\end{equation}
and decompose it into the two interlaced sublattices
\begin{equation}
\label{eq:XY-def}
X=\{x_j=\zeta^{2j}:j\in[L]\},
\qquad
Y=\{y_j=\zeta^{2j-1}:j\in[L]\}.
\end{equation}
Thus \(X\) and \(Y\) are respectively the roots of \(z^L-1\) and
\(z^L+1\), and
\(\Omega_{2L}=X\sqcup Y\). The term ``doubled'' refers to the fact that
\(\Omega_{2L}\) contains \(2L\) sites, twice the number of sites in either
sublattice.

For a finite set \(A=\{a_1,\ldots,a_q\}\), we use the Vandermonde
notation
\begin{equation}
\label{eq:Vandermonde-definition-section3}
\Delta(A)
=
\prod_{1\leq r<s\leq q}(a_s-a_r).
\end{equation}
The relation between \(x_j-y_k\) and the half-shifted sine kernel rewrites
\(G\), up to diagonal factors of unit modulus, as a Cauchy matrix with
entries \((x_j-y_k)^{-1}\). The Cauchy determinant formula, together with
the elementary root-product identities
\[
\begin{aligned}
\abs{\Delta(X)}
&=
\abs{\Delta(Y)}
=
L^{L/2},
\\
\prod_{y\in Y}\abs{x-y}
&=
\prod_{x\in X}\abs{y-x}
=
2.
\end{aligned}
\]
then converts every minor of \(G\) into a Vandermonde weight on
\(\Omega_{2L}\).

More precisely, let \(S,T\subseteq[L]\) satisfy
\(|S|=|T|=k\), and define
\[
X_S=\{x_j:j\in S\},
\qquad
Y_{T^c}=\{y_j:j\in T^c\}.
\]
The corresponding half-filled subset of the doubled root lattice is
\begin{equation}
\label{eq:U-map}
U(S,T)=X_S\cup Y_{T^c}.
\end{equation}
The construction and its inverse are illustrated schematically in
Fig.~\ref{fig:doubled-root-map}. Because \(X_S\) contains \(k\) roots and \(Y_{T^c}\) contains \(L-k\)
roots, \(U(S,T)\) is an \(L\)-element subset of the \(2L\)-site lattice:
\[
|U(S,T)|=L,
\qquad
|U(S,T)\cap X|=k.
\]
The map \((S,T)\mapsto U(S,T)\) is therefore a bijection between balanced
minor labels and half-filled subsets of \(\Omega_{2L}\), where
``half-filled'' means that exactly \(L\) of the \(2L\) roots are occupied.
The Cauchy reduction gives the termwise identity
\begin{equation}
\label{eq:termwise-main}
\abs{\Delta\!\left(U(S,T)\right)}
=
L^{L/2}\abs{\det G[S,T]}.
\end{equation}

Substituting Eq.~\eqref{eq:termwise-main} into
Eq.~\eqref{eq:centralZ} yields
\begin{equation}
\label{eq:discrete-selberg}
\cZ_{\alpha,L}(u)
=
L^{-\alpha L}
\sum_{\substack{U\subseteq\Omega_{2L}\\|U|=L}}
u^{|U\cap X|}
\abs{\Delta(U)}^{2\alpha},
\qquad
\alpha>0.
\end{equation}
This is a discrete Selberg or Vandermonde-gas representation of the TFI
all-minors problem. The particles occupy half of the doubled
root-of-unity lattice and interact through the positive weight
\(\abs{\Delta(U)}^{2\alpha}\). The fugacity distinguishes the two
interlaced sublattices: each occupied site in \(X\) contributes a factor
of \(u\), whereas occupied sites in \(Y\) contribute no additional
factor. For \(u>0\), it therefore acts as a checkerboard or staggered
chemical potential. Unlike the finite constant-term representation
derived in the next section, Eq.~\eqref{eq:discrete-selberg} is valid for
every real \(\alpha>0\).

The doubled-root formulation also makes complement symmetry immediate.
For every half-filled subset \(U\subseteq\Omega_{2L}\), its complement
\(U^c\) is again half-filled and satisfies
\[
\abs{\Delta(U^c)}
=
\abs{\Delta(U)},
\qquad
|U^c\cap X|
=
L-|U\cap X|.
\]
Writing
\[
\cZ_{\alpha,L}(u)
=
\sum_{k=0}^{L}W_{\alpha,L}(k)u^k,
\]
we consequently obtain
\begin{equation}
\label{eq:palindromic}
\cZ_{\alpha,L}(u)
=
u^L\cZ_{\alpha,L}(u^{-1}),
\qquad
W_{\alpha,L}(k)
=
W_{\alpha,L}(L-k).
\end{equation}
The endpoint sectors contain only the complete sublattices \(Y\) and
\(X\), respectively, and hence
\[
W_{\alpha,L}(0)
=
W_{\alpha,L}(L)
=
1.
\]
These relations will be used in the degree-resolved full-counting
analysis.

\section{Aliased Dyson--Morris constant-term representation}
\label{sec:constantterm}

The discrete Selberg representation of Sec.~\ref{sec:cauchy} is valid for
every real \(\alpha>0\). When \(\alpha\) is a positive integer, the
Vandermonde weight becomes a Laurent polynomial with finite Fourier
bandwidth. The discrete root-of-unity measure can then be replaced, inside
constant-term extraction, by the finite set of Fourier aliases visible to
that polynomial. This yields an exact constant-term representation built from the Dyson Laurent polynomial and a finite one-body alias kernel.

\subsection{Root-of-unity measure and Fourier aliases}

Recall the decomposition
\(\Omega_{2L}=X\sqcup Y\), where \(X\) and \(Y\) are respectively the
roots of \(z^L-1\) and \(z^L+1\). Assigning fugacity \(u\) to an occupied
root in \(X\) and unit weight to one in \(Y\) is equivalent to the
one-root weight
\begin{equation}
\label{eq:wu-polynomial}
w_u(z)
=
\frac{1+u}{2}
+
\frac{u-1}{2}z^L,
\qquad
z\in\Omega_{2L}.
\end{equation}
Indeed, \(z^L=1\) on \(X\) and \(z^L=-1\) on \(Y\), so that
\[
\prod_{z\in U}w_u(z)=u^{|U\cap X|}
\]
for every half-filled subset \(U\subseteq\Omega_{2L}\).

Because the Vandermonde vanishes whenever two coordinates coincide, the
subset sum in Eq.~\eqref{eq:discrete-selberg} can be replaced by an
ordered root sum and encoded by the weighted periodic Dirac comb
\begin{equation}
\label{eq:weighted-comb}
\mathscr A_{L,u}(\theta)
=
2\pi
\sum_{a=0}^{2L-1}
w_u(\zeta_a)\,
\delta_{2\pi}
\left(
\theta-\frac{\pi a}{L}
\right),
\qquad
\zeta_a=e^{\pi i a/L}.
\end{equation}
The subset sum may therefore be written as an unrestricted ordered sum,
with the factor \(1/L!\) removing the permutations of each
\(L\)-element subset:
\begin{equation}
\label{eq:ordered-root-sum}
\begin{aligned}
\cZ_{\alpha,L}(u)
={}&
\frac{L^{-\alpha L}}{L!}
\sum_{a_1,\ldots,a_L=0}^{2L-1}
\left[
\prod_{j=1}^{L}w_u(\zeta_{a_j})
\right]
\\
&\times
\prod_{1\leq i<j\leq L}
\abs{\zeta_{a_i}-\zeta_{a_j}}^{2\alpha}.
\end{aligned}
\end{equation}

It is convenient to normalize the weighted comb as
\begin{equation}
\label{eq:normalized-weighted-comb}
\widehat{\mathscr A}_{L,u}(\theta)
=
\frac{\mathscr A_{L,u}(\theta)}{2L}.
\end{equation}
Using Eq.~\eqref{eq:weighted-comb}, the ordered root sum becomes the
exact circular integral
\begin{equation}
\label{eq:normalized-comb-integral-main}
\begin{aligned}
\cZ_{\alpha,L}(u)
={}&
\frac{2^L}{L!\,L^{(\alpha-1)L}}
\int_{0}^{2\pi}
\prod_{j=1}^{L}
\left[
\frac{\dd\theta_j}{2\pi}\,
\widehat{\mathscr A}_{L,u}(\theta_j)
\right]
\\
&\times
\prod_{1\leq i<j\leq L}
\abs{e^{\ii\theta_i}-e^{\ii\theta_j}}^{2\alpha}.
\end{aligned}
\end{equation}
Equation~\eqref{eq:normalized-comb-integral-main} is a circular
Selberg-type integral with a normalized weighted root-of-unity measure.
It is exactly equivalent to the discrete subset sum for every real
\(\alpha>0\). The restriction to positive integer \(\alpha\) enters only
when the Vandermonde weight is subsequently converted into a finite
Laurent polynomial and treated by constant-term extraction.

The Fourier expansion of the normalized comb is

\begin{equation}
\label{eq:comb-Fourier-series}
\widehat{\mathscr A}_{L,u}(\theta)
=
\sum_{q\in\mathbb Z}
c_q(u)e^{iqL\theta},
\end{equation}
with
\begin{equation}
\label{eq:cq-repeated}
c_q(u)
=
\begin{cases}
\dfrac{1+u}{2}, & q\ \mathrm{even},\\[2mm]
\dfrac{u-1}{2}, & q\ \mathrm{odd}.
\end{cases}
\end{equation}
Thus the root-of-unity measure contains aliases at momenta \(qL\), with
the parity of \(q\) distinguishing the uniform and staggered components
of the checkerboard weight.

\newcommand{\tablecellcolor}[2]{%
  \begingroup
  \setlength{\fboxsep}{3pt}%
  \colorbox{#1}{%
    \parbox[c]{\dimexpr\linewidth-2\fboxsep\relax}{#2}%
  }%
  \endgroup
}

\begin{table*}[t]
\centering
\scriptsize
\setlength{\tabcolsep}{4.5pt}
\renewcommand{\arraystretch}{1.55}

\begingroup
\setlength{\fboxsep}{4pt}

\colorbox{TFILightBlue}{%
\begin{tabular}{
@{}
p{0.055\textwidth}
p{0.315\textwidth}
p{0.235\textwidth}
p{0.305\textwidth}
@{}
}

\toprule

\textcolor{TFIBlue}{\bfseries \(\alpha\)}
&
\textcolor{TFIBlue}{\bfseries \(\cZ_{\alpha,L}(u)\)}
&
\textcolor{TFIBlue}{\bfseries \(\cZ_{\alpha,L}(1)\)}
&
\textcolor{TFIBlue}{\bfseries Exact mechanism}
\\

\midrule


\centering
\textcolor{TFIViolet}{\(\mathbf{0}\)}
&
\tablecellcolor{TableAccent!5}{%
\centering
\(\displaystyle
\sum_{k=0}^{L}
\binom{L}{k}^{2}u^k
\)
}
&
\tablecellcolor{TableAccent!5}{%
\centering
\(\displaystyle
\binom{2L}{L}
\)
}
&
\tablecellcolor{TableAccent!5}{%
Formal counting endpoint.
}
\\[0.8ex]


\centering
\raisebox{-1.80\baselineskip}[0pt][0pt]{%
\textcolor{TFIViolet}{\(\mathbf{\frac12}\)}
}
&
\tablecellcolor{TFIBlue!7}{%
\centering
\(\displaystyle
\prod_{j=1}^{L/2}
\left[
1+u^2
+
2u\sec\frac{\pi(2j-1)}{2L}
\right]
\)
\cite{KhassehRamirezTrinoRajabpour2026}
}
&
\tablecellcolor{TFIBlue!7}{%
\centering
\(\displaystyle
2^{L/2}
\prod_{j=1}^{L/2}
\left[
1+\sec\frac{\pi(2j-1)}{2L}
\right]
\)
\cite{RamirezTrinoRajabpour2026}
}
&
\tablecellcolor{TFIBlue!7}{%
\textcolor{TableAccent}{\textit{Closed-product route:}}
ordered-root de Bruijn Pfaffian and checkerboard
\(2\times2\)-block factorization.
}
\\[0.8ex]


&
\tablecellcolor{TFIViolet!6}{%
\centering
\(\displaystyle
(-1)^L
\Pf\!\left[
\begin{smallmatrix}
u\,\mathbb A(G) & I_{2L}
\\
-I_{2L} & -\mathbb J^{\mathrm{PBC}}_{2L}
\end{smallmatrix}
\right]
\)
}
&
\tablecellcolor{TFIViolet!6}{%
\centering
\(\displaystyle
(-1)^L
\Pf\!\left[
\begin{smallmatrix}
\mathbb A(G) & I_{2L}
\\
-I_{2L} & -\mathbb J^{\mathrm{PBC}}_{2L}
\end{smallmatrix}
\right]
\)
}
&
\tablecellcolor{TFIViolet!6}{%
\textcolor{TableAccent}{\textit{Direct-minor route:}}
\(4L\times4L\) balanced-all-minors Pfaffian with the
periodic selector.
}
\\[0.8ex]


\centering
\textcolor{TFIViolet}{\(\mathbf{1}\)}
&
\tablecellcolor{TableAccent!4}{%
\centering
\(\displaystyle
(1+u)^L
\)
}
&
\tablecellcolor{TableAccent!4}{%
\centering
\(\displaystyle
2^L
\)
}
&
\tablecellcolor{TableAccent!4}{%
Cauchy--Binet determinant and root orthogonality for the
discrete \(\beta=2\) ensemble.
}
\\[0.8ex]


\centering
\textcolor{TFIViolet}{\(\mathbf{2}\)}
&
\tablecellcolor{TFIBlue!4}{%
\centering
\(\displaystyle
\prod_{r=1}^{L/2}
\left[
(1+u)^2
-
4u\left(\frac{2r-1}{L}\right)^2
\right]
\)
}
&
\tablecellcolor{TFIBlue!4}{%
\centering
\(\displaystyle
\frac{(2L)!}{L!\,L^L}
\)
\cite{Stephan2014}
}
&
\tablecellcolor{TFIBlue!4}{%
Multiplicity-two confluent de Bruijn Pfaffian for the
discrete \(\beta=4\) ensemble.
}
\\[0.8ex]


\centering
\textcolor{TFIViolet}{\(\mathbf{4}\)}
&
\tablecellcolor{TableAccent!5}{%
\centering
Eq.~\eqref{eq:alpha4-jack-formula}
}
&
\tablecellcolor{TableAccent!5}{%
\centering
\(\displaystyle
\frac{2^L}{L^{2L}}
\bigl[(2L-1)!!\bigr]^2
=
2^{-L}\cZ_{2,L}(1)^2
\)
\cite{Stephan2014}
}
&
\tablecellcolor{TableAccent!5}{%
Generic \(u\): neutral seven-charge shifted-Dyson and rectangular
inverse Jack--Kostka representation. At \(u=1\): complementary
middle-minor collapse.
}
\\

\bottomrule

\end{tabular}%
}

\endgroup

\caption{
Special indices of the fugacity-resolved discrete Selberg problem.
Closed products occur at \(\alpha=\tfrac12,1,2\).
At \(\alpha=\tfrac12\), the ordered-root de Bruijn route yields the
closed product, while an independent direct \(4L\times4L\)
balanced-all-minors Pfaffian represents the complete fugacity polynomial.
At \(\alpha=4\), the generic-fugacity representation is exact but is
neither a closed product nor an ordinary-Pfaffian compression, whereas
the unit-fugacity result collapses to a product.
Citations in the second and third column refer to previously known unrefined
counterparts; the fugacity-resolved formulas and the derivations
presented here are results of this work.
}
\label{tab:solvable-indices}
\end{table*}

\subsection{Finite alias kernel and constant-term representation}

We now assume \(\alpha\in\mathbb Z_{>0}\). Writing
\(x_j=e^{i\theta_j}\), the circular Vandermonde weight becomes the Dyson
Laurent polynomial
\begin{equation}
\label{eq:Vandermonde-Dalpha}
D_\alpha(x)
:=
\prod_{\substack{i,j=1\\i\neq j}}^L
\left(
1-\frac{x_i}{x_j}
\right)^\alpha.
\end{equation}
For each variable \(x_j\), every monomial in \(D_\alpha(x)\) has exponent
in the range
\begin{equation}
\label{eq:bandwidth}
-\alpha(L-1)
\leq m_j\leq
\alpha(L-1).
\end{equation}
The normalized comb contributes powers \(x_j^{qL}\). If
\(|q|\geq\alpha\), then
\[
|qL|
\geq
\alpha L
>
\alpha(L-1),
\]
so that such a mode cannot be cancelled by any monomial of
\(D_\alpha(x)\) and gives no constant-term contribution. The infinite
Fourier series therefore truncates exactly to the finite alias kernel
\begin{equation}
\label{eq:finite-alias-kernel}
K_{\alpha,u}(x)
=
\sum_{q=-(\alpha-1)}^{\alpha-1}
c_q(u)x^{qL}.
\end{equation}
This replacement is exact only inside constant-term extraction:
\(K_{\alpha,u}\) is the part of the Dirac comb visible to
\(D_\alpha\), not a pointwise representation of the comb itself.

Applying the truncation independently to all \(L\) variables gives the
main constant-term representation
\begin{equation}
\label{eq:CTMain}
\cZ_{\alpha,L}(u)
=
\frac{2^L}{L!\,L^{(\alpha-1)L}}
\CT_{x_1,\ldots,x_L}
\left[
D_\alpha(x)
\prod_{j=1}^L
K_{\alpha,u}(x_j)
\right],
\end{equation}
where $\alpha\in\mathbb Z_{>0}$ and \(\CT_{x_1,\ldots,x_L}\) extracts the coefficient of
\(x_1^0\cdots x_L^0\). The prefactor follows from the normalization of
the comb and the ordered-root representation; the derivation is given in
Appendix~\ref{app:aliased-dyson}.

Equation~\eqref{eq:CTMain} differs from the ordinary Dyson identity
\begin{equation}
\label{eq:ordinary-Dyson-CT}
\CT_{x_1,\ldots,x_L}D_\alpha(x)
=
\frac{(\alpha L)!}{(\alpha!)^L}
\end{equation}
through the additional one-body factor
\(\prod_jK_{\alpha,u}(x_j)\). At unit fugacity, the odd aliases vanish,
and
\begin{equation}
\label{eq:unrefined-alias-kernel}
K_{\alpha,1}(x)
=
\sum_{\substack{
q=-(\alpha-1)\\
q\ {\rm even}
}}^{\alpha-1}
x^{qL}.
\end{equation}
Consequently,
\[
K_{1,1}(x)=K_{2,1}(x)=1,
\]
whereas
\[
K_{3,1}(x)=K_{4,1}(x)
=
1+x^{2L}+x^{-2L}.
\]
The ordinary Dyson identity therefore evaluates the unrefined problem
directly only when the visible-alias kernel is constant.

\subsection{Shifted Dyson coefficients and charge neutrality}

Expanding the finite kernels in Eq.~\eqref{eq:CTMain} gives a finite sum
of shifted Dyson coefficients. Define
\begin{equation}
\label{eq:shifted-Dyson-coefficient}
\mathscr C_{\alpha,L}(q_1,\ldots,q_L)
=
\CT_{x_1,\ldots,x_L}
\left[
D_\alpha(x)
\prod_{j=1}^L x_j^{q_jL}
\right].
\end{equation}
Then
\begin{equation}
\label{eq:shifted-Dyson-expansion}
\begin{aligned}
\cZ_{\alpha,L}(u)
&=
\frac{2^L}{L!\,L^{(\alpha-1)L}}
\sum_{q_1,\ldots,q_L=-(\alpha-1)}^{\alpha-1}
\\[-1mm]
&\quad\times
\left[
\prod_{j=1}^L c_{q_j}(u)
\right]
\mathscr C_{\alpha,L}(q_1,\ldots,q_L).
\end{aligned}
\end{equation}
Only neutral alias configurations contribute:
\begin{equation}
\label{eq:alias-neutrality}
\mathscr C_{\alpha,L}(q_1,\ldots,q_L)=0
\qquad\text{unless}\qquad
\sum_{j=1}^Lq_j=0.
\end{equation}
Indeed, \(D_\alpha(x)\) is invariant under the common rescaling
\(x_j\mapsto t x_j\), whereas the inserted monomial acquires the factor
\(t^{L\sum_jq_j}\). A nonzero constant term is therefore possible only
at zero total alias charge.

At unit fugacity, the case \(\alpha=2\) contains only the zero-alias
sector and reduces to the ordinary Dyson identity. For \(\alpha=4\), the
charges \(q_j\in\{-2,0,2\}\) remain visible, and nontrivial neutral
configurations containing both \(+2\) and \(-2\) aliases survive. These
sectors obstruct a direct ordinary-Dyson evaluation and motivate the
additional structures developed in Sec.~\ref{sec:solvable-indices}.

Further details, including the Fourier derivation of the comb,
the finite alias truncation, trigonometric forms of the alias kernel,
and the first small-\(\alpha\) sector decompositions, are collected in
Appendix~\ref{app:aliased-dyson}.


\section{Special indices and exact finite-size formulas}
\label{sec:solvable-indices}

We now collect the exact finite-size results at the special indices where
the fugacity-resolved discrete Selberg problem simplifies. Unless stated
otherwise, the product formulas are written for positive even \(L\). We
distinguish three levels of exactness: closed products, polynomial-size
determinant or Pfaffian compressions, and exact structural
representations, such as neutral alias sums and Jack--Kostka expansions,
which do not by themselves imply efficient evaluation. Since the
Vandermonde exponent is \(\beta=2\alpha\), the indices
\(\alpha=\tfrac12,1,2\) correspond to the classical values
\(\beta=1,2,4\), respectively. The results are summarized in
Table~\ref{tab:solvable-indices}.

\subsection{The formal counting endpoint
\texorpdfstring{\(\alpha=0\)}{alpha=0}}

At \(\alpha=0\), the generating function records only the support of the
balanced-minor distribution. Since every balanced minor of the Cauchy
matrix in Eq.~\eqref{eq:Gdefinition} is nonzero,
\begin{equation}
\label{eq:alpha0-formula}
\cZ_{0,L}(u)
=
\sum_{k=0}^{L}\binom{L}{k}^{2}u^k.
\end{equation}
At \(u=1\), Vandermonde's identity gives
\[
\cZ_{0,L}(1)=\binom{2L}{L}.
\]
Thus this formal endpoint counts balanced minor labels without resolving
their weights.

\subsection{Two Pfaffian routes at
\texorpdfstring{\(\alpha=\tfrac12\)}{alpha=1/2}}
\label{subsec:alpha-half-special}

At \(\alpha=\tfrac12\), the fugacity-resolved all-minors polynomial
admits two complementary ordinary-Pfaffian representations. The first
acts on the doubled-root ensemble and leads directly to the closed
product formula. The second acts on the balanced minors themselves and
provides an independent direct Pfaffian compression of the full
fugacity polynomial.

For the doubled-root route, ordering the selected roots around the circle
turns the absolute Vandermonde into a single determinant. The finite
de Bruijn identity then compresses the subset sum to an ordinary
Pfaffian. For the checkerboard weight, the Pfaffian decomposes into
reflected \(2\times2\) blocks, yielding, for positive even \(L\),
\begin{equation}
\label{eq:alpha-half-product}
\cZ_{1/2,L}(u)
=
\prod_{j=1}^{L/2}
\left[
1+u^2
+
2u\sec\left(\frac{\pi(2j-1)}{2L}\right)
\right].
\end{equation}
The unrefined result follows by setting \(u=1\). The generic-weight
ordered-root Pfaffian and its checkerboard reduction are derived in
Appendix~\ref{app:generic-weight}.

There is also a direct all-minors Pfaffian representation that does not
require passing to the doubled-root ensemble. For each \(j\in[L]\),
introduce two lifted labels
\[
\mathsf r_j=2j,
\qquad
\mathsf c_j=2j+1,
\]
ordered as
\[
(\mathsf r_0,\mathsf c_0,\mathsf r_1,\mathsf c_1,\ldots,
\mathsf r_{L-1},\mathsf c_{L-1}).
\]
Define the antisymmetric lift \(\mathbb A(G)\) of the half-shift matrix by
\begin{equation}
\label{eq:alpha-half-lift-main}
\mathbb A(G)_{\mathsf r_j,\mathsf c_k}=G_{jk},
\qquad
\mathbb A(G)_{\mathsf c_k,\mathsf r_j}=-G_{jk},
\end{equation}
with all \(\mathsf r\)-\(\mathsf r\) and
\(\mathsf c\)-\(\mathsf c\) entries equal to zero.

Let \(\mathbb J^{(0)}_{2L}\) be the standard antisymmetric selector,
\begin{equation}
\label{eq:alpha-half-standard-selector-main}
\left(\mathbb J^{(0)}_{2L}\right)_{\mu\nu}
=
\begin{cases}
+1,&\mu<\nu,\\
0,&\mu=\nu,\\
-1,&\mu>\nu,
\end{cases}
\end{equation}
and introduce the pair-swap and diagonal sign matrices
\begin{equation}
\label{eq:alpha-half-selector-data-main}
\mathsf P_{\rm sw}
=
\bigoplus_{j=0}^{L-1}
\begin{pmatrix}
0&1\\
1&0
\end{pmatrix},
\qquad
D_{\rm P}
=
\bigoplus_{j=0}^{L-1}
\begin{pmatrix}
(-1)^{j+1}&0\\
0&(-1)^j
\end{pmatrix}.
\end{equation}
The periodic selector is
\begin{equation}
\label{eq:alpha-half-periodic-selector-main}
\mathbb J^{\rm PBC}_{2L}
=
D_{\rm P}\,
\mathsf P_{\rm sw}
\mathbb J^{(0)}_{2L}
\mathsf P_{\rm sw}^{T}
D_{\rm P}.
\end{equation}
Then the full fugacity-resolved polynomial has the exact direct
representation
\begin{equation}
\cZ_{1/2,L}(u)
=
(-1)^L
\Pf
\begin{pmatrix}
u\,\mathbb A(G)&I_{2L}\\
-I_{2L}&-\mathbb J^{\rm PBC}_{2L}
\end{pmatrix}.
\label{eq:alpha-half-direct-pfaffian}
\end{equation}
The Pfaffian in Eq.~\eqref{eq:alpha-half-direct-pfaffian} has dimension
\(4L\times4L\). Unlike the product form
\eqref{eq:alpha-half-product}, the direct Pfaffian identity itself does
not require \(L\) to be even.

The role of the factor \(u\) is transparent. A nonvanishing principal
sector contains \(k\) lifted row labels and \(k\) lifted column labels,
and hence has dimension \(2k\). By Pfaffian homogeneity,
\[
\Pf\!\left[
\bigl(u\mathbb A(G)\bigr)_{\mathscr I}
\right]
=
u^k
\Pf\!\left[
\mathbb A(G)_{\mathscr I}
\right].
\]
Thus the factor multiplying \(\mathbb A(G)\) generates precisely the
balanced Majorana degree appearing in the definition of
\(\cZ_{\alpha,L}(u)\). The proof, including the sign structure of the periodic Cauchy minors and
the construction of the periodic selector, is given in
Appendix~\ref{app:alpha-half-direct}.

\subsection{The determinant case
\texorpdfstring{\(\alpha=1\)}{alpha=1}}

At \(\alpha=1\), Cauchy--Binet turns the root-subset sum into an
\(L\times L\) moment determinant,
\begin{equation}
\label{eq:alpha1-moment-det}
\cZ_{1,L}(u)
=
L^{-L}\det M(u),
\qquad
M_{rs}(u)
=
\sum_{z\in\Omega_{2L}}w_u(z)\,z^{r-s},
\end{equation}
with \(r,s=0,\ldots,L-1\). Because \(|r-s|\leq L-1\), the staggered
Fourier component of \(w_u\) does not contribute. Root orthogonality gives
\[
M_{rs}(u)=L(1+u)\delta_{rs},
\]
and therefore
\begin{equation}
\label{eq:alpha1-formula}
\cZ_{1,L}(u)
=
(1+u)^L.
\end{equation}

\subsection{The confluent Pfaffian case
\texorpdfstring{\(\alpha=2\)}{alpha=2}}

At \(\alpha=2\), the fourth power of the Vandermonde admits a
multiplicity-two confluent de Bruijn organization. For the checkerboard
weight, the corresponding moment matrix decomposes into reflected
\(4\times4\) blocks. Evaluating these blocks gives
\begin{equation}
\label{eq:alpha2-product}
\cZ_{2,L}(u)
=
\prod_{r=1}^{L/2}
\left[
(1+u)^2
-
4u\left(\frac{2r-1}{L}\right)^2
\right].
\end{equation}
At unit fugacity,
\begin{equation}
\label{eq:alpha2-u1}
\cZ_{2,L}(1)
=
\frac{(2L)!}{L!\,L^L}
=
\frac{2^L(2L-1)!!}{L^L}.
\end{equation}
After accounting for the normalization convention, this agrees with the
half-filled discrete Dyson-gas result of Ref.~\cite{Stephan2014}. The
generic-weight \(\beta=4\) Pfaffian is derived in
Appendix~\ref{app:generic-weight}. Its checkerboard block factorization,
together with an independent direct all-minors Pfaffian compression, is
given in Appendix~\ref{app:alpha-two}.

\subsection{The exceptional index
\texorpdfstring{\(\alpha=4\)}{alpha=4}}
\label{subsec:alpha4-special}

The index \(\alpha=4\) is exceptional. At generic checkerboard fugacity,
it does not reduce to a closed product or to an ordinary Pfaffian.
Instead, the aliased Dyson representation gives an exact finite sum over
neutral alias sectors. At \(u=1\), however, a complementary middle-minor
identity collapses the answer to a product.

\subsubsection{Generic fugacity: neutral seven-charge representation}

At \(\alpha=4\), the visible aliases carry charges
\(q=-3,\ldots,3\). Charge neutrality organizes the expansion of
Eq.~\eqref{eq:CTMain} by occupation vectors
\begin{equation}
\label{eq:neutral-set}
\mathcal N_L
=
\left\{
\nu\in\mathbb Z_{\ge0}^{7}
\;\middle|\;
\substack{
\displaystyle\sum_{q=-3}^{3}\nu_q=L,\\[-1mm]
\displaystyle\sum_{q=-3}^{3}q\nu_q=0
}
\right\},
\qquad
\nu=(\nu_{-3},\ldots,\nu_3).
\end{equation}
To each neutral sector, associate the partition
\begin{equation}
\label{eq:lambda-nu}
\lambda(\nu)
=
\operatorname{sort}
\bigl(
\underbrace{0,\ldots,0}_{\nu_{-3}},
\underbrace{L,\ldots,L}_{\nu_{-2}},
\ldots,
\underbrace{6L,\ldots,6L}_{\nu_{3}}
\bigr).
\end{equation}
Neutrality implies
\[
|\lambda(\nu)|=3L^2,
\]
which equals the size of the rectangle \((3L)^L\). Identifying the
shifted Dyson coefficients with rectangular inverse Jack--Kostka
coefficients gives
\begin{equation}
\label{eq:alpha4-jack-formula}
\begin{aligned}
\cZ_{4,L}(u)
&=
\frac{2^L}{L^{3L}}
\frac{(4L)!}{L!(4!)^L}
\sum_{\nu\in\mathcal N_L}
\left[
\prod_{q=-3}^{3}c_q(u)^{\nu_q}
\right]
\\[-1mm]
&\quad\times
K^{-1}_{\lambda(\nu),(3L)^L}(1/4).
\end{aligned}
\end{equation}
Here
\(K^{-1}_{\lambda(\nu),(3L)^L}(1/4)\) is the inverse Jack--Kostka
coefficient in the convention \(\alpha_{\rm Jack}=1/4\).
Equation~\eqref{eq:alpha4-jack-formula} is exact for arbitrary fugacity
\(u\), but it is neither a product formula nor an ordinary-Pfaffian
compression. Its derivation is given in
Appendix~\ref{app:alpha-four-jack}.

\subsubsection{Unit fugacity: complementary middle-minor collapse}

At \(u=1\), the checkerboard field disappears. The complementary
middle-minor identity derived in Appendix~\ref{app:middle-minor} gives
\begin{equation}
\label{eq:alpha4-u1}
\cZ_{4,L}(1)
=
\frac{2^L}{L^{2L}}
\bigl[(2L-1)!!\bigr]^2
=
2^{-L}\cZ_{2,L}(1)^2.
\end{equation}
Thus \(\alpha=4\) is product-solvable only at unit fugacity: the generic
problem remains a neutral shifted-Dyson/Jack--Kostka sum, whereas the
unrefined result collapses to the square of the \(\alpha=2\) answer, up
to the factor \(2^{-L}\). After accounting for the normalization
convention, Eq.~\eqref{eq:alpha4-u1} agrees with the half-filled discrete
Dyson-gas result of Ref.~\cite{Stephan2014}.

The three classical indices therefore admit closed product formulas,
whereas \(\alpha=4\) is product-solvable only at unit fugacity. The next
section compares the determinant, Pfaffian, complementary-minor, and
hyperpfaffian mechanisms underlying these results.


\section{Solvability mechanisms and computational status}
\label{sec:solvability-mechanisms}

Section~\ref{sec:solvable-indices} identified the special indices at which
the fugacity-resolved all-minors problem simplifies. We now organize the
algebraic mechanisms behind those formulas and determine which structures
extend beyond the isolated product-solvable points.

Two hierarchies are relevant. The first is the confluent perfect-square
sequence \(2\alpha=q^2\), which contains the Pfaffian points
\(\alpha=\tfrac12\) and \(\alpha=2\) and extends to hyperpfaffian
representations. The second is the positive-integer complementary-minor
and shifted-Dyson hierarchy, which contains the exceptional
\(\alpha=4\) structure even though \(\alpha=4\) does not belong to the
perfect-square sequence. The determinant point \(\alpha=1\) remains
separate, while \(\alpha=0\) is only a formal counting endpoint.

Statements about unavailable reductions refer only to the mechanisms
developed here; their computational implications are discussed separately
in Sec.~\ref{subsec:computational-status}. The arbitrary-weight determinant
and Pfaffian compressions at the classical points are proved in
Appendix~\ref{app:generic-weight}.
At \(\alpha=\tfrac12\),
Appendix~\ref{app:alpha-half-direct} gives an independent direct
balanced-all-minors \(4L\times4L\) Pfaffian for the complete fugacity
polynomial. At \(\alpha=2\),
Appendix~\ref{app:alpha-two} gives both an independent direct all-minors
\(4L\times4L\) Pfaffian compression and the checkerboard
\(4\times4\)-block factorization of the weighted-root Pfaffian.

\subsection{Perfect-square powers and the confluent hierarchy}
\label{subsec:confluent-hierarchy}

The Pfaffian construction at \(\alpha=2\) is the second member of the
perfect-square sequence
\begin{equation}
\label{eq:sec6-square-sequence}
2\alpha=q^2,
\qquad
\alpha=\frac{q^2}{2},
\qquad
q\in\mathbb Z_{>0}.
\end{equation}
At these indices,
\(\abs{\det G[S,T]}^{2\alpha}=\abs{\det G[S,T]}^{q^2}\), and the power
\(q^2\) can be linearized by a multiplicity-\(q\) confluent Cauchy
construction.

For \(S\subseteq[L]\), let \(S^{[q]}\) denote its confluent lift, obtained
by replacing each \(s\in S\) with
\((s,0),\ldots,(s,q-1)\); define \(T^{[q]}\) analogously. The full
confluent matrix \(\mathbb M^{(q)}\) has size \(qL\times qL\), while
\(\mathbb M^{(q)}[S^{[q]},T^{[q]}]\) has size \(qk\times qk\) when
\(\abs{S}=\abs{T}=k\). With the ordering and gauge convention fixed in
Appendix~\ref{app:perfect-square-hyperpfaffian},
\begin{equation}
\label{eq:sec6-confluent-linearization}
\det \mathbb M^{(q)}[S^{[q]},T^{[q]}]
=
\bigl(\det G[S,T]\bigr)^{q^2}.
\end{equation}
For even \(q\), this is exactly $\abs{\det G[S,T]}^{q^2}$.
For odd \(q\), it retains the sign of the real minor. The positive
absolute-value sum is recovered instead from the equivalent
absolute-Vandermonde root-subset representation and its phase-corrected
exterior-algebra construction, as detailed in
Appendix~\ref{app:perfect-square-hyperpfaffian}.

For even \(L\), the resulting exterior-algebra compression depends on the
parity of \(q\):
\begin{equation}
\label{eq:sec6-hyperpfaffian-parity}
\begin{aligned}
q\ \mathrm{even}:&\qquad \text{exterior degree }q,\\
q\ \mathrm{odd}:&\qquad \text{exterior degree }2q.
\end{aligned}
\end{equation}
For even \(q\), the resulting partition function is the top-form
coefficient of an exterior \(q\)-form and hence admits a
\(q\)-hyperpfaffian representation. For odd \(q\) and even \(L\), the
odd-degree node forms must first be paired, producing an exterior
\(2q\)-form and therefore a \(2q\)-hyperpfaffian representation. The
cases \(q=1\) and \(q=2\) reduce to ordinary Pfaffians, whereas
\(q\geq3\) gives genuine higher hyperpfaffians.

The precise exterior forms, their checkerboard weights, the circular
phase factor, and the resulting hyperpfaffian formulas are defined and
derived in Appendix~\ref{app:perfect-square-hyperpfaffian}. Related
hyperpfaffian representations for square-\(\beta\) ensembles were
developed in Refs.~\cite{Sinclair2012,SinclairWells}; here they are
adapted to the half-filled discrete root ensemble and to the
absolute-value structure of the minors.

\begin{table}[t]
\centering
\scriptsize
\setlength{\tabcolsep}{1.5pt}
\renewcommand{\arraystretch}{1.28}

\begingroup
\setlength{\fboxsep}{3pt}

\colorbox{TFILightBlue}{%
\begin{tabular}{
@{}
p{0.07\columnwidth}
p{0.16\columnwidth}
p{0.18\columnwidth}
p{0.38\columnwidth}
@{}
}
\toprule

\centering\textcolor{TFIBlue}{\bfseries \(q\)}
&
\centering\textcolor{TFIBlue}{\bfseries \(\alpha=q^2/2\)}
&
\centering\textcolor{TFIBlue}{\bfseries Exterior degree}
&
\textcolor{TFIBlue}{\bfseries Natural exact object}
\tabularnewline
\midrule

\centering\textcolor{TFIViolet}{\(\mathbf{1}\)}
&
\centering\(\displaystyle \frac12\)
&
\centering\(2\)
&
Ordinary Pfaffian.
\tabularnewline[0.6mm]

\centering\textcolor{TFIViolet}{\(\mathbf{2}\)}
&
\centering\(2\)
&
\centering\(2\)
&
Ordinary Pfaffian.
\tabularnewline[0.6mm]

\centering\textcolor{TFIViolet}{\(\mathbf{3}\)}
&
\centering\(\displaystyle \frac92\)
&
\centering\(6\)
&
Hyperpfaffian of a six-form.
\tabularnewline[0.6mm]

\centering\textcolor{TFIViolet}{\(\mathbf{4}\)}
&
\centering\(8\)
&
\centering\(4\)
&
Hyperpfaffian of a four-form.
\tabularnewline[0.6mm]

\centering\textcolor{TFIViolet}{\(\mathbf{5}\)}
&
\centering\(\displaystyle \frac{25}{2}\)
&
\centering\(10\)
&
Hyperpfaffian of a ten-form.
\tabularnewline

\bottomrule
\end{tabular}%
}

\endgroup

\caption{
First members of the perfect-square confluent hierarchy for even \(L\).
Exterior degree two gives an ordinary Pfaffian; higher exterior degree
gives a hyperpfaffian representation.
}
\label{tab:sec6-square-hierarchy}
\end{table}

The determinant point \(\alpha=1\) and the exceptional point
\(\alpha=4\) do not belong to this perfect-square hierarchy.

\subsection{Complementary-minor hierarchy and the exceptional
\texorpdfstring{\(\alpha=4\)}{alpha=4} point}
\label{subsec:complementary-hierarchy}

A second organization of the positive integer indices follows from
complementary middle minors. For an \(L\)-element subset
\(U\subseteq\Omega_{2L}\), define
\begin{equation}
\label{eq:sec6-MU-def}
M_U
:=
\frac{\abs{\Delta(U)}^4}{(2L)^L}
=
\left|
\det_{u\in U,\,v\in U^c}
\frac{1}{u-v}
\right|.
\end{equation}
Thus \(M_U\) is the absolute value of the normalized Cauchy middle minor
associated with the bipartition
\(U\sqcup U^c=\Omega_{2L}\). The complementary Cauchy reduction is derived
in Appendix~\ref{app:middle-minor}.

The even positive integers are moments of \(M_U\):
\begin{equation}
\label{eq:sec6-even-complementary-hierarchy}
\cZ_{2p,L}(u)
=
\frac{2^{pL}}{L^{pL}}
\sum_{\substack{U\subseteq\Omega_{2L}\\ \abs{U}=L}}
u^{\abs{U\cap X}}M_U^p,
\qquad
p\geq1.
\end{equation}
For odd positive integers, one additional Vandermonde square remains:
\begin{equation}
\label{eq:sec6-odd-complementary-hierarchy}
\cZ_{2p+1,L}(u)
=
\frac{2^{pL}}{L^{(p+1)L}}
\sum_{\substack{U\subseteq\Omega_{2L}\\ \abs{U}=L}}
u^{\abs{U\cap X}}
\abs{\Delta(U)}^2M_U^p,
~~
p\geq0.
\end{equation}

The first two even moments correspond to \(\alpha=2\) and
\(\alpha=4\), respectively. At generic fugacity, the latter is the
neutral shifted-Dyson/Jack--Kostka problem of
Sec.~\ref{subsec:alpha4-special}, whereas at unit fugacity the
complementary-minor identity gives Eq.~\eqref{eq:alpha4-u1}. Thus
\(\alpha=4\) belongs to the integer complementary-minor hierarchy but not
to the perfect-square confluent hierarchy.

The unit-fugacity identity is global and must not be interpreted as the
fugacity-resolved factorization
\(\cZ_{4,L}(u)=2^{-L}\cZ_{2,L}(u)^2\). The square of
\(\cZ_{2,L}(u)\) contains two independently selected subsets, whereas
\(\cZ_{4,L}(u)\) requires the same subset in both replicas. A direct local
doubling of the \(\alpha=2\) Pfaffian would therefore require a diagonal
replica projector enforcing equality of the two subsets. This projector
is not a matchgate tensor: it violates the four-leg Pl\"ucker relation.
Hence this particular local Gaussian construction does not produce a
generic-\(u\) Pfaffian compression. This obstruction does not exclude
nonlocal transformations or enlarged auxiliary constructions; its precise
scope is discussed in Appendix~\ref{app:no-go-pfaffian}.

\subsection{Position of the seven-charge family}
\label{subsec:dyson-jack-literature}

For \(\alpha=4\), the shifted coefficients defined in
Eq.~\eqref{eq:shifted-Dyson-coefficient} involve neutral charges
\[
q_i\in\{-3,-2,-1,0,1,2,3\},
\qquad
\sum_{i=1}^{L}q_i=0.
\]
The Dyson exponent remains fixed at four, whereas the monomial shifts
scale linearly with \(L\). Grouping charge sequences by their
multiplicities gives the seven-charge family in
Eq.~\eqref{eq:alpha4-jack-formula}.

The zero-shift sector is Dyson's constant term
\cite{Dyson1962,Wilson1962}. Kadell-type identities, fixed-layer
coefficients, recursions, and generalized \(q\)-Dyson identities provide
closely related results
\cite{Kadell2000,SillsZeilberger2006,LvXinZhou2009,EkhadZeilberger2013,
KLW2015,Zhou2020,Zhou2021,HuangJiangZhou2026}. Rectangular Jack and
Macdonald theory, together with hyperdeterminantal representations,
provide the natural symmetric-function setting
\cite{Macdonald,Stanley1989,Cai2014,LuqueThibon2003,Matsumoto2006}.

The family required here is more strongly scaled than the standard
fixed-shift cases. Its partitions satisfy
\begin{equation}
\label{eq:sec6-seven-level-family}
\lambda(\nu)
\in
\{0,L,2L,3L,4L,5L,6L\}^{L},
\qquad
\abs{\lambda(\nu)}=3L^2.
\end{equation}
Thus both the partition parts and the monomial shifts grow with the number
of variables. Existing product theorems do not directly evaluate this
seven-charge rectangular family. The Jack--Kostka expression should
therefore be regarded as an exact structural reduction, not as a closed
product or a proved polynomial-time evaluation.

\subsection{Computational status and limitations}
\label{subsec:computational-status}

By Eq.~\eqref{eq:alpha0-formula}, direct enumeration contains
\(\binom{2L}{L}\) balanced-minor terms and is exponential in \(L\). The
special formulas should be measured against this baseline.

At \(\alpha=\tfrac12\) and \(\alpha=2\), the checkerboard product formulas
require only \(O(L)\) arithmetic operations. Their generic-weight
determinant or ordinary-Pfaffian counterparts have the usual polynomial
dense-linear-algebra cost, while the \(\alpha=1\) answer is elementary.

For fixed positive integer \(\alpha\), the number of neutral
alias-multiplicity sectors grows polynomially with \(L\). This fact alone
does not make the resulting representation efficient. At \(\alpha=4\),
the remaining task is the evaluation of one inverse Jack--Kostka
coefficient for each neutral sector. Algorithms that evaluate Jack
polynomials at specified arguments do not automatically solve this
coefficient-extraction problem \cite{DemmelKoev}. Conversely, general
\(\#\mathrm P\)-completeness results for Kostka and
Littlewood--Richardson coefficients do not establish hardness for the
highly structured rectangular family appearing here
\cite{Narayanan2006}. We therefore claim neither a polynomial-time
algorithm nor a hardness theorem for the generic-fugacity
\(\alpha=4\) formula.

The same caution applies to the perfect-square hierarchy.
Hyperpfaffians provide exact finite representations, but exterior degree
greater than two generally leaves a combinatorial expansion. Checkerboard
Fourier sparsity may permit further reductions, but sparsity alone is not
an algorithmic theorem.

Finally, evaluating \(\cZ_{\alpha,L}(u)\) at one fixed numerical value of
\(u\) is different from producing all coefficients of the degree-\(L\)
fugacity polynomial. Arithmetic-operation counts must also be
distinguished from bit-complexity estimates. These conventions are
summarized in Appendix~\ref{app:complexity-conventions}.

The unrefined product at \(\alpha=4\) should therefore be understood as a
special consequence of half filling and complementary middle minors, not
as evidence for generic \(\beta=8\) Pfaffian solvability.


\section{Degree-resolved Pauli spectrum and full counting statistics}
\label{sec:fcs}

The stabilizer R\'enyi entropy retains only the total R\'enyi weight
\(\cZ_{\alpha,L}(1)\), whereas the fugacity-resolved polynomial determines
how that weight is distributed among balanced Majorana degrees. It
therefore reveals the center, width, and response of the degree shell,
information that cannot be reconstructed from the unrefined entropy
alone. At the exactly solvable tilts
\(\alpha=\tfrac12,1,2\), complement symmetry fixes the common center at
\(L/2\), while the fluctuation scale depends strongly on the R\'enyi
index. In particular, the widths at \(\alpha=\tfrac12\) and
\(\alpha=2\) scale as \(\sqrt L\) and \(\sqrt{L\log L}\), respectively.

\begin{figure*}[t]
    \centering
    \includegraphics[width=\textwidth]
    {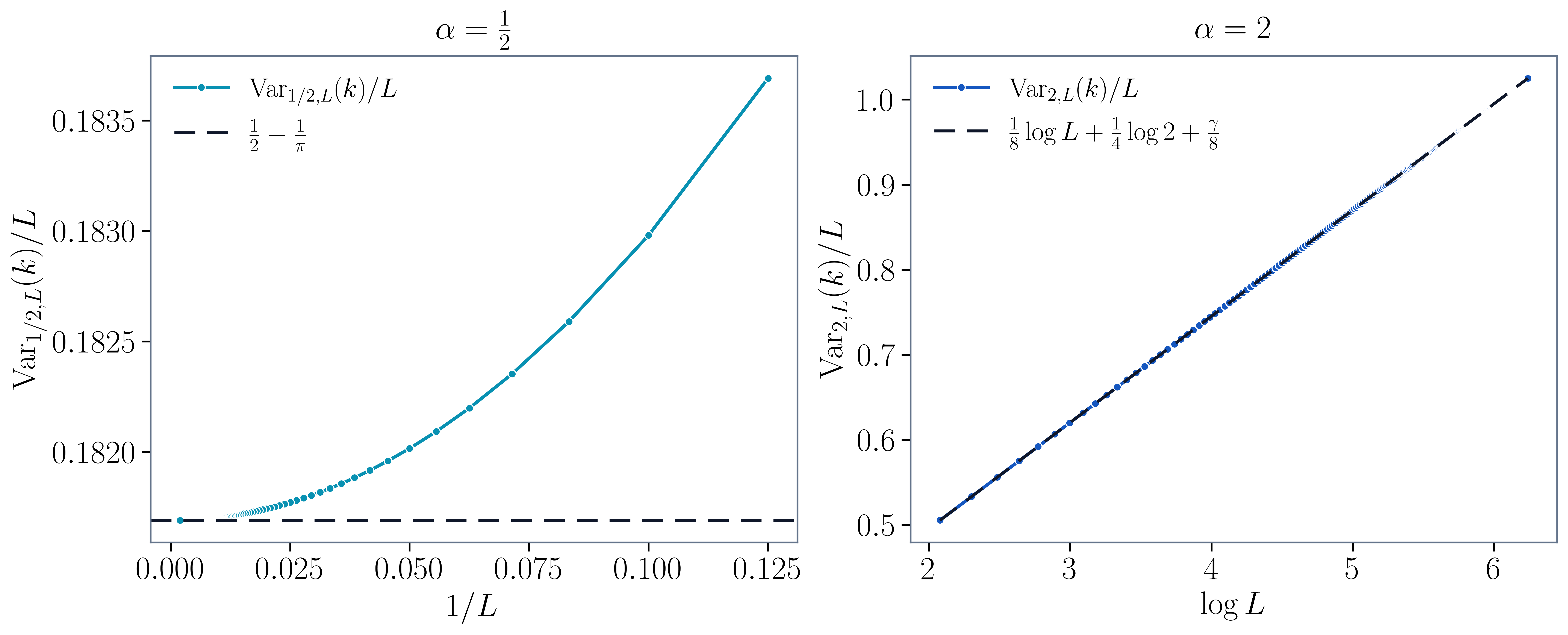}
    \caption{
    Finite-size scaling of the degree fluctuations. Left:
    \(\operatorname{Var}_{1/2,L}(k)/L\) approaches
    \(\frac12-\frac1\pi\), shown by the dashed line. Right:
    \(\operatorname{Var}_{2,L}(k)/L\) is plotted against \(\log L\);
    the dashed line is
    \(\frac18\log L+\frac14\log2+\frac{\gamma}{8}\).
    The comparison shows conventional extensive fluctuations at
    \(\alpha=\frac12\) and a marginal logarithmic enhancement at
    \(\alpha=2\).
    }
    \label{fig:fcs-variance-scaling}
\end{figure*}

\subsection{R\'enyi-tilted Pauli distributions}
\label{subsec:fcs-tilted-pauli}

For a pure state \(\ket{\psi_L}\), define the normalized Pauli distribution
\begin{equation}
\label{eq:FCS-pauli-distribution}
p_L(P)
=
\frac{1}{2^L}
\abs{\bra{\psi_L}P\ket{\psi_L}}^2,
\qquad
\sum_{P\in\mathcal P_L}p_L(P)=1.
\end{equation}
Its R\'enyi moments are related to the all-minors sum by
\begin{equation}
\label{eq:FCS-pauli-moment-Z}
\sum_{P\in\mathcal P_L}p_L(P)^\alpha
=
2^{-\alpha L}\cZ_{\alpha,L}(1).
\end{equation}
For \(\alpha>0\), the corresponding R\'enyi escort distribution is
\begin{equation}
\label{eq:FCS-escort-distribution}
\Pi_{\alpha,L}(P)
=
\frac{p_L(P)^\alpha}
{\displaystyle\sum_{P'\in\mathcal P_L}p_L(P')^\alpha}.
\end{equation}
Changing \(\alpha\) changes how the same Pauli spectrum is sampled. At
\(\alpha=\tfrac12\),
\(\Pi_{1/2,L}(P)\propto\abs{\langle P\rangle}\), so a broad set of
nonzero Pauli coefficients contributes. At \(\alpha=2\),
\(\Pi_{2,L}(P)\propto p_L(P)^2\), so large Pauli probabilities are
enhanced. Equivalently, if two strings are sampled independently from
\(p_L\) and conditioned to coincide, their common value is distributed
according to \(\Pi_{2,L}\).

For the critical TFI Gaussian ground state, every Pauli string with
nonzero expectation value admits a balanced Majorana representation. We
denote by \(d(P)=k\) the number of Majoranas of each species in this
representation. This balanced Majorana degree is not, in general, the
ordinary number of nonidentity one-site Pauli operators.

Writing
\begin{equation}
\label{eq:FCS-Z-coefficients}
\cZ_{\alpha,L}(u)
=
\sum_{k=0}^{L}W_{\alpha,L}(k)u^k,
~~
W_{\alpha,L}(k)
=
\sum_{P:\,d(P)=k}
\abs{\langle P\rangle}^{2\alpha},
\end{equation}
the degree marginal of the escort distribution is
\begin{equation}
\label{eq:FCS-degree-distribution}
\mathbb P_{\alpha,L}(k)
=
\sum_{P:\,d(P)=k}\Pi_{\alpha,L}(P)
=
\frac{W_{\alpha,L}(k)}{\cZ_{\alpha,L}(1)}.
\end{equation}
Thus \(\mathbb P_{\alpha,L}(k)\) is not the fraction of strings at degree
\(k\); it is the fraction of the total \(\alpha\)-tilted Pauli weight
carried by that degree sector.

The endpoint sectors contain the identity and the global parity string,
with degrees \(0\) and \(L\), respectively. Both have unit
expectation-value magnitude, whereas the number of balanced strings is
largest near \(k=L/2\). The competition between sector multiplicity and
individual Pauli weight will be analyzed in
Sec.~\ref{subsec:fcs-comparison}. Unbalanced strings, such as a single
\(\sigma_j^x\), have zero expectation value and do not contribute.

In the doubled-root ensemble, the same degree variable is
\begin{equation}
\label{eq:FCS-root-degree}
k=\abs{U\cap X},
\qquad
Q=2k-L
=
\abs{U\cap X}-\abs{U\cap Y}.
\end{equation}
The centered variable \(Q\) is the staggered occupation imbalance between
the two interlaced root sublattices. Since
\[
u^k
=
u^{L/2}
\exp\!\left(\frac{\log u}{2}Q\right),
\]
the field conjugate to \(Q\) is \(h=\tfrac12\log u\).

\subsection{Generating functions and complement symmetry}
\label{subsec:fcs-generating-functions}

The moment-generating and cumulant-generating functions are
\begin{align}
\label{eq:FCS-moment-generating}
\chi_{\alpha,L}(\lambda)
&=
\left\langle e^{\lambda k}\right\rangle_{\alpha,L}
=
\frac{\cZ_{\alpha,L}(e^\lambda)}{\cZ_{\alpha,L}(1)},
\\
\label{eq:FCS-cumulant-generating}
K_{\alpha,L}(\lambda)
&=
\log\chi_{\alpha,L}(\lambda).
\end{align}
The degree cumulants follow from
\begin{equation}
\label{eq:FCS-cumulants}
\kappa_n^{(\alpha,L)}
=
\left.
\frac{\partial^n K_{\alpha,L}}{\partial\lambda^n}
\right|_{\lambda=0}
=
\left.
\left(u\frac{\partial}{\partial u}\right)^n
\log\cZ_{\alpha,L}(u)
\right|_{u=1}.
\end{equation}

Palindromicity, Eq.~\eqref{eq:palindromic}, implies
\begin{equation}
\label{eq:FCS-probability-symmetry}
\mathbb P_{\alpha,L}(k)
=
\mathbb P_{\alpha,L}(L-k),
\qquad
\chi_{\alpha,L}(\lambda)
=
e^{\lambda L}\chi_{\alpha,L}(-\lambda).
\end{equation}
Consequently,
\begin{equation}
\label{eq:FCS-general-mean}
\left\langle k\right\rangle_{\alpha,L}
=
\frac{L}{2},
\end{equation}
and all odd centered cumulants vanish. The mean is therefore fixed by
symmetry and cannot distinguish the different R\'enyi tilts; the first
informative quantity is the variance.

We define the staggered degree susceptibility by
\begin{equation}
\label{eq:FCS-degree-susceptibility}
\mathcal X_{\alpha,L}
=
\frac{\operatorname{Var}_{\alpha,L}(Q)}{L}
=
\frac{4\operatorname{Var}_{\alpha,L}(k)}{L}.
\end{equation}
A finite limit corresponds to conventional \(\sqrt L\) fluctuations,
whereas a divergence signals an anomalously broad degree shell.

For the thermodynamic counting statistics, let
\begin{equation}
\label{eq:FCS-scaled-CGF}
\phi_\alpha(\lambda)
=
\lim_{L\to\infty}
\frac{1}{L}K_{\alpha,L}(\lambda),
\end{equation}
whenever the limit exists. Complement symmetry gives
\[
\phi_\alpha(\lambda)
=
\lambda+\phi_\alpha(-\lambda),
\]
so the centered part
\(\phi_\alpha(\lambda)-\lambda/2\) is even. The exact product formulas of
Sec.~\ref{sec:solvable-indices} determine
\(\phi_\alpha\) explicitly at \(\alpha=\tfrac12\) and \(\alpha=2\).

\begin{figure*}[t]
    \centering
    \includegraphics[width=\textwidth]
    {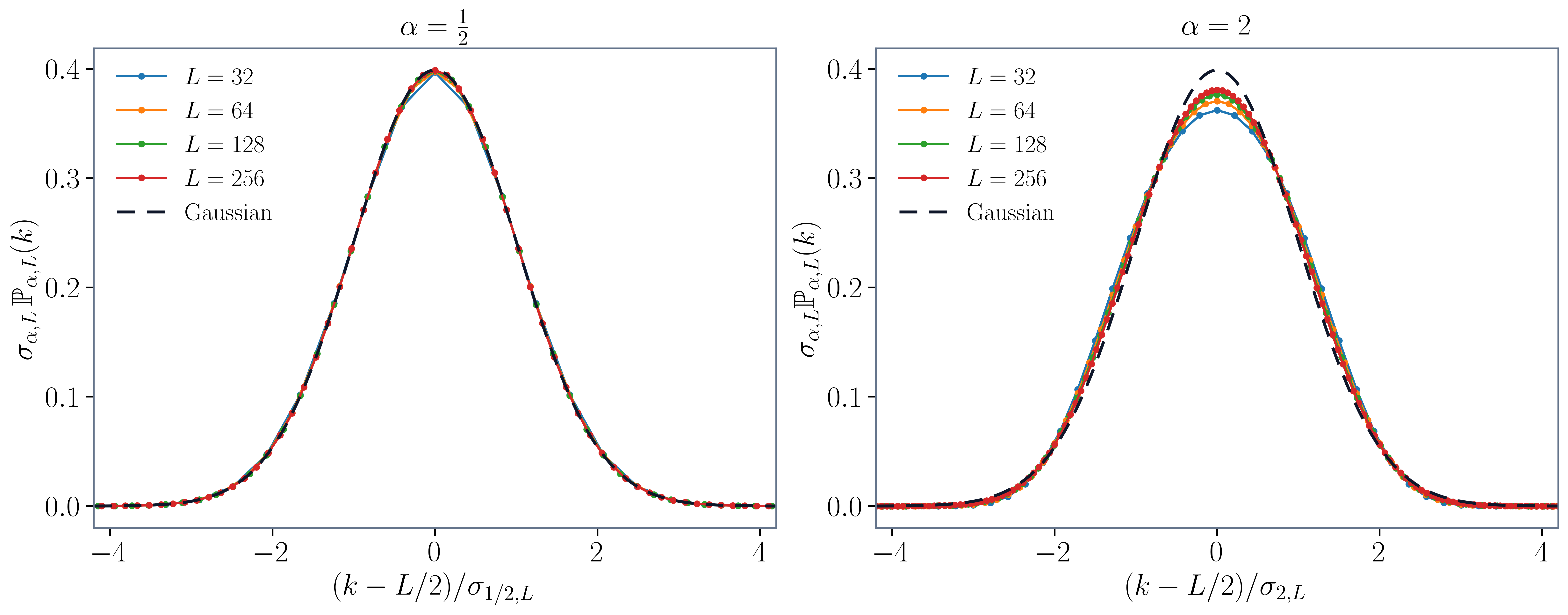}
    \caption{
    Gaussian scaling collapse of the degree distributions at
    \(\alpha=\frac12\) (left) and \(\alpha=2\) (right). We plot
    \(\sigma_{\alpha,L}\mathbb P_{\alpha,L}(k)\) against
    \((k-L/2)/\sigma_{\alpha,L}\) for
    \(L=32,64,128,\) and \(256\). The dashed curve is the
    standard-normal density \(e^{-x^2/2}/\sqrt{2\pi}\). At
    \(\alpha=\frac12\),
    \(\sigma_{1/2,L}^2\sim(\frac12-\frac1\pi)L\), whereas at
    \(\alpha=2\),
    \(\sigma_{2,L}^2
    =\frac{L}{4}(H_{2L}-\frac12H_L)
    \sim L\log L/8\). Both rescaled distributions approach the
    standard-normal form, with larger finite-size corrections at
    \(\alpha=2\).
    }
    \label{fig:gaussian-degree-shell-collapse}
\end{figure*}

\subsection{Pauli-amplitude weighting at
\texorpdfstring{\(\alpha=\tfrac12\)}{alpha=1/2}}
\label{subsec:fcs-alpha-half}

For positive even \(L\), define
\[
\theta_{j,L}
=
\frac{\pi(2j-1)}{2L},
\qquad
a_{j,L}
=
\sec\theta_{j,L}.
\]
Using Eq.~\eqref{eq:alpha-half-product} and normalizing at \(u=1\) gives
\begin{equation}
\label{eq:FCS-half-chi}
\chi_{1/2,L}(\lambda)
=
\prod_{j=1}^{L/2}
\frac{1+e^{2\lambda}+2a_{j,L}e^\lambda}
{2(1+a_{j,L})}.
\end{equation}
Each factor is a probability-generating function. Hence
\begin{equation}
\label{eq:FCS-half-independent-sum}
k
=
\sum_{j=1}^{L/2}X_{j,L},
\end{equation}
where the independent variables \(X_{j,L}\in\{0,1,2\}\) satisfy
\begin{equation}
\label{eq:FCS-half-mode-probabilities}
\begin{aligned}
\Pr(X_{j,L}=0)
&=
\Pr(X_{j,L}=2)
=
\frac{1}{2(1+a_{j,L})},
\\[1mm]
\Pr(X_{j,L}=1)
&=
\frac{a_{j,L}}{1+a_{j,L}}.
\end{aligned}
\end{equation}
Each mode has mean one and variance \(1/(1+a_{j,L})\). Therefore,
\begin{equation}
\label{eq:FCS-half-variance}
\operatorname{Var}_{1/2,L}(k)
=
\sum_{j=1}^{L/2}
\frac{1}{1+a_{j,L}}
=
\sum_{j=1}^{L/2}
\frac{\cos\theta_{j,L}}
{1+\cos\theta_{j,L}}.
\end{equation}
The midpoint Riemann sum gives
\begin{align}
\label{eq:FCS-half-variance-density}
\lim_{L\to\infty}
\frac{\operatorname{Var}_{1/2,L}(k)}{L}
&=
\frac{1}{\pi}
\int_0^{\pi/2}
\frac{\cos\theta}{1+\cos\theta}\,\dd\theta
\nonumber\\
&=
\frac12-\frac{1}{\pi}.
\end{align}
Thus
\begin{equation}
\label{eq:FCS-half-variance-asymptotic}
\operatorname{Var}_{1/2,L}(k)
=
\left(\frac12-\frac1\pi\right)L+o(L),
\qquad
\mathcal X_{1/2,L}
\longrightarrow
2-\frac4\pi.
\end{equation}

The thermodynamic scaled cumulant-generating function is
\begin{equation}
\label{eq:FCS-half-SCGF}
\phi_{1/2}(\lambda)
=
\frac{\lambda}{2}
+
\frac{1}{\pi}
\int_0^{\pi/2}
\log
\left[
\frac{\cosh\lambda+\sec\theta}
{1+\sec\theta}
\right]
\dd\theta.
\end{equation}
Its centered part is analytic at the origin, with
\[
\phi_{1/2}''(0)
=
\frac12-\frac1\pi.
\]
Since Eq.~\eqref{eq:FCS-half-independent-sum} is a triangular array of
bounded independent variables with variance proportional to \(L\), the
Lindeberg condition is automatic and
\begin{equation}
\label{eq:FCS-half-CLT}
\frac{k-L/2}
{\sqrt{L\left(\frac12-\frac1\pi\right)}}
\;\Longrightarrow\;
\mathcal N(0,1).
\end{equation}
The amplitude-weighted degree distribution is therefore centered at
\(L/2\), has width \(O(\sqrt L)\), and concentrates at half degree in
relative units. Its staggered susceptibility remains finite.

\subsection{Binomial weighting at
\texorpdfstring{\(\alpha=1\)}{alpha=1}}
\label{subsec:fcs-alpha-one}

At \(\alpha=1\), the escort distribution is the physical Pauli
distribution itself. Equation~\eqref{eq:alpha1-formula} gives
\begin{equation}
\label{eq:FCS-alpha1-binomial}
\mathbb P_{1,L}(k)
=
2^{-L}\binom{L}{k}.
\end{equation}
Hence
\[
\left\langle k\right\rangle_{1,L}
=
\frac{L}{2},
\qquad
\operatorname{Var}_{1,L}(k)
=
\frac{L}{4},
\]
and the standard binomial central limit theorem gives
\begin{equation}
\label{eq:FCS-alpha1-CLT}
\frac{k-L/2}{\sqrt{L/4}}
\;\Longrightarrow\;
\mathcal N(0,1).
\end{equation}

\subsection{Collision weighting at
\texorpdfstring{\(\alpha=2\)}{alpha=2}}
\label{subsec:fcs-alpha-two}

Let
\[
b_{r,L}
=
\frac{2r-1}{L}.
\]
Using Eq.~\eqref{eq:alpha2-product}, the normalized generating function
takes the centered form
\begin{equation}
\label{eq:FCS-alpha2-chi}
\chi_{2,L}(\lambda)
=
e^{\lambda L/2}
\prod_{r=1}^{L/2}
\frac{\cosh^2(\lambda/2)-b_{r,L}^2}
{1-b_{r,L}^2}.
\end{equation}
Unlike the factors at \(\alpha=\tfrac12\), the individual quadratic
factors in Eq.~\eqref{eq:alpha2-product} are not all
probability-generating functions. The product is an exact algebraic
factorization of the full positive distribution, but it does not
represent \(k\) as a sum of independent classical variables.

Differentiating Eq.~\eqref{eq:FCS-alpha2-chi} gives
\begin{equation}
\label{eq:FCS-alpha2-variance-sum}
\operatorname{Var}_{2,L}(k)
=
\frac12
\sum_{r=1}^{L/2}
\frac{1}{1-b_{r,L}^2}.
\end{equation}
A partial-fraction evaluation yields the exact harmonic-number formula
\begin{equation}
\label{eq:FCS-alpha2-variance-harmonic}
\operatorname{Var}_{2,L}(k)
=
\frac{L}{4}
\left(
H_{2L}-\frac12H_L
\right),
\end{equation}
where \(H_n=\sum_{j=1}^n j^{-1}\). Consequently,
\begin{equation}
\label{eq:FCS-alpha2-variance-asymptotic}
\operatorname{Var}_{2,L}(k)
=
\frac{L}{8}\log L
+
\frac{L}{4}\log2
+
\frac{\gamma L}{8}
+
O(L^{-1}),
\end{equation}
and
\begin{equation}
\label{eq:FCS-alpha2-susceptibility}
\mathcal X_{2,L}
=
H_{2L}-\frac12H_L
=
\frac12\log L+\log2+\frac{\gamma}{2}+o(1).
\end{equation}
The logarithm originates from the soft endpoint
\(b_{r,L}\to1\), where the summand in
Eq.~\eqref{eq:FCS-alpha2-variance-sum} develops a harmonic tail.

The distinct finite-size growth laws in
Eqs.~\eqref{eq:FCS-half-variance-asymptotic} and
\eqref{eq:FCS-alpha2-variance-asymptotic} are compared in
Fig.~\ref{fig:fcs-variance-scaling}.

The thermodynamic scaled cumulant-generating function is
\begin{equation}
\label{eq:FCS-alpha2-SCGF-integral}
\phi_2(\lambda)
=
\frac{\lambda}{2}
+
\frac12
\int_0^1
\log
\left[
\frac{\cosh^2(\lambda/2)-x^2}
{1-x^2}
\right]
\dd x.
\end{equation}
Writing \(a=\cosh(\lambda/2)\), the integral evaluates to
\begin{equation}
\label{eq:FCS-alpha2-SCGF-closed}
\begin{aligned}
\phi_2(\lambda)
={}&
\frac{\lambda}{2}
+
\frac12
\Big[
(a+1)\log(a+1)
\\
&\qquad
-(a-1)\log(a-1)
-2\log 2
\Big].
\end{aligned}
\end{equation}
Near the origin,
\begin{equation}
\label{eq:FCS-alpha2-SCGF-small-lambda}
\phi_2(\lambda)
=
\frac{\lambda}{2}
+
\frac{\lambda^2}{16}
\left[
1+\log16-2\log\abs{\lambda}
\right]
+
O\!\left(\lambda^4\log\abs{\lambda}\right).
\end{equation}
Thus every finite-size generating function is analytic, but the
thermodynamic limit is not twice differentiable at \(\lambda=0\). This
nonanalytic curvature is the counting-field signature of the
logarithmically divergent variance density.

The Gaussian central limit follows directly from the characteristic
function. Let
\[
\sigma_{2,L}^2
=
\operatorname{Var}_{2,L}(k),
\qquad
A_{r,L}
=
\frac{1}{1-b_{r,L}^2}.
\]
Equation~\eqref{eq:FCS-alpha2-chi} gives
\begin{equation}
\label{eq:FCS-alpha2-centered-characteristic}
\left\langle
e^{it(k-L/2)/\sigma_{2,L}}
\right\rangle_{2,L}
=
\prod_{r=1}^{L/2}
\left[
1-
A_{r,L}
\sin^2\left(\frac{t}{2\sigma_{2,L}}\right)
\right].
\end{equation}
Here
\[
\sum_rA_{r,L}
=
2\sigma_{2,L}^2,
~~
\max_rA_{r,L}
=
O(L),
~~
\sum_rA_{r,L}^2
=
O(L^2).
\]
Since
\(\sigma_{2,L}^2\sim L\log L/8\), expanding the logarithm of
Eq.~\eqref{eq:FCS-alpha2-centered-characteristic} gives
\[
\log
\left\langle
e^{it(k-L/2)/\sigma_{2,L}}
\right\rangle_{2,L}
=
-\frac{t^2}{2}
+
O\!\left(\frac{1}{\log^2L}\right).
\]
Therefore,
\begin{equation}
\label{eq:FCS-alpha2-CLT}
\frac{k-L/2}{\sigma_{2,L}}
\;\Longrightarrow\;
\mathcal N(0,1).
\end{equation}

The collision-weighted distribution remains concentrated at half degree
in relative units because
\(\operatorname{Var}_{2,L}(k)/L^2\to0\), but its absolute width is
enhanced from \(O(\sqrt L)\) to \(O(\sqrt{L\log L})\). Despite this
marginal broadening, the centered and variance-rescaled distribution
remains asymptotically Gaussian.

The Gaussian limits at \(\alpha=\tfrac12\) and \(\alpha=2\) are
illustrated in Fig.~\ref{fig:gaussian-degree-shell-collapse}.

\subsection{Comparison and evolution with R\'enyi tilt}
\label{subsec:fcs-comparison}

The three exactly analyzed tilts are summarized in
Table~\ref{tab:fcs-physical-summary}. Their common center is fixed by
complement symmetry, whereas their widths distinguish the corresponding
degree-space regimes.

\begin{table}[t]
\centering
\small
\setlength{\tabcolsep}{2pt}
\renewcommand{\arraystretch}{1.35}

\begingroup
\setlength{\fboxsep}{3pt}

\colorbox{TFILightBlue}{%
\begin{tabular}{
@{}
p{0.08\columnwidth}
p{0.20\columnwidth}
p{0.24\columnwidth}
p{0.32\columnwidth}
@{}
}
\toprule

\centering\textcolor{TFIBlue}{\bfseries \(\alpha\)}
&
\centering\textcolor{TFIBlue}{\bfseries Weight}
&
\centering\textcolor{TFIBlue}{\bfseries Variance}
&
\textcolor{TFIBlue}{\bfseries Degree shell}
\tabularnewline
\midrule

\centering\textcolor{TFIViolet}{\(\mathbf{\tfrac12}\)}
&
\(\abs{\langle P\rangle}\)
&
\(\displaystyle
\left(\frac12-\frac1\pi\right)L
\)
&
Centered at \(L/2\), with \(O(\sqrt L)\) width.
\tabularnewline[0.8mm]

\centering\textcolor{TFIViolet}{\(\mathbf{1}\)}
&
\(p_L(P)\)
&
\(\displaystyle \frac{L}{4}\)
&
Exactly binomial, with \(O(\sqrt L)\) width.
\tabularnewline[0.8mm]

\centering\textcolor{TFIViolet}{\(\mathbf{2}\)}
&
\(p_L(P)^2\)
&
\(\displaystyle
\frac{L}{8}\log L+O(L)
\)
&
Centered at \(L/2\), with \(O(\sqrt{L\log L})\) width.
\tabularnewline

\bottomrule
\end{tabular}%
}

\endgroup

\caption{
Balanced-degree statistics under three R\'enyi escort distributions.
Complement symmetry fixes the common center at \(L/2\), while the tilt
controls the width of the degree shell.
}
\label{tab:fcs-physical-summary}
\end{table}

\begin{figure}[t]
    \centering
    \includegraphics[width=\linewidth]
    {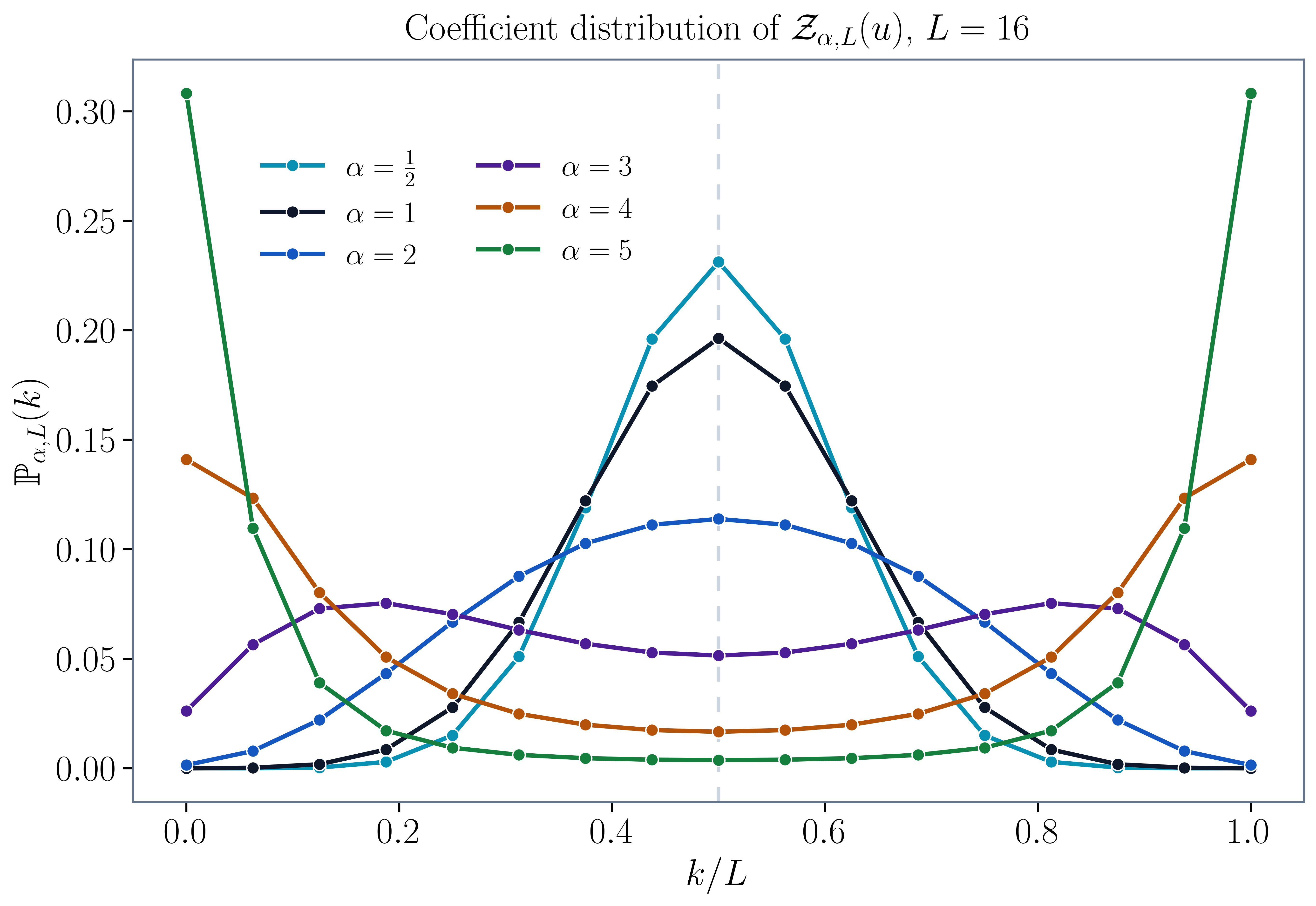}
    \caption{
    Normalized balanced-degree distribution
    \(\mathbb P_{\alpha,L}(k)\) at \(L=16\) for several R\'enyi tilts.
    Complement symmetry enforces reflection about \(k/L=1/2\).
    Increasing \(\alpha\) changes the finite-size profile from a central
    peak to a symmetric bimodal form and eventually to strong endpoint
    dominance. The curves at \(\alpha>2\) show a finite-size evolution
    and do not by themselves establish a sharp thermodynamic transition.
    }
    \label{fig:fcs-alpha-evolution}
\end{figure}

Figure~\ref{fig:fcs-alpha-evolution} shows how the full degree profile
evolves beyond these exact cases at fixed size \(L=16\). The distribution
is centrally peaked for \(\alpha=\tfrac12\) and \(\alpha=1\), and is
substantially broader at \(\alpha=2\). At \(\alpha=3\), the central
maximum has split into two symmetric maxima, while for \(\alpha=4\) and
\(\alpha=5\) the weight is increasingly concentrated near the endpoint
sectors \(k=0\) and \(k=L\).

This evolution reflects a competition between sector multiplicity and
individual coefficient magnitude. The central sectors contain
\(\binom{L}{k}^2\) balanced pairs and therefore have their largest
multiplicity near \(k=L/2\). By contrast, each endpoint sector contains
only one string: the identity at \(k=0\) and the global parity string at
\(k=L\). Both endpoint strings have expectation-value magnitude one. By contrast,
every nontrivial proper minor has magnitude strictly smaller than one.
Indeed, \(G\) is orthogonal and has no vanishing entries. For a proper
column set \(T\subsetneq[L]\), each restricted row \(G_{j,T}\) has
Euclidean norm strictly smaller than one, and Hadamard's inequality gives
\[
\abs{\det G[S,T]}
\leq
\prod_{j\in S}\|G_{j,T}\|_2
<1
\]
for \(0<|S|=|T|<L\). Therefore, at fixed \(L\),
\begin{equation}
\label{eq:FCS-large-alpha-endpoint-limit}
\mathbb P_{\alpha,L}(k)
\underset{\alpha\to\infty}{\longrightarrow}
\frac12\delta_{k,0}
+
\frac12\delta_{k,L}.
\end{equation}

The full-counting analysis therefore contains information that cannot be
reconstructed from the stabilizer R\'enyi entropy alone: the entropy fixes
only the total R\'enyi weight, whereas the fugacity-resolved polynomial
reveals how that weight is distributed among balanced-degree sectors.
Extensions to the remaining exact algebraic representations and to the
complex fugacity plane are discussed in the conclusion.


\section{Conclusion and Outlook}
\label{sec:conclusion}

In this work, we identified a common finite-size algebraic structure
underlying several critical stabilizer and computational-basis
Shannon--R\'enyi quantities. The critical transverse-field Ising
stabilizer entropy provides the central building block linking the
established stabilizer decimations, the half-filled \(XX\) chain, and,
at order two, the Haldane--Shastry escort relation. Beyond these known
connections, we proved that, under the stated finite-size and
boundary-sector conditions, the complete computational-basis distribution
of the range-\(m\) \(XX\) family factorizes over \(m\) squeezed
sublattices. This yields an exact Shannon--R\'enyi decimation and extends
the reach of the common Ising building block throughout the compatible
range-\(m\) family.

We then analyzed the all-minors structure underlying this building block.
By introducing a fugacity that resolves the balanced Majorana degree, we
derived a termwise mapping, valid for every positive real R\'enyi index, to
a checkerboard-weighted discrete Selberg ensemble on a half-filled doubled
root lattice. For positive integer indices, finite Fourier bandwidth
further produces a neutral aliased Dyson--Morris constant term. The
surviving aliases distinguish the finite root lattice from the continuous
Selberg problem and show that the ordinary Dyson evaluation reproduces the
complete answer only when the nonzero alias sectors collapse, as they do
at special indices.

At \(\alpha=\tfrac12,1,2\), the classical determinant and Pfaffian
structures provide polynomial-size compressions for arbitrary one-body
weights on the doubled root lattice. For the checkerboard fugacity
considered here, its sparse Fourier support reduces these finite objects
further to explicit product formulas, whose unrefined limits reproduce
the known finite-size values. At \(\alpha=4\), we obtained an exact
generic-fugacity seven-charge inverse Jack--Kostka representation, while
the unit-fugacity value follows from a complementary middle-minor identity,
$
\cZ_{4,L}(1)
=
2^{-L}\cZ_{2,L}(1)^2$.
More generally, the indices satisfying \(2\alpha=q^2\), with
\(q\in\mathbb Z_{>0}\), admit a confluent hyperpfaffian hierarchy, whereas
positive integer indices admit complementary-minor and shifted-Dyson
organizations. These representations are exact, although only some lead
to closed products or polynomial-size evaluations.

The fugacity refinement also determines the full counting statistics of
the balanced Majorana degree. Complement symmetry fixes its mean at half
the system size. At \(\alpha=\tfrac12\), the fluctuations have the
conventional square-root scale, while at \(\alpha=1\) the finite-size
distribution is exactly binomial. At \(\alpha=2\), the width is enhanced
to the scale \(\sqrt{L\log L}\), and the staggered susceptibility diverges
logarithmically. Nevertheless, at all three solved indices the properly
centered and variance-rescaled distributions have Gaussian scaling limits.
Thus, at \(\alpha=2\), emphasizing the largest Pauli probabilities broadens
the shell around half degree rather than concentrating the weight in
low-degree sectors.

Beyond these solvable points, an important question is whether Gaussian
scaling persists as the R\'enyi tilt is increased. Fixed-size results at
stronger tilt reveal a qualitative evolution from a central peak to a
symmetric bimodal profile and eventually to endpoint dominance. Whether
this evolution produces a non-Gaussian thermodynamic scaling limit,
whether there is a sharp threshold in \(\alpha\), and what the limiting
distribution might be remain open.

On the algebraic side, it remains to determine whether the unit-fugacity
complementary middle-minor collapse at \(\alpha=4\) has a weighted analogue
that evaluates \(\cZ_{4,L}(u)\) at generic fugacity, whether the rectangular
inverse Jack--Kostka coefficients of the seven-charge representation obey
closed recurrences or admit an efficient evaluation, and whether
checkerboard Fourier sparsity factorizes the higher hyperpfaffians or
produces polynomial-cost recurrences. Further isolated indices or
fugacity values may also admit determinant, Pfaffian, or product collapses.
Finally, the zeros of the fugacity polynomial in the complex plane provide
a complementary route to these questions and may clarify how competing
alias sectors reorganize in the thermodynamic limit.

{\it Acknowledgements:}
We thank Fabian Ballar and Markus Heyl for useful comments. M.A.R. acknowledges partial support from CNPq and FAPERJ (grant number E-26/210.062/2023). R.K. gratefully acknowledges the resources on the LiCCA HPC cluster of the University of Augsburg, co-funded by the Deutsche Forschungsgemeinschaft (DFG, German Research Foundation)–Project-ID 499211671. We thank the Abdus Salam International Centre for Theoretical Physics (ICTP) for its hospitality during the completion of this work.

\newpage

\appendix


\section{Taxonomy of the fugacity-resolved all-minors problem}
\label{app:taxonomy}

Throughout the paper, the fugacity-resolved all-minors quantity is
expressed in several mathematically equivalent forms. Each representation
uses different microscopic variables and highlights a distinct physical
or algebraic structure, but all describe the same finite-size partition
function. This appendix fixes the terminology used in the main text,
specifies the domain of validity of each representation, and distinguishes
exact structural reductions from computationally efficient evaluations.

The primary name for the object defined in Eq.~\eqref{eq:centralZ} is the
\emph{fugacity-resolved balanced all-minors partition function},
\begin{equation}
\label{eq:tax-master}
\cZ_{\alpha,L}(u)
=
\sum_{k=0}^{L}u^k
\sum_{\substack{S,T\subseteq[L]\\|S|=|T|=k}}
\abs{\det G[S,T]}^{2\alpha}.
\end{equation}
The adjective ``balanced'' refers to the equality \(|S|=|T|\).
For the TFI ground state, this common cardinality is the balanced
Majorana degree: the associated nonvanishing Pauli string contains
\(k\) Majoranas of each species, hence \(2k\) Majoranas in total.
This degree is generally different from ordinary one-site Pauli support.
Algebraically, Eq.~\eqref{eq:tax-master} is a degree-\(L\) polynomial in
complex \(u\). For \(\alpha>0\) and \(u\geq0\), it is also a positive
partition function. The endpoint \(\alpha=0\) is treated separately as
the combinatorial support-counting limit.

The same object may be viewed as a balanced Pauli moment, a balanced
all-minors sum, a neutral two-component defect gas, a half-filled
checkerboard log gas, or a discrete-measure Selberg integral. For positive
integer \(\alpha\), the last form admits a further reduction to an aliased
Dyson constant term. Determinants, Pfaffians, complementary minors, Jack
coefficients, and hyperpfaffians are not additional partition functions;
they are parameter-dependent algebraic representations of the same
finite-size object.

\subsection{Quantum-information and counting-statistics forms}
\label{app:taxonomy-quantum}

Let \(P_{S,T}\) denote the Hermitian Pauli string associated, under the
fixed Jordan--Wigner and Majorana-ordering convention, with a balanced
pair \((S,T)\). The Pauli--minor correspondence of
Sec.~\ref{sec:cauchy} gives
\begin{equation}
\label{eq:tax-wick}
\abs{\langle P_{S,T}\rangle}^{2}
=
\abs{\det G[S,T]}^{2}.
\end{equation}
Consequently,
\begin{equation}
\label{eq:tax-pauli-form}
\cZ_{\alpha,L}(u)
=
\sum_{P\in\mathcal P_L^{\rm bal}}
u^{k(P)}
\abs{\langle P\rangle}^{2\alpha},
\end{equation}
where \(\mathcal P_L^{\rm bal}\) is the set of Pauli strings with
nonzero expectation value in the chosen Gaussian ground state and
\(k(P)\) is their balanced Majorana degree. In this form,
\(\cZ_{\alpha,L}(u)\) is the
\emph{\(\alpha\)-tilted balanced Pauli-string partition function}.
The word ``tilted'' refers to reweighting the Pauli probabilities by
their \(\alpha\)th powers; the common factor \(2^{\alpha L}\) is absorbed
into the unnormalized partition function.

At unit fugacity, \(\cZ_{\alpha,L}(1)\) is the unnormalized Pauli moment
entering the TFI stabilizer R\'enyi entropy through
Eq.~\eqref{eq:SRE-Z-section}. Thus the phrase
\emph{TFI stabilizer R\'enyi moment} refers specifically to the
unrefined specialization \(u=1\), not to the full fugacity polynomial.

Writing
\begin{equation}
\label{eq:tax-pgf}
\frac{\cZ_{\alpha,L}(u)}{\cZ_{\alpha,L}(1)}
=
\sum_{k=0}^{L}\mathbb P_{\alpha,L}(k)u^k,
\qquad
\mathbb P_{\alpha,L}(k)
=
\frac{W_{\alpha,L}(k)}{\cZ_{\alpha,L}(1)},
\end{equation}
with
\(
\cZ_{\alpha,L}(u)=\sum_{k=0}^{L}W_{\alpha,L}(k)u^k,
\)
one obtains the probability-generating function of balanced Majorana
degree in the \(\alpha\)-escort Pauli ensemble. With \(u=e^\lambda\),
derivatives of
\(
\log[\cZ_{\alpha,L}(e^\lambda)/\cZ_{\alpha,L}(1)]
\)
generate the degree cumulants. With \(u=e^{i\chi}\),
\[
\frac{\cZ_{\alpha,L}(e^{i\chi})}{\cZ_{\alpha,L}(1)}
=
\sum_{k=0}^{L}
\mathbb P_{\alpha,L}(k)e^{i\chi k}
\]
is the corresponding characteristic function. Accordingly,
\emph{balanced-degree full counting statistics} is precise provided
that the counted degree is stated explicitly.

At \(u=1\), the TFI--\(XX\) correspondence gives the
computational-basis participation moment of the half-filled \(XX\) chain,
\begin{equation}
\label{eq:tax-xx-moment}
\cZ_{\alpha,L}(1)
=
2^{\alpha L}
\sum_{\mathcal C}
\bigl[p_{\mathcal C}^{XX}(2L)\bigr]^{\alpha},
\end{equation}
where \(p_{\mathcal C}^{XX}(2L)\) is the computational-basis probability
of the half-filled configuration \(\mathcal C\) in the \(2L\)-site
\(XX\)-chain ground state. The name
\emph{\(XX\)-chain Shannon--R\'enyi moment} is therefore appropriate
only at unit fugacity.

\subsection{Defect-gas and one-component root-gas forms}
\label{app:taxonomy-gases}

Before passing to the half-filled root subset, the Cauchy determinant
formula organizes the same object as a neutral two-component gas. With
the interlaced root sets \(X=\{x_j\}\) and \(Y=\{y_j\}\) of
Eq.~\eqref{eq:XY-def},
\begin{equation}
\label{eq:tax-defect-gas}
\begin{aligned}
\cZ_{\alpha,L}(u)
={}&
\sum_{k=0}^{L}
\left[u\left(\frac{2}{L}\right)^{2\alpha}\right]^k
\sum_{\substack{S,T\subseteq[L]\\|S|=|T|=k}}
\\[-1mm]
&\times
\frac{
\displaystyle
\prod_{\substack{i<j\\i,j\in S}}
\abs{x_i-x_j}^{2\alpha}
\prod_{\substack{i<j\\i,j\in T}}
\abs{y_i-y_j}^{2\alpha}
}{
\displaystyle
\prod_{\substack{i\in S\\j\in T}}
\abs{x_i-y_j}^{2\alpha}
}.
\end{aligned}
\end{equation}
The selected \(X\)-sites may be viewed as charge-\(+1\) particles and
the selected \(Y\)-sites as charge-\(-1\) defects relative to the
reference configuration in which all \(Y\)-sites are occupied. The
constraint \(|S|=|T|\) is charge neutrality. We therefore call
Eq.~\eqref{eq:tax-defect-gas} a
\emph{neutral two-component lattice log gas on interlaced sublattices}.

The bijection
\begin{equation}
\label{eq:tax-bijection}
U(S,T)=X_S\cup Y_{T^c}
\end{equation}
maps a balanced pair \((S,T)\) to an \(L\)-element subset of the doubled
root lattice and preserves the fugacity,
\(
|U\cap X|=|S|.
\)
Equation~\eqref{eq:discrete-selberg} then becomes
\begin{equation}
\label{eq:tax-root-gas}
\cZ_{\alpha,L}(u)
=
L^{-\alpha L}
\sum_{\substack{U\subset\Omega_{2L}\\|U|=L}}
u^{|U\cap X|}
\abs{\Delta(U)}^{2\alpha}.
\end{equation}
With \(\beta=2\alpha\), this is a
\emph{checkerboard-weighted half-filled discrete circular
\(\beta\)-ensemble}, or equivalently a
\emph{half-filled discrete circular log gas in a staggered chemical
potential}. At \(u=1\), the checkerboard field disappears and the
historically grounded name is the
\emph{half-filled discrete Dyson--Gaudin gas}
\cite{Gaudin1973,MehtaMehta1975,Stephan2014}.
For \(u\neq1\), the phrase
\emph{checkerboard-deformed Dyson--Gaudin gas} is useful descriptive
terminology, but it does not denote a separate standard ensemble.

For \(u>0\), introducing occupations \(n_a\in\{0,1\}\), with
\(\sum_{a=0}^{2L-1}n_a=L\), gives the equivalent classical lattice-gas
form
\begin{equation}
\label{eq:tax-occupation-gas}
\begin{aligned}
\cZ_{\alpha,L}(u)
={}&
L^{-\alpha L}
\sum_{\substack{n_a\in\{0,1\}\\\sum_a n_a=L}}
\exp\Bigg[
2\alpha\sum_{a<b}n_an_b
\log\abs{z_a-z_b}
\\[-1mm]
&\hspace{36mm}
+(\log u)\sum_{a\,\mathrm{even}}n_a
\Bigg].
\end{aligned}
\end{equation}
This form makes explicit the logarithmic pair interaction and the
staggered field conjugate to the \(X\)-sublattice occupation. Equivalently,
after \(s_a=2n_a-1\), it becomes a constrained zero-magnetization Ising
model with long-range logarithmic couplings and a staggered field.

\subsection{Selberg and constant-term forms}
\label{app:taxonomy-selberg-ct}

The root sum in Eq.~\eqref{eq:tax-root-gas} is a finite
root-of-unity analogue of a circular Selberg integral. When this
algebraic structure is emphasized, we call it the
\emph{checkerboard-resolved discrete Selberg sum}. Its equivalent
normalized Dirac-comb integral is given in
Eq.~\eqref{eq:normalized-comb-integral-main}; we refer to that form as a
\emph{circular Selberg-type integral with a weighted root-of-unity
measure}. Both representations are valid for every real \(\alpha>0\).

For \(\alpha\in\mathbb Z_{>0}\), finite Laurent bandwidth allows the
Dirac comb to be replaced, inside constant-term extraction only, by the
finite alias kernel of Eq.~\eqref{eq:finite-alias-kernel}. The result is
\begin{equation}
\label{eq:tax-aliased-ct}
\begin{aligned}
\cZ_{\alpha,L}(u)
={}&
\frac{2^L}{L!\,L^{(\alpha-1)L}}
\\[-1mm]
&\times
\CT_{x_1,\ldots,x_L}
\left[
D_\alpha(x)
\prod_{j=1}^{L}K_{\alpha,u}(x_j)
\right].
\end{aligned}
\end{equation}
We call Eq.~\eqref{eq:tax-aliased-ct} the
\emph{aliased Dyson constant-term representation}. Here ``aliased''
means that the root-of-unity measure has Fourier modes separated by
multiples of \(L\), while the finite-bandwidth Dyson polynomial sees only
a finite subset of them. The kernel \(K_{\alpha,u}\) is therefore not a
second measure and is not pointwise equal to the Dirac comb; the
replacement is exact only under the constant-term operation. The
ordinary Dyson identity evaluates the full problem only when all visible
nonzero aliases vanish.

Expanding the kernels gives the
\emph{neutral shifted-Dyson expansion},
\begin{equation}
\label{eq:tax-shifted-expansion}
\begin{aligned}
\cZ_{\alpha,L}(u)
={}&
\frac{2^L}{L!\,L^{(\alpha-1)L}}
\sum_{q_1,\ldots,q_L=-(\alpha-1)}^{\alpha-1}
\left[\prod_{j=1}^{L}c_{q_j}(u)\right]
\\[-1mm]
&\times
\mathscr C_{\alpha,L}(q_1,\ldots,q_L),
\end{aligned}
\end{equation}
where \(\mathscr C_{\alpha,L}\) is defined in
Eq.~\eqref{eq:shifted-Dyson-coefficient} and vanishes unless
\(\sum_jq_j=0\). The term ``shifted Dyson'' records the inserted exponent
vector; closely related quantities are also called disturbed Dyson
coefficients in the constant-term literature
\cite{Kadell2000,SillsZeilberger2006}.

\begin{figure*}[t]
\centering
\includegraphics[width=\textwidth]
{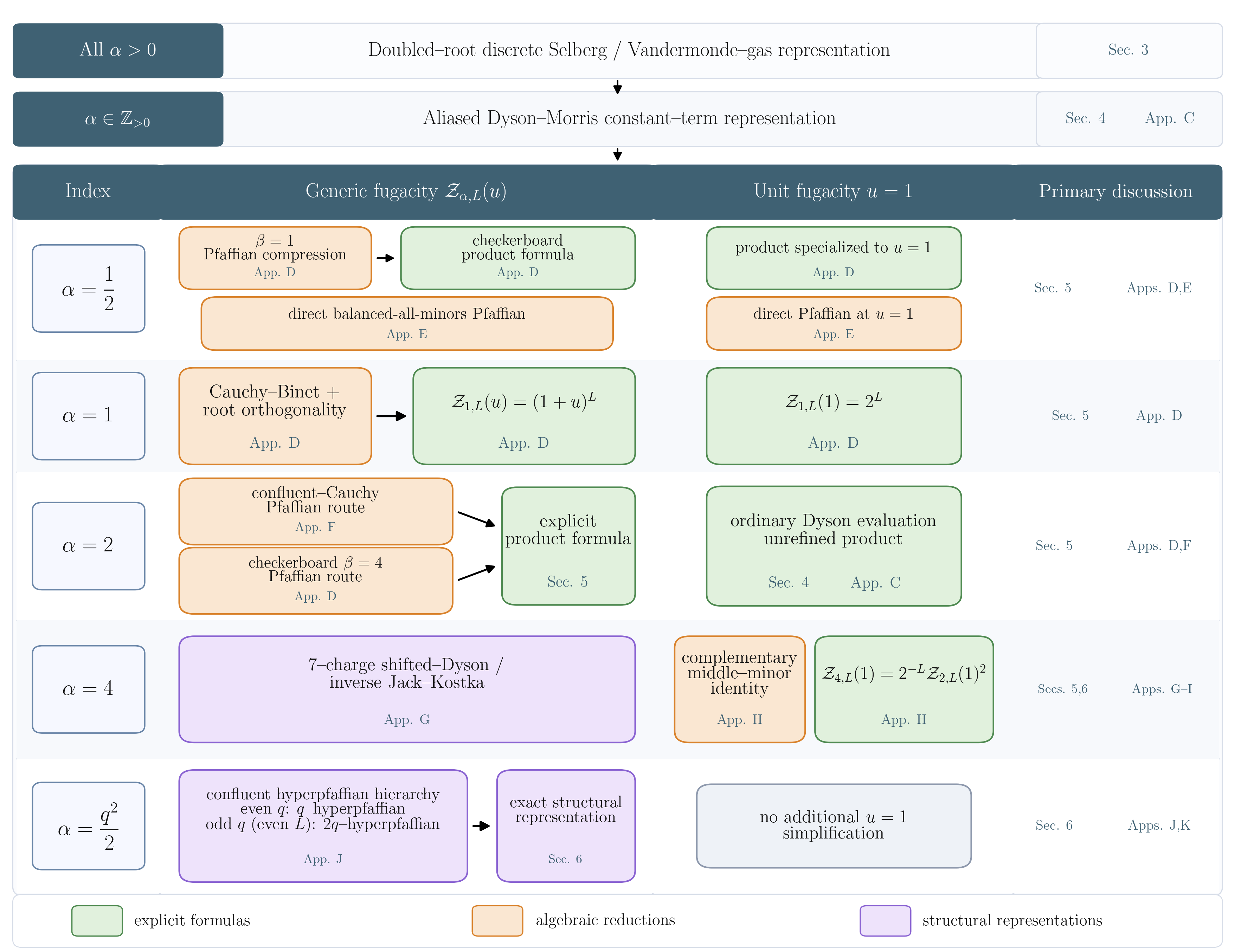}
\caption{
\textbf{Algebraic routes at special R\'enyi indices.}
The doubled-root discrete Selberg representation applies for all
$\alpha>0$, whereas the aliased Dyson--Morris constant-term
representation applies for positive integer $\alpha$.
At $\alpha=\tfrac12,1,2$, Pfaffian, determinant, and
root-orthogonality reductions lead to explicit formulas.
At $\alpha=4$, the generic-fugacity problem admits an exact
shifted-Dyson/Jack--Kostka representation, while the unit-fugacity
specialization collapses through a complementary middle-minor
identity.
For the perfect-square hierarchy $2\alpha=q^2$, the generic-fugacity
polynomial admits an exact hyperpfaffian representation, with no
additional universal simplification presently identified at $u=1$.
Colors distinguish algebraic reductions, explicit formulas, and
structural representations. The indicated sections and appendices
give the corresponding derivations.
}
\label{fig:algebraic-routes}
\end{figure*}

\subsection{Special algebraic representations and exactness levels}
\label{app:taxonomy-algebraic}

The representations above exist at the level of the full partition
function. Further algebraic compressions occur only in restricted
parameter regimes.

\paragraph{Two ordinary Pfaffian representations,
\(\alpha=\tfrac12\).}
For an arbitrary one-body weight on the doubled root lattice, the
partition function is an ordinary \(L\times L\) Pfaffian of skew moments.
For the checkerboard weight, Fourier sparsity reduces this ordered-root
Pfaffian to the product in Eq.~\eqref{eq:alpha-half-product}.
Independently, for the periodic TFI half-shift matrix,
Appendix~\ref{app:alpha-half-direct} gives a direct
\(4L\times4L\) balanced-all-minors Pfaffian for the complete fugacity
polynomial. The first representation exposes the root-ensemble
structure, while the second performs the absolute-minor sum directly
through a periodic sign selector.

\paragraph{Fourier-moment determinant, \(\alpha=1\).}
For an arbitrary one-body weight, the partition function is an
\(L\times L\) Toeplitz moment determinant. For the checkerboard weight,
root orthogonality diagonalizes the moment matrix and gives
\(\cZ_{1,L}(u)=(1+u)^L\).

\paragraph{Confluent de Bruijn Pfaffian, \(\alpha=2\).}
For an arbitrary one-body weight, the partition function is an ordinary
\(2L\times2L\) Pfaffian of value--derivative moments. For the
checkerboard weight, the sparse moment matrix decomposes into finite
blocks and gives the product in Eq.~\eqref{eq:alpha2-product}.

\paragraph{Complementary middle-minor representation.}
Positive integer indices can also be organized as moments of
complementary Cauchy or skew-Hilbert minors. In general, this is an exact
structural representation rather than an efficient evaluation. At
\(\alpha=4\) and \(u=1\), the middle-minor identity collapses to the
product in Eq.~\eqref{eq:alpha4-u1}.

\paragraph{Rectangular inverse Jack--Kostka representation.}
At \(\alpha=4\) and generic fugacity, the partition function becomes a
finite sum over neutral charge multiplicities and inverse Jack--Kostka
coefficients. The representation is exact, but it is neither a closed
product nor a proved polynomial-time evaluation.

\paragraph{Confluent hyperpfaffian representation.}
When \(2\alpha=q^2\), with \(q\in\mathbb Z_{>0}\), a multiplicity-\(q\)
confluent construction gives a finite higher-degree exterior-algebra
representation. The cases \(q=1,2\) reduce to ordinary Pfaffians,
whereas for \(q\geq3\) the natural object is a genuine hyperpfaffian and
need not be efficiently evaluable.

We use the following hierarchy of exactness. A \emph{closed product}
gives an immediate finite-size evaluation. A
\emph{polynomial-size determinant or Pfaffian} replaces the exponential
subset sum by ordinary polynomial-size linear algebra. An
\emph{exact structural representation}, such as a constant term, neutral
alias sum, complementary-minor moment, Jack--Kostka expansion, or
hyperpfaffian, is fully exact but may still lack both a closed product and
a proved polynomial-time evaluation.
\subsection{Terminology across representations}
\label{app:taxonomy-policy}

The terminology used throughout the paper reflects the representation of
the same underlying finite-size object:

\begin{itemize}[leftmargin=*,label={},itemsep=5pt]

\item
\textbf{All-minors and Pauli forms.}
In its original formulation, we refer to
\(\cZ_{\alpha,L}(u)\) as the
\emph{fugacity-resolved balanced all-minors partition function}.
Through the Pauli--minor correspondence, the same quantity is the
\emph{\(\alpha\)-tilted balanced Pauli-string partition function}.
After normalization by \(\cZ_{\alpha,L}(1)\), its fugacity polynomial
generates the \emph{balanced-degree full counting statistics}.

\item
\textbf{Log-gas and Selberg forms.}
Before the root-subset bijection, the all-minors sum has the form of a
\emph{neutral two-component defect log gas} on the two interlaced
sublattices. After the bijection, it becomes a
\emph{checkerboard-weighted half-filled discrete circular log gas}, or
equivalently a \emph{checkerboard-resolved discrete Selberg sum}. At
\(u=1\), the checkerboard field disappears, leaving the
\emph{half-filled discrete Dyson--Gaudin gas}. Its normalization to the
half-filled \(XX\)-chain Shannon--R\'enyi moment is stated explicitly
whenever this interpretation is used.

\item
\textbf{Constant-term forms.}
For positive integer \(\alpha\), finite Fourier bandwidth yields the
\emph{aliased Dyson constant-term representation}. The qualifier
``aliased'' records the nonzero Fourier sectors inherited from the
root-of-unity measure. We use \emph{ordinary Dyson constant term} only
when the visible alias kernel reduces to a constant.

\item
\textbf{Special algebraic reductions.}
Determinant, Pfaffian, complementary-minor, Jack--Kostka, and
hyperpfaffian terminology refers to the parameter regimes in which the
corresponding representations have been established. These structures
have different computational implications: a closed product gives an
immediate evaluation, a polynomial-size determinant or Pfaffian gives an
efficient finite compression, whereas a Jack--Kostka expansion or a
higher hyperpfaffian may remain an exact structural representation
without a known efficient evaluation.

\end{itemize}

Figure~\ref{fig:algebraic-routes} summarizes how these general
representations lead, at special R\'enyi indices, to algebraic
compressions, explicit formulas, or exact structural representations.

\section{Shannon factorization of the $XX_m$ chains}
\label{app:XXm-factorization}

In this appendix we prove the Shannon--R\'enyi decimation identity used in
Sec.~\ref{sec:web},
\begin{equation}
\label{eq:app-XXm-H-factorization-goal}
H_\alpha^{XX_m}(L)
=
m\,H_\alpha^{XX}(L/m),
\end{equation}
under the finite-size and boundary-sector assumptions stated below.  The
proof is independent of the stabilizer decimation identities of
Ref.~\cite{RamirezTrinoRajabpour2026}.  It is a direct consequence of the fact that
the range-$m$ hopping Hamiltonian decomposes into $m$ independent
squeezed $XX$ chains.

\subsection{Range-\texorpdfstring{$m$}{m} hopping and squeezed sublattices}

We work in a fixed fermionic boundary sector,
\begin{equation}
\label{eq:appB-boundary-twist}
c_{j+L}
=
e^{i\phi}c_j ,
\end{equation}
and consider the number-conserving range-$m$ hopping Hamiltonian
\begin{equation}
\label{eq:appB-XXm-fermion}
H_m
=
-\sum_{j=0}^{L-1}
\left(
c_j^\dagger c_{j+m}
+
c_{j+m}^\dagger c_j
\right),
\end{equation}
where indices outside the interval $0,\ldots,L-1$ are understood using
Eq.~\eqref{eq:appB-boundary-twist}.  We assume that $L=m\ell$.  The
lattice can then be decomposed into $m$ residue classes modulo $m$.  For
each residue $r=0,\ldots,m-1$, define the squeezed-chain fermions
\begin{equation}
\label{eq:appB-squeezed-fermions}
d_{r,s}
=
c_{r+ms},
\qquad
s=0,\ldots,\ell-1 .
\end{equation}
For $s=0,\ldots,\ell-2$, hopping by $m$ sites becomes nearest-neighbor
hopping in the squeezed coordinate, namely
$c_{r+ms+m}=d_{r,s+1}$.  The last hop is fixed by the
boundary condition and gives $d_{r,\ell}=e^{i\phi}d_{r,0}$.  Thus each squeezed chain
inherits the same fermionic boundary twist $e^{i\phi}$.

The Hamiltonian decomposes as
\begin{equation}
\label{eq:appB-H-decomposition}
H_m
=
\sum_{r=0}^{m-1}
H_{XX}^{(r)},
\end{equation}
where
\begin{equation}
\label{eq:appB-H-XX-r}
H_{XX}^{(r)}
=
-\sum_{s=0}^{\ell-1}
\left(
d_{r,s}^\dagger d_{r,s+1}
+
d_{r,s+1}^\dagger d_{r,s}
\right),
\qquad
d_{r,\ell}=e^{i\phi}d_{r,0}.
\end{equation}
Therefore the range-$m$ hopping chain is exactly a direct sum of $m$
ordinary $XX$ chains of length $\ell=L/m$, after reordering the sites into
the $m$ squeezed sublattices.

Equivalently, let $\Pi_m$ be the permutation that groups the sites by their
residue modulo $m$, namely
\[
\{\,r+jm : j=0,1,2,\ldots\,\},
\qquad
r=0,1,\ldots,m-1.
\]
Then the one-particle Hamiltonian satisfies
\begin{equation}
\label{eq:appB-one-particle-block}
\Pi_m h_m \Pi_m^{\mathsf T}
=
\bigoplus_{r=0}^{m-1}
h_{XX}^{(r,\phi)}(\ell).
\end{equation}
Here $h_{XX}^{(r,\phi)}(\ell)$ denotes the one-particle hopping matrix of
the ordinary $XX$ chain living on the $r$-th squeezed sublattice. 
\subsection{Ground state and correlation matrix}

Let $\ket{\Psi_m}$ be the Gaussian ground state of $H_m$ in the fixed
boundary sector with twist $\phi$.  If no allowed momentum lies exactly at
the Fermi level, this Gaussian ground state is unique in that sector.  If
zero modes are present, the statements below refer to the product Gaussian
occupation prescription inherited from the squeezed $XX$ chains.

Since the Hamiltonian is a direct sum of $m$ independent squeezed chains,
the Gaussian ground state factorizes as
\begin{equation}
\label{eq:appB-ground-state-product}
\ket{\Psi_m}
=
\bigotimes_{r=0}^{m-1}
\ket{\Psi_{XX}^{(r)}} .
\end{equation}
Here $\ket{\Psi_{XX}^{(r)}}$ is the ground state of the ordinary critical
$XX$ chain on the $r$-th squeezed sublattice.  Each squeezed chain has
length $\ell=L/m$ and carries the same boundary twist $\phi$.

Let
\begin{equation}
\label{eq:appB-C-def}
C_{jk}
=
\bra{\Psi_m} c_j^\dagger c_k \ket{\Psi_m}
\end{equation}
be the one-body correlation matrix of the $XX_m$ state on the original
length-$L$ chain.  In the squeezed basis, the factorized ground state gives
\begin{equation}
\label{eq:appB-C-block-components}
\left\langle
d_{r,s}^\dagger d_{r',t}
\right\rangle
=\delta_{r,r'} C^{XX}_{st},
\end{equation}
where $C^{XX}$ is the correlation matrix of one ordinary squeezed $XX$
chain.  Thus there are no correlations between different squeezed
sublattices.

Equivalently, after reordering the original sites by the permutation
matrix $\Pi_m$, the correlation matrix becomes block diagonal:
\begin{equation}
\label{eq:appB-C-block}
\Pi_m C \Pi_m^{\mathsf T}
=
\bigoplus_{r=0}^{m-1}
C_{XX}^{(r,\phi)}(\ell).
\end{equation}
This means that the same $XX$ correlation matrix appears once for each of
the $m$ squeezed chains.

For a number-conserving Gaussian state, the matrix entering the
computational-basis probability formula is
\begin{equation}
\label{eq:appB-G-def}
G
=
I_L-2C .
\end{equation}
Correspondingly, for one squeezed chain we define
\begin{equation}
\label{eq:appB-GXX-def}
G^{XX}
=
I_\ell-2C^{XX}.
\end{equation}
Using the block form of $C$, we obtain
\begin{equation}
\label{eq:appB-G-block}
\Pi_m G \Pi_m^{\mathsf T}
=
\bigoplus_{r=0}^{m-1}
G_{XX}^{(r,\phi)}(\ell).
\end{equation}
Here $G_{XX}^{(r,\phi)}(\ell)$ denotes the matrix $I_\ell-2C_{XX}^{(r,\phi)}(\ell)$
for the ordinary $XX$ chain living on the $r$-th squeezed sublattice.

This is the only matrix factorization needed for the Shannon--R\'enyi
identity.  Notice that $G$ here is simply the usual matrix $I_L-2C$ of the
number-conserving Gaussian ground state.  No additional indexed
``Shannon'' or ``stabilizer'' matrix is being introduced.

\subsection{Factorization of computational-basis probabilities}

Let $\mathbf n=(n_0,n_1,\ldots,n_{L-1})$, with $n_j\in\{0,1\}$, be a
computational-basis configuration of the original length-$L$ chain.  Define
\begin{equation}
\label{eq:appB-I-n}
I_{\mathbf n}
=
\operatorname{diag}
(2n_0-1,2n_1-1,\ldots,2n_{L-1}-1).
\end{equation}
For a number-conserving Gaussian state, the probability of the configuration
$\mathbf n$ is
\begin{equation}
\label{eq:appB-prob-det}
p_{\mathbf n}
=
\det
\left(
\frac{I_L-I_{\mathbf n}G}{2}
\right).
\end{equation}

We now split the configuration into its $m$ squeezed-sublattice
configurations.  For each residue class $r=0,\ldots,m-1$, define
\begin{equation}
\label{eq:appB-n-r-def}
\mathbf n^{(r)}
=
\left(
n_r,
n_{r+m},
n_{r+2m},
\ldots,
n_{r+(\ell-1)m}
\right).
\end{equation}
Thus $\mathbf n^{(r)}$ is the configuration seen by the $r$-th squeezed
chain, which has length $\ell=L/m$.

After applying the same site-reordering permutation $\Pi_m$, the diagonal
matrix $I_{\mathbf n}$ also becomes block diagonal:
\begin{equation}
\label{eq:appB-In-block}
\Pi_m I_{\mathbf n} \Pi_m^{\mathsf T}
=
\bigoplus_{r=0}^{m-1}
I_{\mathbf n^{(r)}} .
\end{equation}
Here $I_{\mathbf n^{(r)}}$ is the diagonal matrix associated with the
configuration $\mathbf n^{(r)}$ on the $r$-th squeezed chain.

Using this block form together with
Eq.~\eqref{eq:appB-G-block}, the determinant in
Eq.~\eqref{eq:appB-prob-det} factorizes into $m$ independent block
determinants:
\begin{equation}
\label{eq:appB-prob-det-factorized}
\begin{aligned}
p_{\mathbf n}^{XX_m}(L)
&=
\det
\left[
\frac{
I_L-I_{\mathbf n}G
}{2}
\right]
\\
&=
\prod_{r=0}^{m-1}
\det
\left[
\frac{
I_\ell
-
I_{\mathbf n^{(r)}}G_{XX}^{(r,\phi)}(\ell)
}{2}
\right].
\end{aligned}
\end{equation}
The determinant in the $r$-th factor is precisely the computational-basis
probability of the configuration $\mathbf n^{(r)}$ in the $r$-th squeezed
$XX$ chain.  Therefore
\begin{equation}
\label{eq:appB-prob-factorization}
p_{\mathbf n}^{XX_m}(L)
=
\prod_{r=0}^{m-1}
p_{\mathbf n^{(r)}}^{XX,(r,\phi)}(\ell).
\end{equation}

All squeezed chains are identical as $XX$ chains, but they act on different
sublattices.  Thus, when the sublattice label is not needed explicitly, one
may also write the last identity as
\begin{equation}
\label{eq:appB-prob-factorization-short}
p_{\mathbf n}^{XX_m}(L)
=
\prod_{r=0}^{m-1}
p_{\mathbf n^{(r)}}^{XX,\phi}(\ell).
\end{equation}
This is a factorization of the full computational-basis probability
distribution, not merely a factorization of the Hamiltonian spectrum.

\subsection{R\'enyi partition sums and entropy}

For a state $\ket{\Psi}$ with computational-basis probabilities
$p_{\mathbf n}$, define the Shannon--R\'enyi partition sum by
\begin{equation}
\label{eq:appB-Zalpha-def}
Z_\alpha[\Psi]
=
\sum_{\mathbf n}
p_{\mathbf n}^{\alpha}.
\end{equation}
For the $XX_m$ state, the probability factorization in
Eq.~\eqref{eq:appB-prob-factorization} gives
\begin{equation}
\label{eq:appB-Zalpha-factorization-derivation}
\begin{aligned}
Z_\alpha^{XX_m,\phi}(L)
&=
\sum_{\mathbf n}
\left[
p_{\mathbf n}^{XX_m,\phi}(L)
\right]^\alpha
\\
&=
\sum_{\mathbf n^{(0)},\ldots,\mathbf n^{(m-1)}}
\prod_{r=0}^{m-1}
\left[
p_{\mathbf n^{(r)}}^{XX,(r,\phi)}(\ell)
\right]^\alpha
\\
&=
\prod_{r=0}^{m-1}
\sum_{\mathbf n^{(r)}}
\left[
p_{\mathbf n^{(r)}}^{XX,(r,\phi)}(\ell)
\right]^\alpha .
\end{aligned}
\end{equation}
The second line uses the fact that specifying the full configuration
$\mathbf n$ is equivalent to specifying the $m$ squeezed-chain
configurations
$\mathbf n^{(0)},\ldots,\mathbf n^{(m-1)}$.

Since all squeezed chains are identical as $XX$ chains with the same twist
$\phi$, each factor in the last line is the same.  Therefore
\begin{equation}
\label{eq:appB-Z-factorization}
Z_\alpha^{XX_m,\phi}(L)
=
\left[
Z_\alpha^{XX,\phi}(\ell)
\right]^m,
\qquad
\ell=L/m .
\end{equation}
When the boundary sector is fixed and no confusion can arise, we will omit
the superscript $\phi$.

For $\alpha\neq1$, the Shannon--R\'enyi entropy is
\begin{equation}
\label{eq:appB-Halpha-def}
H_\alpha
=
\frac{1}{1-\alpha}
\log Z_\alpha .
\end{equation}
Using Eq.~\eqref{eq:appB-Z-factorization}, we obtain
\begin{equation}
\label{eq:appB-H-factorization}
H_\alpha^{XX_m,\phi}(L)
=
m\,H_\alpha^{XX,\phi}(\ell),
\qquad
\ell=L/m .
\end{equation}
Equivalently, after suppressing the fixed boundary-sector label,
\begin{equation}
\label{eq:appB-H-factorization-short}
H_\alpha^{XX_m}(L)
=
m\,H_\alpha^{XX}(L/m).
\end{equation}

The Shannon entropy at $\alpha=1$ follows either by taking the limit
$\alpha\to1$, or directly from the additivity of the entropy of a product
probability distribution:
\begin{equation}
\label{eq:appB-H1-factorization}
H_1^{XX_m}(L)
=
m\,H_1^{XX}(L/m).
\end{equation}
\subsection{Relation to the TFI stabilizer entropy}

The result above is purely a Shannon--R\'enyi identity.  It follows from
the factorization of the computational-basis probability distribution of
the $XX_m$ chain, and does not rely on the stabilizer-minor decimation
theorem.

We now combine this Shannon identity with the TFI--XX relation of
Ref.~\cite{RamirezTrinoRajabpour2026}.  With the stabilizer entropy convention used in
this work, this relation reads
\begin{equation}
\label{eq:appB-TFI-XX-relation}
H_\alpha^{XX}(2N)
=
\mathcal M_\alpha^{\rm TFI}(N)
+
N\log 2 .
\end{equation}
Taking $L=2mN$ in Eq.~\eqref{eq:appB-H-factorization-short}, we obtain
\begin{equation}
\label{eq:appB-XXm-TFI-final}
H_\alpha^{XX_m}(2mN)
=
m\,H_\alpha^{XX}(2N)
=
m\,\mathcal M_\alpha^{\rm TFI}(N)
+
mN\log 2 .
\end{equation}

Thus the critical TFI stabilizer entropy controls the
computational-basis Shannon--R\'enyi entropy of the full $XX_m$ family,
up to the additive normalization term fixed by the stabilizer entropy
convention.  The mechanism is different from the stabilizer decimation
discussed in Ref.~\cite{RamirezTrinoRajabpour2026}: here the result follows from the
factorization of the $XX_m$ probability distribution into $m$ squeezed
$XX$ chains.
\section{Details of the aliased Dyson--Morris representation}
\label{app:aliased-dyson}

This appendix supplies the technical details of the aliased
Dyson--Morris representation developed in
Sec.~\ref{sec:constantterm}.  We derive the Fourier expansion of the
weighted root-of-unity comb, prove the finite alias truncation, and record
useful trigonometric forms, shifted-Dyson sectors, and special-fugacity
checks.

The discrete Selberg sum and its Dirac-comb integral representation are
valid for every real $\alpha>0$.  The finite alias kernel and the resulting
constant-term formula require
$\alpha\in\mathbb Z_{>0}$, for which the Vandermonde weight is a Laurent
polynomial of finite Fourier bandwidth. The derivation follows the sequence summarized below.

\begin{center}
\setlength{\fboxsep}{4pt}
\renewcommand{\arraystretch}{0.85}

\begin{tabular}{c}

\colorbox{TFILightBlue}{%
\parbox{0.78\columnwidth}{\centering
\textcolor{TFIBlue}{\bfseries 1}\enspace
Weighted root-of-unity comb}}
\\[-0.2mm]
{\color{TFIBlue}\(\Big\downarrow\)}
\\[-0.2mm]

\colorbox{TFILightBlue}{%
\parbox{0.78\columnwidth}{\centering
\textcolor{TFIBlue}{\bfseries 2}\enspace
Fourier alias expansion}}
\\[-0.2mm]
{\color{TFIBlue}\(\Big\downarrow\)}
\\[-0.2mm]

\colorbox{TFILightBlue}{%
\parbox{0.78\columnwidth}{\centering
\textcolor{TFIBlue}{\bfseries 3}\enspace
Finite-bandwidth truncation}}
\\[-0.2mm]
{\color{TFIBlue}\(\Big\downarrow\)}
\\[-0.2mm]

\colorbox{TFILightBlue}{%
\parbox{0.78\columnwidth}{\centering
\textcolor{TFIBlue}{\bfseries 4}\enspace
Neutral shifted-Dyson sectors}}
\\[-0.2mm]
{\color{TFIBlue}\(\Big\downarrow\)}
\\[-0.2mm]

\parbox{0.78\columnwidth}{\centering
\textcolor{TFIBlue}{\bfseries 5}\enspace
Finite aliased constant-term formula}

\end{tabular}
\end{center}

\subsection{Fourier coefficients of the weighted Dirac comb}

Let $\theta_a=\pi a/L$ and $\zeta_a=e^{i\theta_a}$, with
$a=0,\ldots,2L-1$.  Since $\zeta_a^L=(-1)^a$, the one-root weight on the
doubled root lattice can be written as
\begin{equation}
\label{eq:appC-wa}
w_u(\zeta_a)
=
\frac{1+u}{2}
+
\frac{u-1}{2}(-1)^a .
\end{equation}
This gives weight $u$ on the even sublattice and weight $1$ on the odd
sublattice.

The weighted Dirac comb is
\begin{equation}
\label{eq:appC-comb-def}
\mathscr A_{L,u}(\theta)
=
2\pi
\sum_{a=0}^{2L-1}
w_u(\zeta_a)
\delta_{2\pi}(\theta-\theta_a).
\end{equation}
Using the Fourier expansion of the periodic delta function,
\begin{equation}
\label{eq:appC-periodic-delta}
\delta_{2\pi}(\theta-\theta_a)
=
\frac{1}{2\pi}
\sum_{n\in\mathbb Z}
e^{in(\theta-\theta_a)},
\end{equation}
we obtain
\begin{equation}
\label{eq:appC-comb-fourier-start}
\mathscr A_{L,u}(\theta)
=
\sum_{n\in\mathbb Z}
e^{in\theta}
\sum_{a=0}^{2L-1}
w_u(\zeta_a)e^{-in\pi a/L}.
\end{equation}

The two root-of-unity sums needed here are
\begin{equation}
\label{eq:appC-root-sums}
\sum_{a=0}^{2L-1}e^{-in\pi a/L}
=
\begin{cases}
2L, & n\equiv0 \pmod{2L},\\
0, & \text{otherwise},
\end{cases}
\end{equation}
and
\begin{equation}
\label{eq:appC-root-sums-shifted}
\begin{aligned}
\sum_{a=0}^{2L-1}(-1)^a e^{-in\pi a/L}
&=
\sum_{a=0}^{2L-1}e^{-i(n-L)\pi a/L}
\\
&=
\begin{cases}
2L, & n\equiv L \pmod{2L},\\
0,  & \text{otherwise}.
\end{cases}
\end{aligned}
\end{equation}
Therefore only Fourier modes $n=qL$ survive.  Even $q$ comes from the
uniform part of the weight, while odd $q$ comes from the staggered part.
Thus
\begin{equation}
\label{eq:appC-comb-fourier}
\mathscr A_{L,u}(\theta)
=
2L
\sum_{q\in\mathbb Z}
c_q(u)e^{iqL\theta},
\end{equation}
with
\begin{equation}
\label{eq:appC-cell}
c_q(u)
=
\begin{cases}
\dfrac{1+u}{2}, & q\ \mathrm{even},\\[2mm]
\dfrac{u-1}{2}, & q\ \mathrm{odd}.
\end{cases}
\end{equation}
This proves Eq.~\eqref{eq:comb-Fourier-series} of the main text.

\subsection{Finite alias truncation for integer \texorpdfstring{$\alpha$}{alpha}}

The finite truncation of the comb is the only step where the assumption
$\alpha\in\mathbb Z_{>0}$ is needed.  For integer $\alpha$, the circular
Vandermonde weight becomes the Laurent polynomial
\begin{equation}
\label{eq:appC-Dalpha}
D_\alpha(x)
=
\prod_{\substack{i,j=1\\ i\neq j}}^L
\left(
1-\frac{x_i}{x_j}
\right)^\alpha .
\end{equation}
In every monomial of \(D_\alpha(x)\), the exponent \(m_j\) of any fixed
variable \(x_j\) satisfies
\begin{equation}
\label{eq:appC-bandwidth}
-\alpha(L-1)
\le m_j\le
\alpha(L-1).
\end{equation}
On the other hand, the Fourier expansion of the weighted comb contains
only powers of the form $x_j^{qL}$.  If $|q|\ge\alpha$, then
$|qL|\ge\alpha L$, which is outside the exponent range available from
$D_\alpha(x)$ in the variable $x_j$.  Such a mode can therefore never be
cancelled to produce a constant term.  Hence, for the purpose of
constant-term extraction, the infinite Fourier series of the comb may be
replaced exactly by the finite alias kernel
\begin{equation}
\label{eq:appC-K-def}
K_{\alpha,u}(x)
=
\sum_{q=-(\alpha-1)}^{\alpha-1}
c_q(u)x^{qL}.
\end{equation}
This is the finite alias kernel of Eq.~\eqref{eq:finite-alias-kernel}.

For non-integer $\alpha$, the Vandermonde weight is not a Laurent
polynomial.  The finite-bandwidth argument above is then unavailable, and
the Dirac-comb representation cannot be reduced to this finite
constant-term form.

\subsection{Trigonometric and Dirichlet-kernel forms of the alias kernel}

The finite alias kernel also admits useful trigonometric forms.  Set
$a=\alpha-1$ and $t=L\theta$.  Since
$c_q(u)=(u+(-1)^q)/2$, the kernel can be written as
\begin{equation}
\label{eq:appC-K-Dirichlet}
K_{\alpha,u}(e^{i\theta})
=
\frac12
\left[
u\,\mathcal D_{\alpha-1}(L\theta)
+
\mathcal D_{\alpha-1}(L\theta+\pi)
\right],
\end{equation}
where
\begin{equation}
\label{eq:appC-Dirichlet}
\mathcal D_n(t)
=
\sum_{q=-n}^{n}e^{iqt}
=
\frac{\sin\!\left[(n+\frac12)t\right]}{\sin(t/2)}
\end{equation}
is the finite Dirichlet kernel, with the usual limiting value at zeros of
the denominator.

Equivalently, separating the even and odd alias sectors gives the manifestly
real form
\begin{equation}
\label{eq:appC-K-cosine}
\begin{aligned}
K_{\alpha,u}(e^{i\theta})
={}&
\frac{1+u}{2}
\left[
1+
2\sum_{m=1}^{\lfloor(\alpha-1)/2\rfloor}
\cos(2mL\theta)
\right]
\\
&+
(u-1)
\sum_{m=0}^{\lfloor(\alpha-2)/2\rfloor}
\cos\!\left[(2m+1)L\theta\right].
\end{aligned}
\end{equation}
Thus, for real $u$, the alias kernel is real on the unit circle.

At the unrefined point $u=1$, only even aliases survive.  Hence
\begin{equation}
\label{eq:appC-K-u1-cosine}
K_{\alpha,1}(e^{i\theta})
=
1+
2\sum_{m=1}^{\lfloor(\alpha-1)/2\rfloor}
\cos(2mL\theta).
\end{equation}
Equivalently,
\begin{equation}
\label{eq:appC-K-u1-sine}
K_{\alpha,1}(e^{i\theta})
=
\frac{
\sin\!\left(\nu_\alpha L\theta\right)
}{
\sin(L\theta)
},
\end{equation}
where
\begin{equation}
\label{eq:appC-nu-alpha}
\nu_\alpha
=
2\left\lfloor\frac{\alpha-1}{2}\right\rfloor+1
=
\begin{cases}
\alpha, & \alpha\ \mathrm{odd},\\
\alpha-1, & \alpha\ \mathrm{even}.
\end{cases}
\end{equation}
At points where $\sin(L\theta)=0$, the quotient is understood by
continuity.

For the first few integer indices one obtains
\begin{equation}
\label{eq:appC-K-small-u}
\begin{aligned}
K_{1,u}(e^{i\theta})
&=
\frac{1+u}{2},
\\
K_{2,u}(e^{i\theta})
&=
\frac{1+u}{2}
+
(u-1)\cos(L\theta),
\\
K_{3,u}(e^{i\theta})
&=
\frac{1+u}{2}
\left[
1+2\cos(2L\theta)
\right]
+
(u-1)\cos(L\theta),
\\
K_{4,u}(e^{i\theta})
&=
\frac{1+u}{2}
\left[
1+2\cos(2L\theta)
\right]
\\&+
(u-1)
\left[
\cos(L\theta)+\cos(3L\theta)
\right],
\\
K_{5,u}(e^{i\theta})
&=
\frac{1+u}{2}
\left[
1+2\cos(2L\theta)+2\cos(4L\theta)
\right]
\\
&\hspace{1cm}
+
(u-1)
\left[
\cos(L\theta)+\cos(3L\theta)
\right].
\end{aligned}
\end{equation}

At $u=1$, these reduce to
\begin{equation}
\label{eq:appC-K-small-u1}
\begin{aligned}
K_{1,1}(x)&=1,
\\
K_{2,1}(x)&=1,
\\
K_{3,1}(x)&=1+x^{2L}+x^{-2L},
\\
K_{4,1}(x)&=1+x^{2L}+x^{-2L},
\\
K_{5,1}(x)
&=
1+x^{2L}+x^{-2L}+x^{4L}+x^{-4L}.
\end{aligned}
\end{equation}
Thus the unrefined kernel collapses to one only for $\alpha=1$ and
$\alpha=2$.

\subsection{Ordinary Dyson versus aliased Dyson}

For integer $\alpha$, inserting the finite alias kernel into the
constant-term representation gives
\begin{equation}
\label{eq:appC-CT}
\cZ_{\alpha,L}(u)
=
\frac{2^L}{L!\,L^{(\alpha-1)L}}
\CT_{x_1,\ldots,x_L}
\left[
D_\alpha(x)
\prod_{j=1}^L K_{\alpha,u}(x_j)
\right].
\end{equation}
The prefactor comes from the product of the $L$ Fourier-comb factors,
namely $L^{-\alpha L}(2L)^L=2^L/L^{(\alpha-1)L}$.

This should be distinguished from the ordinary Dyson identity, which
evaluates only the constant term of $D_\alpha(x)$:
\begin{equation}
\label{eq:appC-Dyson}
\CT_{x_1,\ldots,x_L} D_\alpha(x)
=
\frac{(\alpha L)!}{(\alpha!)^L}.
\end{equation}
The present expression is different because it contains the additional
alias factor $\prod_{j=1}^L K_{\alpha,u}(x_j)$.  Therefore the ordinary
Dyson identity gives the full answer only when this alias factor collapses
to a constant.

Two useful checks follow immediately.  First, for $\alpha=1$, the alias
kernel contains only the zero mode, $K_{1,u}(x)=c_0(u)=(1+u)/2$.  Since
$\CT D_1(x)=L!$, Eq.~\eqref{eq:appC-CT} gives
\begin{equation}
\label{eq:appC-Z-alpha1}
\cZ_{1,L}(u)
=
\frac{2^L}{L!}
\left(
\frac{1+u}{2}
\right)^L
L!
=
(1+u)^L.
\end{equation}
This agrees with the Cauchy--Binet result.

Second, for $\alpha=2$ at the unrefined point $u=1$, the alias kernel also
collapses, $K_{2,1}(x)=1$.  Therefore Eq.~\eqref{eq:appC-CT} reduces to
the ordinary Dyson constant term:
\begin{equation}
\label{eq:appC-Z-alpha2}
\cZ_{2,L}(1)
=
\frac{2^L}{L!\,L^L}
\CT D_2(x)
=
\frac{2^L}{L!\,L^L}
\frac{(2L)!}{2^L}
=
\frac{(2L)!}{L!\,L^L}.
\end{equation}
These are the two cases in which the unrefined integer-index problem is
evaluated directly by the ordinary Dyson identity.  For higher integer
indices, nontrivial alias sectors remain, and the ordinary Dyson identity
alone is no longer sufficient.

\subsection{Shifted Dyson coefficients and small alias sectors}

Expanding the alias kernels gives a finite sum of shifted Dyson
coefficients.  For alias charges $q_1,\ldots,q_L$, define
\begin{equation}
\label{eq:appC-shifted-C}
\mathscr C_{\alpha,L}(q_1,\ldots,q_L)
=
\CT_{x_1,\ldots,x_L}
\left[
D_\alpha(x)
\prod_{j=1}^L x_j^{q_j L}
\right].
\end{equation}
Thus the aliased constant-term formula can be written as
\begin{equation}
\label{eq:appC-shifted-expansion}
\begin{aligned}
\cZ_{\alpha,L}(u)
&=
\frac{2^L}{L!\,L^{(\alpha-1)L}}
\sum_{q_1,\ldots,q_L=-(\alpha-1)}^{\alpha-1}
\\
&\quad\times
\left[
\prod_{j=1}^L c_{q_j}(u)
\right]
\mathscr C_{\alpha,L}
(q_1,\ldots,q_L).
\end{aligned}
\end{equation}
The shifted coefficient vanishes unless the aliases are neutral:
\begin{equation}
\label{eq:appC-alias-neutrality}
\mathscr C_{\alpha,L}(q_1,\ldots,q_L)=0
\qquad
\text{unless}
\qquad
\sum_{j=1}^L q_j=0.
\end{equation}
Indeed, $D_\alpha(x)$ is invariant under the common scaling
$x_j\mapsto t x_j$, while the inserted monomial
$\prod_{j=1}^L x_j^{q_jL}$ gains the factor
$t^{L\sum_j q_j}$.  A nonzero constant term can therefore occur only when
the total alias charge is zero.

At $u=1$, only even aliases survive.  For $\alpha=1$ and $\alpha=2$, this
leaves only the zero alias, so the answer reduces to the ordinary Dyson
sector.  For $\alpha=3$ and $\alpha=4$, the surviving aliases are
$q_j\in\{-2,0,2\}$.  Neutral sectors are therefore labelled by two disjoint
subsets $A,B\subseteq\{1,\ldots,L\}$ with $|A|=|B|$: the set $A$ carries
the $+2$ aliases, and the set $B$ carries the $-2$ aliases.  Define
\begin{equation}
\label{eq:appC-qAB-def}
q_j(A,B)
=
\begin{cases}
2, & j\in A,\\
-2, & j\in B,\\
0, & j\notin A\cup B.
\end{cases}
\end{equation}
Then, for $\alpha=3$ or $\alpha=4$ at $u=1$,
\begin{equation}
\label{eq:appC-alpha34-sector}
\begin{aligned}
\cZ_{\alpha,L}(1)
&=
\frac{2^L}{L!\,L^{(\alpha-1)L}}
\sum_{r=0}^{\lfloor L/2\rfloor}
\sum_{\substack{
A,B\subseteq\{1,\ldots,L\}\\
A\cap B=\emptyset\\
|A|=|B|=r}}
\\[-1mm]
&\quad\times
\mathscr C_{\alpha,L}
\bigl(q_1(A,B),\ldots,q_L(A,B)\bigr).
\end{aligned}
\end{equation}
The term $r=0$ is the ordinary Dyson sector.  The terms with $r\ge1$ are
the nontrivial alias sectors.  In particular, their presence at
$\alpha=4$ explains why the exact $\alpha=4$ evaluation in the main text
cannot follow from the ordinary Dyson identity alone.

\subsection{Special fugacity values and consistency checks}

The fugacity $u$ keeps track of the number $|U\cap X|$ of selected roots on
the even sublattice.  Several special values are useful.

At $u=1$, one obtains the unrefined all-minors sum relevant for the
stabilizer and Shannon--R\'enyi entropies.  In the alias kernel, this keeps
only even aliases:
\begin{equation}
\label{eq:appC-K-u1-special}
K_{\alpha,1}(x)
=
\sum_{\substack{q=-(\alpha-1)\\ q\ \mathrm{even}}}^{\alpha-1}
x^{qL}.
\end{equation}
Thus the kernel collapses to one for $\alpha=1$ and $\alpha=2$, while
nontrivial aliases survive for higher integer indices.

At $u=0$, only the sector with $|U\cap X|=0$ contributes to the discrete
Selberg sum.  The only such half-filled subset is $U=Y$.  Since
$|\Delta(Y)|^{2\alpha}=L^{\alpha L}$, the prefactor in the discrete
Selberg formula gives
\begin{equation}
\label{eq:appC-u0}
\cZ_{\alpha,L}(0)=1
\end{equation}
for every real $\alpha>0$.  By palindromicity, the coefficient of $u^L$ is
also one.  In the finite alias representation for integer $\alpha$, the
same endpoint value is encoded in the alternating alias kernel
$K_{\alpha,0}(e^{i\theta})
=
\mathcal D_{\alpha-1}(L\theta+\pi)/2$.

At $u=-1$, the generating polynomial becomes the alternating sum over the
degree $|U\cap X|$.  The alias coefficients are $c_q(-1)=0$ for even $q$
and $c_q(-1)=-1$ for odd $q$.  Thus only odd aliases survive in
$K_{\alpha,-1}$.  No closed form follows in general, but this value is a
useful diagnostic of the alias sectors.

These special cases reproduce the solvable checks used in the main text.
For $\alpha=1$, the finite alias formula gives
$\cZ_{1,L}(u)=(1+u)^L$ for arbitrary fugacity.  For $\alpha=2$ at $u=1$,
the alias kernel collapses and Dyson's identity gives
\begin{equation}
\label{eq:appC-alpha2-check}
\cZ_{2,L}(1)
=
\frac{(2L)!}{L!\,L^L}.
\end{equation}
Finally, for every real $\alpha>0$, the discrete Selberg formula gives the
endpoint value $\cZ_{\alpha,L}(0)=1$, with the unit coefficient at $u^L$
following from palindromicity.  For small $L$, the finite constant-term
expression can also be expanded directly and checked term by term against
the original all-minors sum.

\section{Generic one-body weights at
\texorpdfstring{$\beta=1,2,4$}{beta=1,2,4}}
\label{app:generic-weight}

This appendix separates two levels of solvability for the half-filled
discrete circular ensemble.

At the first level, the one-body weights
\(\mathbf w=(w_0,\ldots,w_{2L-1})\) are arbitrary. At the three classical
values \(\beta=1,2,4\), the corresponding weighted ensemble admits,
respectively, an ordinary Pfaffian, a determinant, and a confluent
Pfaffian representation. These are generic-weight polynomial-size
compressions and do not, for a general choice of \(\mathbf w\), imply
closed product formulas.

At the second level, the arbitrary weights are specialized to the
checkerboard choice relevant to the fugacity-resolved TFI problem,
\[
w_a(u)=
\begin{cases}
u, & a\ \mathrm{even},\\
1, & a\ \mathrm{odd}.
\end{cases}
\]
The Fourier moments of this particular weight are highly sparse. This
additional structure diagonalizes the generic determinant at
\(\beta=2\) and reduces the generic Pfaffians at \(\beta=1,4\) to finite
reflected blocks, thereby producing the explicit fugacity-resolved
formulas used in the main text. Generic-weight compressibility and
checkerboard product solvability are therefore distinct statements.

The relation with the notation of the main text is
\[
\beta=2\alpha,
\]
so that \(\beta=1,2,4\) correspond to
\(\alpha=\tfrac12,1,2\). The \(\beta=1\) construction and the
checkerboard product formulas below are stated for positive even \(L\).

\subsection{Generic weighted half-filled root ensemble}
\label{appF:weighted-ensemble}

Let $N=2L$, and
\begin{equation}
\label{eq:appF-roots}
z_a=e^{2\pi\ii a/N}=e^{\pi\ii a/L},
\qquad
a=0,\ldots,N-1 .
\end{equation}
We assign a one-body weight $w_a$ to each root $z_a$ and write
\[
\mathbf w=(w_0,\ldots,w_{N-1}).
\]
For an ordered half-filled subset
\[
A=\{a_1<\cdots<a_L\}\subset\{0,\ldots,N-1\},
\]
define
\begin{equation}
\label{eq:appF-Delta}
\Delta_A
=
\Delta(z_{a_1},\ldots,z_{a_L})
=
\prod_{1\le r<s\le L}(z_{a_s}-z_{a_r}).
\end{equation}
The generic weighted discrete circular ensemble is
\begin{equation}
\label{eq:appF-weighted-Z}
\cZ^{(\beta)}_L[\mathbf w]
=
L^{-\beta L/2}
\sum_{0\le a_1<\cdots<a_L\le N-1}
\left(\prod_{j=1}^{L}w_{a_j}\right)
\abs{\Delta_A}^{\beta}.
\end{equation}
For nonnegative weights this is a positive discrete log-gas partition
function.  The identities below are polynomial identities in the
one-body weights and therefore extend to arbitrary complex $\mathbf w$.

The fugacity-resolved polynomial of the main text is recovered by taking
$\beta=2\alpha$ and choosing the checkerboard weight
\begin{equation}
\label{eq:appF-checkerboard-weight-early}
w_a(u)
=
\begin{cases}
u, & a\ \mathrm{even},\\
1, & a\ \mathrm{odd}.
\end{cases}
\end{equation}
With this specialization,
\begin{equation}
\label{eq:appF-main-identification-early}
\cZ^{(2\alpha)}_L[\mathbf w(u)]
=
\cZ_{\alpha,L}(u).
\end{equation}

We also introduce the discrete Fourier moments
\begin{equation}
\label{eq:appF-Fourier-moments}
\widehat w_m
=
\sum_{a=0}^{N-1}w_a z_a^m,
\qquad
m\in\mathbb Z.
\end{equation}
Since $z_a^N=1$, they satisfy $\widehat w_{m+N}=\widehat w_m$.

\subsection{Generic-weight ordered-root Pfaffian at
\texorpdfstring{$\beta=1$}{beta=1}}
\label{appF:beta-one}

At $\beta=1$, the absolute Vandermonde becomes a single determinant once
the selected roots are placed in circular order.  The ordered subset sum
can then be evaluated by the finite de Bruijn identity.

Introduce the centered set of modes
\begin{equation}
\label{eq:appF-centered-modes}
\mathcal K_L
=
\left\{
-\frac{L-1}{2},
-\frac{L-3}{2},
\ldots,
\frac{L-3}{2},
\frac{L-1}{2}
\right\}.
\end{equation}
For even $L$, these modes are half-integers.  We use the convention
\begin{equation}
z_a^m
=
e^{2\pi\ii a m/N},
\qquad
m\in\mathcal K_L.
\end{equation}

\medskip
\noindent\textbf{Ordered absolute Vandermonde identity.}
\label{lem:appF-ordered-vandermonde}
Let $A=\{a_1<\cdots<a_L\}$ and set
\[
\mu_L=\frac{L(L-1)}{2}.
\]
Then
\begin{equation}
\label{eq:appF-ordered-absolute-vdm}
\det
\left[
z_{a_r}^{m}
\right]_
{\substack{1\le r\le L\\ m\in\mathcal K_L}}
=
\ii^{\mu_L}\abs{\Delta_A},
\end{equation}
where the columns are ordered by increasing $m$.

\medskip
\noindent\textit{Derivation.}
The centered determinant differs from the ordinary Vandermonde determinant
only by a row phase:
\begin{align}
\det
\left[
z_{a_r}^{m}
\right]_
{\substack{1\le r\le L\\ m\in\mathcal K_L}}
&=
\left(
\prod_{r=1}^{L}
z_{a_r}^{-(L-1)/2}
\right)
\det
\left[
z_{a_r}^{s-1}
\right]_{r,s=1}^{L}
\notag\\
&=
\exp\left[
-\frac{\pi\ii}{N}(L-1)
\sum_{r=1}^{L}a_r
\right]
\Delta_A.
\label{eq:appF-centered-to-ordinary}
\end{align}
For $a_r<a_s$,
\[
z_{a_s}-z_{a_r}
=
2\ii\,
\exp\left[
\frac{\pi\ii}{N}(a_r+a_s)
\right]
\sin\left[
\frac{\pi(a_s-a_r)}{N}
\right].
\]
Because $0<a_s-a_r<N$, the sine factor is positive.  Multiplying over all
pairs gives
\[
\Delta_A
=
\ii^{\mu_L}
\exp\left[
\frac{\pi\ii}{N}(L-1)
\sum_{r=1}^{L}a_r
\right]
\abs{\Delta_A}.
\]
Substitution into Eq.~\eqref{eq:appF-centered-to-ordinary} proves the
claim.

For $m\in\mathcal K_L$, define
\begin{equation}
\label{eq:appF-phi}
\phi_m(a)=w_a z_a^m.
\end{equation}

\medskip
\noindent\textbf{Generic-weight Pfaffian at
\texorpdfstring{$\beta=1$}{beta=1}.}
\label{thm:appF-beta1-pfaffian}
Let $\mathsf A^{(1)}[\mathbf w]$ be the $L\times L$ skew-symmetric matrix
indexed by $m,n\in\mathcal K_L$, with entries
\begin{equation}
\label{eq:appF-beta1-skew-moments}
\mathsf A^{(1)}_{mn}[\mathbf w]
=
\sum_{0\le a<b\le N-1}
w_aw_b
\left(
z_a^m z_b^n
-
z_a^n z_b^m
\right).
\end{equation}
Then
\begin{equation}
\label{eq:appF-beta1-pfaffian}
\cZ_L^{(1)}[\mathbf w]
=
L^{-L/2}\ii^{-\mu_L}
\Pf \mathsf A^{(1)}[\mathbf w].
\end{equation}

\medskip
\noindent\textit{Derivation.}
Using Eq.~\eqref{eq:appF-ordered-absolute-vdm},
\begin{equation}
\cZ_L^{(1)}[\mathbf w]
=
L^{-L/2}\ii^{-\mu_L}
\sum_{0\le a_1<\cdots<a_L\le N-1}
\det
\left[
\phi_m(a_r)
\right]_
{\substack{1\le r\le L\\ m\in\mathcal K_L}}.
\end{equation}
Since $L$ is even, the finite de Bruijn identity evaluates the ordered
subset sum as the Pfaffian of the skew moments
\[
\sum_{0\le a<b\le N-1}
\left[
\phi_m(a)\phi_n(b)
-
\phi_n(a)\phi_m(b)
\right].
\]
Using Eq.~\eqref{eq:appF-phi} gives
Eq.~\eqref{eq:appF-beta1-skew-moments}, proving the result.

For the checkerboard choice $\mathbf w=\mathbf w(u)$,
\begin{equation}
\label{eq:appF-beta1-main-specialization}
\cZ_L^{(1)}[\mathbf w(u)]
=
\cZ_{1/2,L}(u).
\end{equation}

\subsection{Generic-weight Andr\'eief determinant at
\texorpdfstring{$\beta=2$}{beta=2}}
\label{appF:beta-two}

At $\beta=2$, the Vandermonde weight is a product of two ordinary
determinants, so the half-filled subset sum is evaluated by the finite
Cauchy--Binet, or Andr\'eief, identity.

Introduce the $L\times L$ Toeplitz moment matrix
\begin{equation}
\label{eq:appF-beta2-moment-matrix}
\mathsf M^{(2)}_{rs}[\mathbf w]
=
\sum_{a=0}^{N-1}
w_a z_a^{r-s}
=
\widehat w_{r-s},
\qquad
r,s=0,\ldots,L-1.
\end{equation}

\medskip
\noindent\textbf{Generic-weight determinant at
\texorpdfstring{$\beta=2$}{beta=2}.}
\label{thm:appF-beta2-determinant}
For arbitrary one-body weights $\mathbf w$,
\begin{equation}
\label{eq:appF-beta2-determinant}
\cZ_L^{(2)}[\mathbf w]
=
L^{-L}
\det \mathsf M^{(2)}[\mathbf w].
\end{equation}

\medskip
\noindent\textit{Derivation.}
For every half-filled subset $A=\{a_1<\cdots<a_L\}$,
\begin{equation}
\label{eq:appF-beta2-two-determinants}
\abs{\Delta_A}^{2}
=
\det
\left[
z_{a_j}^{r}
\right]_
{\substack{0\le r\le L-1\\1\le j\le L}}
\det
\left[
z_{a_j}^{-s}
\right]_
{\substack{1\le j\le L\\0\le s\le L-1}}.
\end{equation}
Absorbing the one-body weights into the first determinant and applying
Cauchy--Binet gives the determinant of the matrix with entries
\[
\sum_{a=0}^{N-1}
w_a z_a^{r-s},
\]
which is precisely $\mathsf M^{(2)}[\mathbf w]$.  Restoring the
normalization $L^{-L}$ establishes the result.

For the checkerboard choice,
\begin{equation}
\label{eq:appF-beta2-main-specialization}
\cZ_L^{(2)}[\mathbf w(u)]
=
\cZ_{1,L}(u).
\end{equation}

\subsection{Generic-weight confluent Pfaffian at
\texorpdfstring{$\beta=4$}{beta=4}}
\label{appF:beta-four}

At $\beta=4$, the fourth power of the Vandermonde is linearized by a
multiplicity-two confluent Vandermonde determinant.  The resulting
half-filled subset sum is compressed to an ordinary Pfaffian.

Let
\begin{equation}
\label{eq:appF-polynomials}
p_r(z)=z^r,
\qquad
p_r'(z)=rz^{r-1},
\qquad
r=0,\ldots,2L-1.
\end{equation}

\medskip
\noindent\textbf{Multiplicity-two confluent Vandermonde identity.}
\label{lem:appF-confluent-vandermonde}
For distinct points $z_1,\ldots,z_L$,
\begin{equation}
\label{eq:appF-confluent-vandermonde}
\begin{aligned}
&\det
\begin{bmatrix}
 p_0(z_1) & p_0'(z_1) & \cdots & p_0(z_L) & p_0'(z_L)\\
 p_1(z_1) & p_1'(z_1) & \cdots & p_1(z_L) & p_1'(z_L)\\
 \vdots & \vdots & & \vdots & \vdots\\
 p_{2L-1}(z_1) & p_{2L-1}'(z_1)
 & \cdots &
 p_{2L-1}(z_L) & p_{2L-1}'(z_L)
\end{bmatrix}
\\[-1mm]
&\qquad=
\Delta(z_1,\ldots,z_L)^4.
\end{aligned}
\end{equation}
\medskip
\noindent\textit{Derivation.}
This is the standard confluent Vandermonde identity.  Starting from the
ordinary Vandermonde determinant in $2L$ variables, group the variables in
pairs and take the confluent limit
\[
z_{j,0},z_{j,1}\to z_j,
\qquad
j=1,\ldots,L,
\]
after dividing by the internal Vandermonde factor of each pair.  Each point
then contributes a value column and a derivative column, while every
inter-pair Vandermonde factor occurs with multiplicity four.

On the unit circle,
\begin{equation}
\label{eq:appF-unit-circle-fourth-power}
\abs{\Delta_A}^4
=
\left(
\prod_{a\in A}z_a^{-2(L-1)}
\right)
\Delta_A^4.
\end{equation}
Define the $2L\times2L$ skew-symmetric matrix
$\mathsf B^{(4)}[\mathbf w]$, indexed by
$r,s=0,\ldots,2L-1$, by
\begin{align}
\label{eq:appF-beta4-skew-moments}
\mathsf B^{(4)}_{rs}[\mathbf w]
&=
\sum_{a=0}^{N-1}
w_a z_a^{-2(L-1)}
\left[
p_r(z_a)p_s'(z_a)
-
p_r'(z_a)p_s(z_a)
\right]
\notag\\
&=
(s-r)
\sum_{a=0}^{N-1}
w_a z_a^{r+s-2L+1}
\notag\\
&=
(s-r)\widehat w_{r+s-2L+1}.
\end{align}
Thus $\mathsf B^{(4)}[\mathbf w]$ is Hankel-like: its Fourier index
depends on $r+s$ rather than on $r-s$.

\medskip
\noindent\textbf{Generic-weight Pfaffian at
\texorpdfstring{$\beta=4$}{beta=4}.}
\label{thm:appF-beta4-pfaffian}
For arbitrary one-body weights $\mathbf w$,
\begin{equation}
\label{eq:appF-beta4-pfaffian}
\cZ_L^{(4)}[\mathbf w]
=
L^{-2L}
\Pf \mathsf B^{(4)}[\mathbf w].
\end{equation}

\medskip
\noindent\textit{Derivation.}
Insert Eqs.~\eqref{eq:appF-unit-circle-fourth-power} and
\eqref{eq:appF-confluent-vandermonde} into
Eq.~\eqref{eq:appF-weighted-Z}.  Each selected root contributes the pair
of confluent columns formed by its value and derivative vectors.  The
Pfaffian minor-summation identity then sums over all choices of these
two-column blocks.  The induced skew pairing between polynomial labels
$r$ and $s$ is exactly Eq.~\eqref{eq:appF-beta4-skew-moments}.
Restoring the normalization $L^{-2L}$ proves the result.

For the checkerboard choice,
\begin{equation}
\label{eq:appF-beta4-main-specialization}
\cZ_L^{(4)}[\mathbf w(u)]
=
\cZ_{2,L}(u).
\end{equation}

The results obtained so far are valid for arbitrary complex one-body
weights \(\mathbf w\). They provide polynomial-size determinant or
Pfaffian representations, but they do not in general yield closed
products.

We now impose the specific checkerboard weight relevant to the
fugacity-resolved TFI problem. Its Fourier moments are supported only at
multiples of \(L\). This additional sparsity converts the generic moment
matrices into diagonal or finite-block forms and is the mechanism behind
the explicit formulas below.

\subsection{Checkerboard specialization and explicit formulas}
\label{appF:checkerboard}

For the checkerboard weight in
Eq.~\eqref{eq:appF-checkerboard-weight-early}, one may equivalently write
\begin{equation}
\label{eq:appF-checkerboard-cd}
w_a(u)
=
c+d z_a^L,
\qquad
c=\frac{1+u}{2},
\qquad
d=\frac{u-1}{2},
\end{equation}
because $z_a^L=(-1)^a$.  Its Fourier moments obey
\begin{equation}
\label{eq:appF-checkerboard-Fourier}
\widehat w_m(u)
=
\sum_{a=0}^{N-1}w_a(u)z_a^m
=
\begin{cases}
L(1+u), & m\equiv0\pmod{2L},\\[1mm]
L(u-1), & m\equiv L\pmod{2L},\\[1mm]
0, & \mathrm{otherwise}.
\end{cases}
\end{equation}
This sparse Fourier support reduces the three generic-weight compressions
to the classical checkerboard formulas.

\subsubsection{Checkerboard reduction of the generic
\texorpdfstring{$\beta=1$}{beta=1} Pfaffian}

For half-integer exponents define
\begin{equation}
\label{eq:appF-SPQ}
S(P,Q)
=
\sum_{0\le a<b\le N-1}
\left(
z_a^Pz_b^Q
-
z_a^Qz_b^P
\right).
\end{equation}
A geometric-series evaluation followed by root orthogonality gives
\begin{equation}
\label{eq:appF-SPQ-support}
S(P,Q)=0
\qquad
\text{unless}
\qquad
P+Q\equiv0\pmod N,
\end{equation}
and, in the nonzero case,
\begin{equation}
\label{eq:appF-SPQ-value}
S(P,Q)
=
N\,\frac{z_1^P+1}{z_1^P-1}.
\end{equation}

Expanding the checkerboard weights in
Eq.~\eqref{eq:appF-beta1-skew-moments} gives
\begin{equation}
\label{eq:appF-beta1-checkerboard-expand}
\begin{aligned}
\mathsf A^{(1)}_{mn}[\mathbf w(u)]
={}&
c^2S(m,n)
+
cdS(m+L,n)
\\
&+
cdS(m,n+L)
+
d^2S(m+L,n+L).
\end{aligned}
\end{equation}
For even $L$, the mixed terms vanish and the only nonzero entries pair
$m=-t$ with $n=t$, where
\[
t=\frac12,\frac32,\ldots,\frac{L-1}{2}.
\]
The corresponding reflected-mode entry is
\begin{equation}
\label{eq:appF-beta1-even-block}
\mathsf A^{(1)}_{-t,t}[\mathbf w(u)]
=
2L\ii
\left[
c^2\cot\left(\frac{\pi t}{2L}\right)
-
d^2\tan\left(\frac{\pi t}{2L}\right)
\right].
\end{equation}
Hence the Pfaffian decomposes into independent $2\times2$ blocks, giving
\begin{equation}
\label{eq:appF-beta1-product}
\cZ_{1/2,L}(u)
=
\prod_{j=1}^{L/2}
\left[
1+u^2
+
2u\sec\left(\frac{\pi(2j-1)}{2L}\right)
\right].
\end{equation}

\subsubsection{Checkerboard reduction of the generic
\texorpdfstring{$\beta=2$}{beta=2} determinant}

Since $r-s\in[-(L-1),L-1]$, only the zero Fourier mode in
Eq.~\eqref{eq:appF-checkerboard-Fourier} contributes to the Toeplitz
moment matrix.  Therefore
\begin{equation}
\label{eq:appF-beta2-checkerboard-matrix}
\mathsf M^{(2)}_{rs}[\mathbf w(u)]
=
L(1+u)\delta_{rs},
\end{equation}
and Eq.~\eqref{eq:appF-beta2-determinant} gives
\begin{equation}
\label{eq:appF-beta2-checkerboard-result}
\cZ_{1,L}(u)
=
(1+u)^L.
\end{equation}

\subsubsection{Checkerboard reduction of the generic
\texorpdfstring{$\beta=4$}{beta=4} Pfaffian}

For $\beta=4$, the Fourier index
$r+s-2L+1$ lies between $-2L+1$ and $2L-1$.  Hence only the modes
$0,\pm L$ survive, and
\begin{equation}
\label{eq:appF-beta4-checkerboard-sparse}
\begin{aligned}
\mathsf B^{(4)}_{rs}[\mathbf w(u)]
&=
L(s-r)\Big[
(1+u)\delta_{r+s,\,2L-1}
\\
&+(u-1)\delta_{r+s,\,L-1}
+(u-1)\delta_{r+s,\,3L-1}
\Big].
\end{aligned}
\end{equation}
Thus the generic confluent Pfaffian is supported on three anti-diagonals.
Its decomposition into reflected $4\times4$ blocks and the resulting
product formula
\begin{equation}
\label{eq:appF-beta4-product}
\cZ_{2,L}(u)
=
\prod_{j=1}^{L/2}
\left[
(1+u)^2
-
4u\left(\frac{2j-1}{L}\right)^2
\right]
\end{equation}
are derived in Appendix~\ref{app:alpha-two-checkerboard}.

\subsection{Generic-weight compressions and checkerboard outcomes}
\label{appF:summary}

The three generic-weight compressions are collected in
Table~\ref{tab:appF-generic-weight-summary}.

\begin{table*}[t]
\centering
\small
\renewcommand{\arraystretch}{1.35}
\setlength{\tabcolsep}{4.5pt}

\begingroup
\setlength{\fboxsep}{6pt}

\colorbox{TFILightBlue}{%
\begin{tabular}{
@{}
p{0.055\textwidth}
p{0.10\textwidth}
p{0.39\textwidth}
p{0.10\textwidth}
p{0.24\textwidth}
@{}
}
\toprule

\centering\textcolor{TFIBlue}{\bfseries \(\beta\)}
&
\centering\textcolor{TFIBlue}{\bfseries \(\alpha=\beta/2\)}
&
\textcolor{TFIBlue}{\bfseries Generic-weight finite object}
&
\centering\textcolor{TFIBlue}{\bfseries Size}
&
\textcolor{TFIBlue}{\bfseries Checkerboard outcome}
\tabularnewline
\midrule

\centering\textcolor{TFIViolet}{\(\mathbf{1}\)}
&
\centering\(\displaystyle \frac12\)
&
Ordered-root de Bruijn Pfaffian with skew moments
\(\mathsf A^{(1)}_{mn}[\mathbf w]\).
&
\centering\(L\times L\)
&
Reflected \(2\times2\) blocks and the product in
Eq.~\eqref{eq:appF-beta1-product}.
\tabularnewline[1mm]

\centering\textcolor{TFIViolet}{\(\mathbf{2}\)}
&
\centering\(\displaystyle 1\)
&
Andr\'eief/Cauchy--Binet determinant of the Toeplitz moment matrix
\(\mathsf M^{(2)}_{rs}[\mathbf w]=\widehat w_{r-s}\).
&
\centering\(L\times L\)
&
Diagonal moment matrix and
\(\cZ_{1,L}(u)=(1+u)^L\).
\tabularnewline[1mm]

\centering\textcolor{TFIViolet}{\(\mathbf{4}\)}
&
\centering\(\displaystyle 2\)
&
Confluent de Bruijn Pfaffian with Hankel-like skew moments
\(\mathsf B^{(4)}_{rs}[\mathbf w]=(s-r)\widehat w_{r+s-2L+1}\).
&
\centering\(2L\times2L\)
&
Three anti-diagonals, reflected \(4\times4\) blocks, and the product in
Eq.~\eqref{eq:appF-beta4-product}.
\tabularnewline

\bottomrule
\end{tabular}%
}

\endgroup

\caption{
The two levels of solvability at the classical values
\(\beta=1,2,4\). For arbitrary one-body weights, the ensemble admits a
polynomial-size determinant or Pfaffian representation. For the
checkerboard weight, sparse Fourier support reduces these generic finite
objects further and produces the explicit fugacity-resolved formulas.
}
\label{tab:appF-generic-weight-summary}
\end{table*}

Table~\ref{tab:appF-generic-weight-summary} makes the distinction
between the two levels explicit. The determinant and Pfaffian
compressions exist for arbitrary one-body weights. The closed
fugacity-resolved formulas require the additional checkerboard Fourier
sparsity: it diagonalizes the \(\beta=2\) moment matrix and reduces the
\(\beta=1,4\) Pfaffians to finite reflected blocks. After the
identification \(\beta=2\alpha\), these are the classical solvable
indices \(\alpha=\tfrac12,1,2\) used in the main text.


\section{A direct all-minors Pfaffian at
\texorpdfstring{\(\alpha=\tfrac12\)}{alpha=1/2}}
\label{app:alpha-half-direct}

This appendix proves the direct Pfaffian representation
\eqref{eq:alpha-half-direct-pfaffian}. Unlike the ordered-root de Bruijn
construction of Appendix~\ref{app:generic-weight}, the present argument
acts directly on the balanced minors of the periodic critical TFI
half-shift matrix. The proof has two ingredients: a periodic selector
whose principal Pfaffians reproduce the signs of the trigonometric Cauchy
minors, and a principal-Pfaffian summation identity that performs the
complete balanced all-minors sum.

\subsection{Antisymmetric lift and periodic selector}

For each \(j\in[L]\), introduce the lifted labels
\(\mathsf r_j=2j\) and \(\mathsf c_j=2j+1\), ordered as
\[
(\mathsf r_0,\mathsf c_0,\mathsf r_1,\mathsf c_1,\ldots,
\mathsf r_{L-1},\mathsf c_{L-1}).
\]
The \(2L\times2L\) antisymmetric lift of \(G\) is defined by
\begin{equation}
\label{eq:app-half-lift-definition}
\mathbb A(G)_{\mathsf r_j,\mathsf c_k}=G_{jk},
\qquad
\mathbb A(G)_{\mathsf c_k,\mathsf r_j}=-G_{jk},
\end{equation}
with vanishing \(\mathsf r\)-\(\mathsf r\) and
\(\mathsf c\)-\(\mathsf c\) blocks. For \(S,T\subseteq[L]\), let
\begin{equation}
\label{eq:app-half-principal-index-set}
\mathscr I(S,T)
=
\{\mathsf r_s:s\in S\}
\cup
\{\mathsf c_t:t\in T\},
\end{equation}
where every principal submatrix inherits the interleaved order above.

Let \(\mathbb J^{(0)}_{2L}\) be the standard antisymmetric selector
\begin{equation}
\label{eq:app-half-J0-definition}
\left(\mathbb J^{(0)}_{2L}\right)_{\mu\nu}
=
\begin{cases}
+1,&\mu<\nu,\\
0,&\mu=\nu,\\
-1,&\mu>\nu.
\end{cases}
\end{equation}
Every even-dimensional principal submatrix of
\(\mathbb J^{(0)}_{2L}\) has Pfaffian one. Define
\begin{equation}
\label{eq:app-half-selector-data}
\mathsf P_{\rm sw}
=
\bigoplus_{j=0}^{L-1}
\begin{pmatrix}
0&1\\
1&0
\end{pmatrix},
\qquad
D_{\rm P}
=
\bigoplus_{j=0}^{L-1}
\begin{pmatrix}
(-1)^{j+1}&0\\
0&(-1)^j
\end{pmatrix}.
\end{equation}
The periodic selector is
\begin{equation}
\label{eq:app-half-JPBC-definition}
\mathbb J^{\rm PBC}_{2L}
=
D_{\rm P}\,
\mathsf P_{\rm sw}
\mathbb J^{(0)}_{2L}
\mathsf P_{\rm sw}^{T}
D_{\rm P}.
\end{equation}
It is real and antisymmetric.

\subsection{Termwise cancellation of the minor signs}

Let
\[
S=\{s_1<\cdots<s_k\},
\qquad
T=\{t_1<\cdots<t_k\},
\]
and define
\begin{equation*}
\label{eq:app-half-N-definitions}
\begin{aligned}
N_{<}(S,T)
&=
\#\bigl\{(s,t)\in S\times T:\ s<t\bigr\},
\\
N_{>}(S,T)
&=
\#\bigl\{(s,t)\in S\times T:\ s>t\bigr\}.
\end{aligned}
\end{equation*}
Set \(m=|S\cap T|\), so that
\begin{equation}
\label{eq:app-half-N-relation}
N_{<}(S,T)+N_{>}(S,T)+m=k^2.
\end{equation}

With
\(x_j=\pi(j+\tfrac12)/L\) and \(y_j=\pi j/L\), the trigonometric
Cauchy identity gives
\begin{equation}
\label{eq:app-half-G-minor-Cauchy}
\begin{aligned}
\det G[S,T]
={}&
\frac{
(-1)^{
\sum_{s\in S}s+\sum_{t\in T}t
}
}{L^k}
\\
&\times
\frac{
\displaystyle
\prod_{p<q}
\sin(x_{s_p}-x_{s_q})
\sin(y_{t_q}-y_{t_p})
}{
\displaystyle
\prod_{p,q}
\sin(x_{s_p}-y_{t_q})
}.
\end{aligned}
\end{equation}
For \(p<q\), the first sine in the numerator is negative and the second
is positive, while
\(\sin(x_s-y_t)<0\) exactly when \(s<t\). Hence
\begin{equation}
\label{eq:app-half-G-minor-sign}
\operatorname{sgn}\det G[S,T]
=
(-1)^{
\sum_{s\in S}s+
\sum_{t\in T}t+
\binom{k}{2}+
N_{<}(S,T)
}.
\end{equation}

To relate the determinant to the lifted Pfaffian, reorder
\(\mathscr I(S,T)\) from the inherited interleaved order to
\[
(\mathsf r_{s_1},\ldots,\mathsf r_{s_k},
 \mathsf c_{t_1},\ldots,\mathsf c_{t_k}).
\]
This requires \(N_{>}(S,T)\) transpositions. Using
\[
\Pf
\begin{pmatrix}
0&X\\
-X^T&0
\end{pmatrix}
=
(-1)^{\binom{k}{2}}\det X,
\]
one obtains
\begin{equation}
\label{eq:app-half-lift-principal-pf}
\Pf\!\left[
\mathbb A(G)_{\mathscr I(S,T)}
\right]
=
(-1)^{\binom{k}{2}+N_{>}(S,T)}
\det G[S,T].
\end{equation}
Combining Eqs.~\eqref{eq:app-half-N-relation},
\eqref{eq:app-half-G-minor-sign}, and
\eqref{eq:app-half-lift-principal-pf} gives
\begin{equation}
\label{eq:app-half-lift-principal-sign}
\operatorname{sgn}
\Pf\!\left[
\mathbb A(G)_{\mathscr I(S,T)}
\right]
=
(-1)^{
\sum_{s\in S}s+
\sum_{t\in T}t+
k-m
}.
\end{equation}

We now evaluate the corresponding principal Pfaffian of the selector.
The pair swap reverses the two selected labels at precisely the
\(m=|S\cap T|\) sites where both are present, contributing \((-1)^m\).
The restriction of \(D_{\rm P}\) contributes
\[
\det\!\left[(D_{\rm P})_{\mathscr I(S,T)}\right]
=
(-1)^{
\sum_{s\in S}s+
\sum_{t\in T}t+
k
}.
\]
Since every even principal Pfaffian of
\(\mathbb J^{(0)}_{2L}\) equals one,
\begin{equation}
\label{eq:app-half-JPBC-principal-pf}
\Pf\!\left[
\left(\mathbb J^{\rm PBC}_{2L}\right)_{\mathscr I(S,T)}
\right]
=
(-1)^{
\sum_{s\in S}s+
\sum_{t\in T}t+
k+m
}.
\end{equation}
Because \(k-m\equiv k+m\pmod 2\), the two signs coincide. Moreover,
Eq.~\eqref{eq:app-half-lift-principal-pf} shows that the lifted
Pfaffian has magnitude \(|\det G[S,T]|\), while the selector Pfaffian has
unit magnitude. Therefore
\begin{equation}
\label{eq:app-half-termwise-absolute}
\Pf\!\left[
\mathbb A(G)_{\mathscr I(S,T)}
\right]
\Pf\!\left[
\left(\mathbb J^{\rm PBC}_{2L}\right)_{\mathscr I(S,T)}
\right]
=
|\det G[S,T]|.
\end{equation}

\subsection{Principal-Pfaffian summation and fugacity}

For arbitrary antisymmetric \(2L\times2L\) matrices \(X\) and \(Y\),
\begin{equation}
\label{eq:app-half-principal-pf-summation}
(-1)^L
\Pf
\begin{pmatrix}
X&I_{2L}\\
-I_{2L}&-Y
\end{pmatrix}
=
\sum_{\substack{
\mathscr I\subseteq\{0,\ldots,2L-1\}\\
|\mathscr I|\ {\rm even}
}}
\Pf(X_{\mathscr I})\Pf(Y_{\mathscr I}),
\end{equation}
with the empty principal Pfaffian equal to one. This follows directly
from the perfect-matching expansion: indices outside
\(\mathscr I\) pair vertically through the identity blocks, while the
same set \(\mathscr I\) is paired internally through \(X\) and \(Y\).

Set \(X=u\,\mathbb A(G)\) and
\(Y=\mathbb J^{\rm PBC}_{2L}\). Since \(\mathbb A(G)\) has only
\(\mathsf r\)-\(\mathsf c\) entries, its principal Pfaffian vanishes
unless the principal set contains the same number of labels of both
types. Every nonzero sector is therefore uniquely
\(\mathscr I(S,T)\) with \(|S|=|T|=k\). Its dimension is \(2k\), so
Pfaffian homogeneity gives
\[
\Pf\!\left[
\bigl(u\mathbb A(G)\bigr)_{\mathscr I(S,T)}
\right]
=
u^k
\Pf\!\left[
\mathbb A(G)_{\mathscr I(S,T)}
\right].
\]
Using Eq.~\eqref{eq:app-half-termwise-absolute} in
Eq.~\eqref{eq:app-half-principal-pf-summation}, we obtain
\begin{equation}
\label{eq:app-half-final-direct-pfaffian}
\begin{aligned}
&(-1)^L
\Pf
\begin{pmatrix}
u\,\mathbb A(G)&I_{2L}\\
-I_{2L}&-\mathbb J^{\rm PBC}_{2L}
\end{pmatrix}
\\
&\qquad=
\sum_{k=0}^{L}u^k
\sum_{\substack{S,T\subseteq[L]\\|S|=|T|=k}}
|\det G[S,T]|
=
\cZ_{1/2,L}(u).
\end{aligned}
\end{equation}
This proves Eq.~\eqref{eq:alpha-half-direct-pfaffian}.

The direct construction and the doubled-root de Bruijn construction are
thus two independent polynomial-size compressions of the same fugacity
polynomial. The former cancels the periodic minor signs sector by sector;
the latter exposes the checkerboard Fourier structure and yields the
closed product in Eq.~\eqref{eq:alpha-half-product}.

\section{Two Pfaffian routes at
\texorpdfstring{$\alpha=2$}{alpha=2}}
\label{app:alpha-two}

This appendix develops two distinct Pfaffian routes for the
fugacity-resolved polynomial \(\cZ_{2,L}(u)\). We assume throughout that
\(L\) is positive and even.

The first route is the checkerboard specialization of the generic-weight
\(\beta=4\) Pfaffian derived in
Appendix~\ref{app:generic-weight}. The generic weighted-root ensemble is
represented by a \(2L\times2L\) confluent de Bruijn Pfaffian. For the
checkerboard one-body weight, the Fourier moments are supported only at
\(0,\pm L\), so the skew moment matrix decomposes into independent
reflected \(4\times4\) blocks. Their Pfaffians yield the explicit product
in Eq.~\eqref{eq:alpha2-product}.

The second route is independent of the weighted-root construction and
works directly with the balanced all-minors definition in
Eq.~\eqref{eq:centralZ}. A multiplicity-two confluent Cauchy identity
linearizes each fourth power \(\abs{\det G[S,T]}^4\), while a paired-index
Pfaffian summation performs the complete sum over balanced subsets. This
produces a \(4L\times4L\) Pfaffian representation of the full polynomial
\(\cZ_{2,L}(u)\). It establishes polynomial-size Pfaffian compressibility
directly at the all-minors level, but does not by itself expose the
closed-product factors.

The two constructions therefore have complementary roles. The
weighted-root route inherited from Appendix~\ref{app:generic-weight}
explains the checkerboard block factorization and the explicit product,
whereas the direct all-minors route provides an independent Pfaffian
compression without first passing to the doubled-root ensemble.

\subsection{Checkerboard block factorization of the weighted
\texorpdfstring{$\beta=4$}{beta=4} Pfaffian}
\label{app:alpha-two-checkerboard}

We begin with the route inherited directly from
Appendix~\ref{app:generic-weight}. There, the arbitrary-weight
\(\beta=4\) ensemble was reduced to the \(2L\times2L\) confluent
de Bruijn Pfaffian
\[
\cZ_L^{(4)}[\mathbf w]
=
L^{-2L}\Pf\mathsf B^{(4)}[\mathbf w].
\]
We now specialize that general construction to the checkerboard weight
\(\mathbf w(u)\). By Eq.~\eqref{eq:appF-checkerboard-Fourier}, only
the Fourier modes \(0,\pm L\) contribute. Writing
\begin{equation}
\label{eq:appG-ab}
a=\frac{1+u}{2},
\qquad
b=\frac{u-1}{2},
\end{equation}
the generic skew matrix takes the form
\[
\mathsf B^{(4)}[\mathbf w(u)]
=
2L\,A^{(L)}(u),
\]
where \(A^{(L)}(u)\) is the \(2L\times2L\) skew-symmetric matrix
\begin{equation}
\label{eq:appG-A-explicit}
\begin{aligned}
A^{(L)}_{mn}(u)
=
(n-m)\Big[
&a\,\delta_{m+n,\,2L-1}
\\
&+b\,\delta_{m+n,\,L-1}
+b\,\delta_{m+n,\,3L-1}
\Big],
\end{aligned}
\end{equation}
with $m,n=0,\ldots,2L-1$. Using the generic normalization
\[
\cZ_{2,L}(u)
=
L^{-2L}
\Pf\mathsf B^{(4)}[\mathbf w(u)]
\]
and the homogeneity
\(\Pf(cA)=c^L\Pf(A)\) for a \(2L\times2L\) skew matrix, we find
\begin{equation}
\label{eq:appG-Z-pf-small}
\cZ_{2,L}(u)
=
L^{-2L}\Pf\!\left(2L\,A^{(L)}(u)\right)
=
\frac{2^L}{L^L}\Pf A^{(L)}(u).
\end{equation}

Introduce the reflected indices
\begin{equation}
\label{eq:appG-reflected-pairs}
f_i=i,
\qquad
g_i=2L-1-i,
\qquad
i=0,\ldots,L-1,
\end{equation}
and define
\begin{equation}
\label{eq:appG-si}
s_i=L-1-2i.
\end{equation}
Up to antisymmetry, the only nonzero matrix elements are
\begin{align}
\label{eq:appG-A-fg}
A_{f_i,g_i}
&=
a(L+s_i),
\\
\label{eq:appG-A-ff}
A_{f_i,f_{L-1-i}}
&=
b\,s_i,
\\
\label{eq:appG-A-gg}
A_{g_i,g_{L-1-i}}
&=
-b\,s_i.
\end{align}

For \(i=0,\ldots,L/2-1\), the reflected pair \(i,L-1-i\) produces one
independent \(4\times4\) block, with \(s_i>0\). In the ordered basis
\[
(f_i,g_i,f_{L-1-i},g_{L-1-i}),
\]
the block is
\begin{equation}
\label{eq:appG-4block}
A_s
=
\begin{pmatrix}
0&a(L+s)&bs&0\\
-a(L+s)&0&0&-bs\\
-bs&0&0&a(L-s)\\
0&bs&-a(L-s)&0
\end{pmatrix},
\end{equation}
with $s=s_i>0$. Its Pfaffian is
\begin{align}
\label{eq:appG-block-pf-simple}
\Pf A_s
&=
a^2(L^2-s^2)+b^2s^2
\notag\\
&=
\frac{L^2(1+u)^2}{4}-u\,s^2,
\end{align}
where \(a^2-b^2=u\). Reordering the complete matrix into these blocks can
only produce an overall, \(u\)-independent Pfaffian sign; the endpoint
condition \(\cZ_{2,L}(0)=1\) fixes this sign to be positive.

For even \(L\), the positive values of \(s_i\) are
\[
1,3,\ldots,L-1.
\]
Substituting the block Pfaffians into
Eq.~\eqref{eq:appG-Z-pf-small} gives
\begin{equation}
\label{eq:appG-alpha2-product-final}
\cZ_{2,L}(u)
=
\prod_{r=1}^{L/2}
\left[
(1+u)^2
-
4u\left(\frac{2r-1}{L}\right)^2
\right],
\end{equation}
which is Eq.~\eqref{eq:alpha2-product}. At \(u=1\),
\begin{equation}
\label{eq:appG-alpha2-u1-from-product}
\cZ_{2,L}(1)
=
\frac{(2L)!}{L!\,L^L}
=
\frac{2^L(2L-1)!!}{L^L},
\end{equation}
in agreement with the ordinary Dyson evaluation in
Sec.~\ref{sec:solvable-indices} and
Appendix~\ref{app:aliased-dyson}.

\subsection{Confluent-Cauchy compression of the all-minors sum}
\label{app:alpha-two-confluent-cauchy}

We now turn to a second, independent Pfaffian organization that starts
directly from the balanced all-minors sum and does not use the
generic-weight root-ensemble Pfaffian of
Appendix~\ref{app:generic-weight}.

We index the sites by \(j,k=0,\ldots,L-1\), identifying \([L]\) with
\(\{0,\ldots,L-1\}\) in this appendix, and use the root sets
\[
X=\{x_j\}_{j=0}^{L-1},
\qquad
Y=\{y_k\}_{k=0}^{L-1},
\]
introduced in Sec.~\ref{sec:cauchy}. Recall the Cauchy factorization
\begin{equation}
\label{eq:appG-G-Cauchy}
G
=
\frac{2\ii}{L}D_X C D_Y,
\qquad
C_{jk}
=
\frac{1}{x_j-y_k}.
\end{equation}
The diagonal phase matrices are explicitly
\begin{equation}
\label{eq:appG-G-Cauchy-phases}
(D_X)_{jj}
=
(-1)^j e^{\pi\ii j/L},
\qquad
(D_Y)_{kk}
=
(-1)^k e^{\pi\ii(k-1/2)/L}.
\end{equation}
All entries of \(D_X\) and \(D_Y\) have unit modulus. Since \(G\) is
real,
\[
(\det G[S,T])^4
=
\abs{\det G[S,T]}^4.
\]

For each site, introduce two confluent labels \(a=0,1\). If
\(S\subseteq[L]\), define
\begin{equation}
\label{eq:appG-doubled-set}
S^{[2]}
=
\{(j,0),(j,1):j\in S\},
\end{equation}
and define \(T^{[2]}\) analogously. In every confluent minor, the labels
are ordered lexicographically as
\((j,0),(j,1)\), with the site indices \(j\) in increasing order.

Let \(\mathbb M\) be the \(2L\times2L\) matrix
\begin{equation}
\label{eq:appG-M-def}
\begin{aligned}
\mathbb M_{(j,a),(k,b)}
={}&
\left(\frac{2}{L}\right)^2
(D_X)_{jj}^{\,2}(D_Y)_{kk}^{\,2}
\\
&\times
\partial_{x_j}^{a}(-\partial_{y_k})^{b}
\frac{1}{x_j-y_k},
\qquad
a,b\in\{0,1\}.
\end{aligned}
\end{equation}
The sign in \((-\partial_{y_k})^b\) is a convenient confluent gauge. For
one pair of variables, the corresponding local block is
\begin{equation}
\label{eq:appG-local-block}
\begin{pmatrix}
\dfrac{1}{x-y}
&
-\dfrac{1}{(x-y)^2}
\\[3mm]
-\dfrac{1}{(x-y)^2}
&
\dfrac{2}{(x-y)^3}
\end{pmatrix},
\end{equation}
whose determinant is \((x-y)^{-4}\).

\medskip
\noindent\textbf{Multiplicity-two confluent Cauchy identity.}
For \(S,T\subseteq[L]\) with \(|S|=|T|\),
\begin{equation}
\label{eq:appG-confluent-C}
\det
\left[
\partial_{x_j}^{a}
(-\partial_{y_k})^{b}
\frac{1}{x_j-y_k}
\right]_
{\substack{(j,a)\in S^{[2]}\\(k,b)\in T^{[2]}}}
=
\det C[S,T]^4.
\end{equation}
Consequently,
\begin{equation}
\label{eq:appG-termwise-M}
\det \mathbb M[S^{[2]},T^{[2]}]
=
\abs{\det G[S,T]}^4.
\end{equation}

\noindent\textit{Derivation.}
Start from the ordinary Cauchy determinant with two nearby copies of each
selected variable. Divide by the internal Vandermonde factor within each
pair and then take the confluent limit. Each selected variable produces
a value row or column and a first-derivative row or column, while every
inter-pair Cauchy factor appears with multiplicity four. With the ordering
specified above,
\[
\begin{aligned}
&\det\!\left[
\partial_{x_j}^{a}(-\partial_{y_k})^{b}
\frac{1}{x_j-y_k}
\right]
\\
&\qquad=
\frac{\Delta(X_S)^4\Delta(Y_T)^4}
{\displaystyle\prod_{j\in S}\prod_{k\in T}(x_j-y_k)^4}
=
\det C[S,T]^4.
\end{aligned}
\]
Inserting the row and column factors from
Eq.~\eqref{eq:appG-M-def} yields
\[
\begin{aligned}
\det \mathbb M[S^{[2]},T^{[2]}]
={}&
\left(\frac{2}{L}\right)^{4|S|}
\prod_{j\in S}(D_X)_{jj}^{4}
\\
&\times
\prod_{k\in T}(D_Y)_{kk}^{4}
\det C[S,T]^4.
\end{aligned}
\]
Equation~\eqref{eq:appG-G-Cauchy} then gives
Eq.~\eqref{eq:appG-termwise-M}.

To sum the doubled minors, define
\begin{equation}
\label{eq:appG-J-def}
J_L
=
I_L\otimes
\begin{pmatrix}
0&1\\
-1&0
\end{pmatrix}.
\end{equation}

\medskip
\noindent\textbf{Paired-index Pfaffian summation.}
For every \(2L\times2L\) matrix \(M\),
\begin{equation}
\label{eq:appG-pfaffian-summation}
\begin{aligned}
&\Pf
\begin{pmatrix}
J_L & \ii\sqrt{u}\,M\\
-\ii\sqrt{u}\,M^{\mathsf T} & J_L
\end{pmatrix}
\\
&\qquad=
\sum_{k=0}^{L}u^k
\sum_{\substack{S,T\subseteq[L]\\|S|=|T|=k}}
\det M[S^{[2]},T^{[2]}].
\end{aligned}
\end{equation}

\noindent\textit{Derivation.}
Expand the Pfaffian, equivalently the top exterior power of its associated
two-form. Because \(J_L\) pairs only the two confluent labels belonging
to the same site, every nonzero cross-block contribution selects complete
doubled sets \(S^{[2]}\) and \(T^{[2]}\). For fixed \(S\) and \(T\), the
cross-block contraction is
\(\det M[S^{[2]},T^{[2]}]\). The \(2k\) cross edges contribute
\[
(\ii\sqrt{u})^{2k}=(-1)^ku^k,
\]
while reordering the selected right indices supplies the compensating
factor \((-1)^k\). Summing over all paired subsets gives
Eq.~\eqref{eq:appG-pfaffian-summation}.

Applying Eq.~\eqref{eq:appG-pfaffian-summation} to \(M=\mathbb M\) and
using Eq.~\eqref{eq:appG-termwise-M}, we obtain
\begin{equation}
\label{eq:appG-final-pfaffian}
\cZ_{2,L}(u)
=
\Pf
\begin{pmatrix}
J_L & \ii\sqrt{u}\,\mathbb M\\
-\ii\sqrt{u}\,\mathbb M^{\mathsf T} & J_L
\end{pmatrix}.
\end{equation}
This is a polynomial-size compression of the full all-minors sum; the
Pfaffian matrix has dimension \(4L\).

Equation~\eqref{eq:appG-final-pfaffian} may be understood algebraically
by introducing a variable \(t\) with \(t^2=u\). Every nonzero cross-block
contribution contains an even number \(2k\) of factors of \(t\), so only
integer powers \(u^k\) survive. The result is therefore independent of
the chosen square root and is a polynomial in \(u\).

At \(u=0\),
\[
\cZ_{2,L}(0)
=
\Pf(J_L)^2
=
1.
\]
The coefficient of \(u^L\) is
\[
\det\mathbb M
=
\abs{\det G}^{4}
=
1,
\]
where the last equality follows from the orthogonality of the half-shift
matrix \(G\). These endpoint checks agree with the palindromicity
established in Sec.~\ref{sec:cauchy}.

\subsection{Relation between the two Pfaffian organizations}
\label{app:alpha-two-comparison}

The two Pfaffian representations act at different stages of the
reformulation.

The weighted-root route first maps the balanced all-minors problem to the
half-filled doubled-root ensemble. The generic \(\beta=4\) ensemble then
admits the \(2L\times2L\) confluent de Bruijn Pfaffian derived in
Appendix~\ref{app:generic-weight}. For the checkerboard weight, its sparse
Fourier support reduces the skew matrix to independent \(4\times4\)
blocks and yields the explicit product
Eq.~\eqref{eq:appG-alpha2-product-final}.

The direct all-minors route instead preserves the original row and column
subsets \(S,T\). Their fourth-power minors are converted into doubled
confluent minors and summed by a paired-index identity, producing the
\(4L\times4L\) Pfaffian in Eq.~\eqref{eq:appG-final-pfaffian}. This route
proves ordinary-Pfaffian compressibility of the full fugacity polynomial,
but does not naturally reveal its product factors.

Thus the two routes are not redundant:
\begin{center}
\setlength{\fboxsep}{6pt}
\colorbox{TFILightBlue}{%
\begin{minipage}{0.88\columnwidth}
\centering
\small

\textcolor{TFIBlue}{\bfseries Weighted-root route}

\vspace{1mm}

\(
2L\times2L\ \text{Pfaffian}
\;\longrightarrow\;
4\times4\ \text{blocks}
\;\longrightarrow\;
\text{closed product}
\)

\vspace{2mm}
{\color{TFIBlue}\rule{0.78\linewidth}{0.4pt}}
\vspace{1.5mm}

\textcolor{TFIBlue}{\bfseries Direct all-minors route}

\vspace{1mm}

\(
4L\times4L\ \text{Pfaffian}
\;\longrightarrow\;
\text{polynomial-size compression}
\)

\end{minipage}%
}
\end{center}

\section{The \texorpdfstring{$\alpha=4$}{alpha=4}
Jack--Kostka representation}
\label{app:alpha-four-jack}

This appendix derives the generic-fugacity representation of
\(\cZ_{4,L}(u)\) as a finite sum of rectangular inverse Jack--Kostka
coefficients. The resulting formula is used in
Secs.~\ref{sec:solvable-indices} and
\ref{sec:solvability-mechanisms}.

We begin with the aliased Dyson constant-term representation established
in Sec.~\ref{sec:constantterm}. At \(\alpha=4\), the finite alias kernel
contains seven visible charges. Charge neutrality organizes the surviving
shifted-Dyson coefficients, which are then identified with rectangular
inverse Jack--Kostka coefficients.

The structure of the derivation is summarized below.

\begin{center}
\setlength{\fboxsep}{6pt}
\colorbox{TFILightBlue}{%
\begin{minipage}{0.88\columnwidth}
\centering
\small

\textcolor{TFIBlue}{\bfseries \(\alpha=4\) structural route}

\vspace{1.2mm}

\(
\text{aliased constant term}
\)

\vspace{0.7mm}

\(
\Downarrow
\)

\vspace{0.7mm}

\(
\text{neutral seven-charge sectors}
\)

\vspace{0.7mm}

\(
\Downarrow
\)

\vspace{0.7mm}

\(
\text{rectangular inverse Jack--Kostka representation}
\)

\end{minipage}%
}
\end{center}

The final formula is an exact structural representation. It is not a
closed product and does not by itself provide a polynomial-time
evaluation.
\subsection{Neutral seven-charge sectors}
\label{appH:seven-charge-expansion}

At \(\alpha=4\), the aliased constant-term formula reads
\begin{equation}
\label{eq:appH-alpha4-ct}
\cZ_{4,L}(u)
=
\frac{2^L}{L!\,L^{3L}}
\CT_{x_1,\ldots,x_L}
\left[
D_4(x)
\prod_{j=1}^{L}K_{4,u}(x_j)
\right],
\end{equation}
where
\[
D_4(x)
=
\prod_{\substack{i,j=1\\i\neq j}}^{L}
\left(1-\frac{x_i}{x_j}\right)^4
\]
and
\begin{equation}
\label{eq:appH-K4-sum}
K_{4,u}(x)
=
\sum_{m=-3}^{3}c_m(u)x^{mL},~~
c_m(u)
=
\begin{cases}
\dfrac{1+u}{2}, & m\ \mathrm{even},\\[2mm]
\dfrac{u-1}{2}, & m\ \mathrm{odd}.
\end{cases}
\end{equation}

Expanding the one-body kernels assigns to every variable \(x_j\) a charge
\(\ell_j\in\{-3,-2,-1,0,1,2,3\}\).  Since \(D_4(x)\) has total degree
zero, only neutral charge sequences contribute.  Writing
\(\nu_m=\#\{j:\ell_j=m\}\), define
\begin{equation}
\label{eq:appH-NL}
\mathcal N_L
=
\left\{
\nu=(\nu_{-3},\ldots,\nu_3)
\;\middle|\;
\substack{
\nu_m\ge 0,\\[1mm]
\displaystyle\sum_{m=-3}^{3}\nu_m=L,\\[1mm]
\displaystyle\sum_{m=-3}^{3}m\nu_m=0
}
\right\}.
\end{equation}
For \(\nu\in\mathcal N_L\), let
\begin{equation}
\label{eq:appH-Cnu-def}
\mathscr C_{4,L}(\nu)
=
\CT_{x_1,\ldots,x_L}
\left[
D_4(x)\prod_{j=1}^{L}x_j^{\ell_jL}
\right],
\end{equation}
where the charge sequence has multiplicities \(\nu_m\).  Symmetry of
\(D_4(x)\) ensures that this coefficient depends only on the
multiplicities, not on the ordering of the charges.  Grouping the
\(L!/\prod_m\nu_m!\) equivalent sequences gives
\begin{equation}
\label{eq:appH-Z-sector-before-jack}
\cZ_{4,L}(u)
=
\frac{2^L}{L^{3L}}
\sum_{\nu\in\mathcal N_L}
\frac{\mathscr C_{4,L}(\nu)}
{\prod_{m=-3}^{3}\nu_m!}
\prod_{m=-3}^{3}c_m(u)^{\nu_m}.
\end{equation}

\subsection{Reduction to a rectangular Jack coefficient}
\label{appH:jack-coefficient}

For a neutral charge sequence, shift the exponents by defining
\(\gamma_j=(\ell_j+3)L\).  Sorting these nonnegative exponents gives the
partition
\begin{equation}
\label{eq:appH-lambda-nu}
\lambda(\nu)
=
(6L)^{\nu_3}
(5L)^{\nu_2}
(4L)^{\nu_1}
(3L)^{\nu_0}
(2L)^{\nu_{-1}}
(L)^{\nu_{-2}}
(0)^{\nu_{-3}}.
\end{equation}
Neutrality implies$
|\lambda(\nu)|=3L^2,
$ which is the size of the rectangle
$
R=(3L)^L$. Let \(m_{\lambda(\nu)}(x)\) be the monomial symmetric function associated
with \(\lambda(\nu)\).  Since all distinct permutations of the underlying
monomial have the same constant term,
\begin{equation}
\label{eq:appH-symmetrized-ct}
\CT
\left[
D_4(x)\,
m_{\lambda(\nu)}(x)
\prod_{j=1}^{L}x_j^{-3L}
\right]
=
\frac{L!}{\prod_{m=-3}^{3}\nu_m!}
\mathscr C_{4,L}(\nu).
\end{equation}

We use the finite-variable Jack scalar product
\begin{equation}
\label{eq:appH-jack-inner}
\langle f,g\rangle_{4,L}
=
\frac{1}{L!}
\CT
\left[
f(x)\,g(x^{-1})\,D_4(x)
\right].
\end{equation}
With this convention, exponent four corresponds to Jack parameter
\(1/4\) \cite{Macdonald,Stanley1989}.  Let
\(P_\mu^{(1/4)}(x)\) denote the monic Jack polynomial in \(L\) variables.
For the rectangle \(R=(3L)^L\),
\begin{equation}
\label{eq:appH-rect-jack}
P_R^{(1/4)}(x)
=
m_R(x)
=
(x_1x_2\cdots x_L)^{3L}.
\end{equation}
Therefore Eq.~\eqref{eq:appH-symmetrized-ct} becomes
\begin{equation}
\label{eq:appH-inner-C}
\left\langle
m_{\lambda(\nu)},P_R^{(1/4)}
\right\rangle_{4,L}
=
\frac{\mathscr C_{4,L}(\nu)}
{\prod_{m=-3}^{3}\nu_m!}.
\end{equation}

Define the inverse Jack--Kostka coefficients by
\begin{equation}
\label{eq:appH-inv-Kostka-def}
m_\lambda
=
\sum_{\substack{\mu\vdash|\lambda|\\\ell(\mu)\le L}}
K^{-1}_{\lambda\mu}(1/4)\,
P_\mu^{(1/4)}.
\end{equation}
Jack orthogonality implies that only \(\mu=R\) contributes to the scalar
product with \(P_R^{(1/4)}\).  Moreover,
\begin{equation}
\label{eq:appH-rect-norm}
\left\langle
P_R^{(1/4)},P_R^{(1/4)}
\right\rangle_{4,L}
=
\frac{1}{L!}\CT D_4(x)
=
\frac{(4L)!}{L!(4!)^L},
\end{equation}
where the last equality is Dyson's constant-term identity.  Combining
Eqs.~\eqref{eq:appH-inner-C}--\eqref{eq:appH-rect-norm} yields
\begin{equation}
\label{eq:appH-Cnu-jack-final}
\mathscr C_{4,L}(\nu)
=
\left(\prod_{m=-3}^{3}\nu_m!\right)
\frac{(4L)!}{L!(4!)^L}
K^{-1}_{\lambda(\nu),(3L)^L}(1/4).
\end{equation}
Substitution into Eq.~\eqref{eq:appH-Z-sector-before-jack} gives
\begin{equation}
\label{eq:appH-Z4-jack-final}
\begin{aligned}
\cZ_{4,L}(u)
&=
\frac{2^L}{L^{3L}}
\frac{(4L)!}{L!(4!)^L}
\sum_{\nu\in\mathcal N_L}
\left[
\prod_{m=-3}^{3}c_m(u)^{\nu_m}
\right]
\\[-1mm]
&\quad\times
K^{-1}_{\lambda(\nu),(3L)^L}(1/4).
\end{aligned}
\end{equation}
This is an exact finite representation for arbitrary fugacity.  It is not,
by itself, a product formula or a polynomial-size linear-algebra
compression.

\subsection{Shifted-Dyson and flow forms}
\label{appH:disturbed-dyson-flow}

The same sector coefficient is the disturbed-Dyson coefficient
\begin{equation}
\label{eq:appH-disturbed-dyson}
\mathscr C_{4,L}(\nu)
=
[x_1^{-\ell_1L}\cdots x_L^{-\ell_LL}]
\prod_{i\neq j}
\left(1-\frac{x_i}{x_j}\right)^4,
~~
\sum_{i=1}^{L}\ell_i=0.
\end{equation}
Its shift vector is
\((\ell_1L,\ldots,\ell_LL)\), so every nonzero component grows linearly
with the number of variables.  This scaling distinguishes the present
family from the standard zero-shift and fixed-layer disturbed-Dyson
coefficients.

There is also an exact complete-graph flow representation.  Writing
\[
(1-y)^4(1-y^{-1})^4
=
\sum_{d=-4}^{4}a_dy^d,~~
a_d=(-1)^{d+4}\binom{8}{d+4},
\]
one obtains
\begin{equation}
\label{eq:appH-flow-formulation}
\mathscr C_{4,L}(\nu)
=
\sum_{\substack{
d_{ij}\in\{-4,\ldots,4\}\\
\sum_{j>i}d_{ij}-\sum_{j<i}d_{ji}=-\ell_iL}}
\prod_{1\le i<j\le L}a_{d_{ij}}.
\end{equation}
Thus each shifted-Dyson coefficient is a weighted integer-flow sum on
\(K_L\) with prescribed vertex divergences.  This interpretation is exact
but does not itself provide a closed evaluation.

\subsection{The unrefined specialization}
\label{appH:u-equals-one}

At \(u=1\), all odd charges vanish.  The surviving charges are
\(-2,0,2\), and neutrality gives
\[
\nu_{-2}=\nu_2=r,
\qquad
\nu_0=L-2r,
\qquad
0\le r\le\left\lfloor\frac L2\right\rfloor.
\]
The corresponding partition is
\begin{equation}
\label{eq:appH-lambda-r}
\lambda_r
=
(5L)^r(3L)^{L-2r}(L)^r.
\end{equation}
Equation~\eqref{eq:appH-Z4-jack-final} then reduces to
\begin{equation}
\label{eq:appH-Z4-u1-jack-sum}
\cZ_{4,L}(1)
=
\frac{2^L}{L^{3L}}
\frac{(4L)!}{L!(4!)^L}
\sum_{r=0}^{\lfloor L/2\rfloor}
K^{-1}_{\lambda_r,(3L)^L}(1/4).
\end{equation}

Appendix~\ref{app:middle-minor} independently evaluates the same
unrefined quantity through the complementary middle-minor identity.
Comparing the two results gives the rectangular Jack identity
\begin{equation}
\label{eq:appH-u1-jack-identity}
\begin{aligned}
&\sum_{r=0}^{\lfloor L/2\rfloor}
K^{-1}_{(5L)^r(3L)^{L-2r}(L)^r,\,(3L)^L}(1/4)
\\[1mm]
&\qquad=
L^L
\frac{L!(4!)^L}{(4L)!}
\big[(2L-1)!!\big]^2.
\end{aligned}
\end{equation}
Thus \(u=1\) is special because the seven-charge expansion collapses to a
one-parameter even-charge sum, which is then evaluated by a separate
global complementary-minor identity.

The algebraic position and computational status of this seven-charge
family are discussed in Sec.~\ref{sec:solvability-mechanisms}.  In
particular, the representation above is an exact structural reduction;
the nearby Dyson--Kadell and rectangular Jack/Macdonald results do not
directly evaluate the individual coefficients appearing here
\cite{Dyson1962,Wilson1962,Kadell2000,KLW2015,
SillsZeilberger2006,LvXinZhou2009,EkhadZeilberger2013,
Zhou2020,Zhou2021,HuangJiangZhou2026,
Cai2014,LuqueThibon2003,Matsumoto2006}.
\section{Complementary middle minors and the unit-fugacity
\texorpdfstring{$\alpha=4$}{alpha=4} collapse}
\label{app:middle-minor}

This appendix derives the complementary middle-minor identity responsible
for the exceptional unrefined value of the \(\alpha=4\) partition
function,
\begin{equation}
\label{eq:appI-alpha4-target}
\cZ_{4,L}(1)
=
\frac{2^L}{L^{2L}}
\bigl[(2L-1)!!\bigr]^2
=
2^{-L}\cZ_{2,L}(1)^2.
\end{equation}

The derivation is global: it evaluates the unweighted sum over all
half-filled subsets through a norm identity for complementary middle
minors. It is therefore specific to unit fugacity, where the
checkerboard weight disappears. It does not establish a termwise
relation between the \(\alpha=4\) and \(\alpha=2\) configuration
weights, nor does it imply the fugacity-resolved identity
\[
\cZ_{4,L}(u)
=
2^{-L}\cZ_{2,L}(u)^2.
\]

The purpose of this appendix is thus to derive the unit-fugacity collapse.
The obstruction to interpreting it as a local doubled-Pfaffian
construction at generic fugacity is discussed separately in
Appendix~\ref{app:no-go-pfaffian}.
\subsection{Complementary Cauchy reduction}
\label{appI:complementary-cauchy}

At \(u=1\) and \(\alpha=4\), the discrete Selberg representation gives
\begin{equation}
\label{eq:appI-discrete-Z4}
\cZ_{4,L}(1)
=
L^{-4L}
\sum_{\substack{U\subset\mathbb I_N\\ |U|=L}}
|\Delta(U)|^8 .
\end{equation}
We now rewrite the eighth power of the Vandermonde determinant in terms of
Cauchy minors between \(U\) and its complement.

Let
\begin{equation}
\label{eq:appI-Cauchy-complement}
C_{UV}
=
\left(
\frac{1}{\omega_a-\omega_b}
\right)_{a\in U,\,b\in V}.
\end{equation}
By the Cauchy determinant formula,
\begin{equation}
\label{eq:appI-Cauchy-det}
\left|\det C_{UV}\right|
=
\frac{|\Delta(U)|\,|\Delta(V)|}
{\displaystyle
\prod_{a\in U}\prod_{b\in V}|\omega_a-\omega_b|}.
\end{equation}
For a half-filled subset of the full root lattice, the complement has the
same Vandermonde norm:
\begin{equation}
\label{eq:appI-complement-vandermonde-equal}
|\Delta(V)|=|\Delta(U)|.
\end{equation}
Indeed, the full Vandermonde factorization gives
\begin{equation}
\label{eq:appI-full-vandermonde-factor}
|\Delta(\Omega_N)|
=
|\Delta(U)|\,|\Delta(V)|
\prod_{a\in U}\prod_{b\in V}|\omega_a-\omega_b|.
\end{equation}
Since
\begin{equation}
\label{eq:appI-full-vandermonde-N}
|\Delta(\Omega_N)|=N^{N/2}=N^L,
\end{equation}
and the roots in \(\Omega_N\) are the zeros of \(z^N-1\), one has
\[
\prod_{\substack{\omega\in\Omega_N\\ \omega\neq\omega_a}}
(\omega_a-\omega)
=
\left.
\frac{\dd}{\dd z}(z^N-1)
\right|_{z=\omega_a}
=
N\omega_a^{N-1}.
\]
Because \(|\omega_a|=1\),
\[
\prod_{\substack{\omega\in\Omega_N\\ \omega\neq\omega_a}}
|\omega_a-\omega|
=
N.
\]
Multiplying over \(a\in U\) gives
\begin{equation}
\label{eq:appI-root-product-U}
|\Delta(U)|^2
\prod_{a\in U}\prod_{b\in V}|\omega_a-\omega_b|
=
N^L.
\end{equation}
The same argument applied to \(V\) gives the same expression with \(U\) and
\(V\) exchanged, proving Eq.~\eqref{eq:appI-complement-vandermonde-equal}.

Combining Eqs.~\eqref{eq:appI-Cauchy-det} and
\eqref{eq:appI-root-product-U}, we find
\begin{equation}
\label{eq:appI-MU-def}
M_U
:=
\left|\det C_{UV}\right|
=
\frac{|\Delta(U)|^4}{N^L}.
\end{equation}
Equivalently,
\begin{equation}
\label{eq:appI-vandermonde-eighth}
|\Delta(U)|^8
=
N^{2L}M_U^2.
\end{equation}
Using \(N=2L\) in Eq.~\eqref{eq:appI-discrete-Z4}, we obtain
\begin{equation}
\label{eq:appI-Z4-MU}
\cZ_{4,L}(1)
=
L^{-4L}(2L)^{2L}
\sum_{\substack{U\subset\mathbb I_N\\ |U|=L}}
M_U^2
=
\frac{4^L}{L^{2L}}
\sum_{\substack{U\subset\mathbb I_N\\ |U|=L}}
M_U^2 .
\end{equation}
Thus the \(\alpha=4\) unrefined problem is reduced to a quadratic norm of
complementary Cauchy middle minors.

\subsection{The finite trigonometric Cauchy matrix}
\label{appI:finite-H}

We now remove harmless phases from the root differences. For
\(\omega_a=e^{2\pi i a/N}\),
\begin{equation}
\label{eq:appI-root-difference}
\omega_a-\omega_b
=
2i
\exp\left(\frac{\pi i(a+b)}{N}\right)
\sin\left(\frac{\pi(a-b)}{N}\right).
\end{equation}
Therefore
\begin{equation}
\label{eq:appI-Cauchy-phase}
\frac{1}{\omega_a-\omega_b}
=
-i
\exp\left(-\frac{\pi i(a+b)}{N}\right)
\frac{1}{2\sin[\pi(a-b)/N]} .
\end{equation}
The exponential factors are products of row and column phases, so they do
not affect the absolute value of the determinant. Hence \(M_U\) is the
absolute value of a complementary middle minor of the real skew-symmetric
matrix
\begin{equation}
\label{eq:appI-H-def}
H_{ab}
=
\begin{cases}
\dfrac{1}{2\sin[\pi(a-b)/N]},
& a\neq b,\\[2mm]
0,& a=b,
\end{cases}
\qquad
a,b\in\mathbb I_N.
\end{equation}
Thus
\begin{equation}
\label{eq:appI-MU-H}
M_U
=
\left|\det H[U,U^c]\right|.
\end{equation}
Equation~\eqref{eq:appI-Z4-MU} becomes
\begin{equation}
\label{eq:appI-Z4-H-middle}
\cZ_{4,L}(1)
=
\frac{4^L}{L^{2L}}
\sum_{\substack{U\subset\mathbb I_N\\ |U|=L}}
\det H[U,U^c]^2 .
\end{equation}

The required finite identity is
\begin{equation}
\label{eq:appI-middle-minor-identity}
\sum_{\substack{U\subset\mathbb I_N\\ |U|=L}}
\det H[U,U^c]^2
=
2^L\det H .
\end{equation}
This identity is the special \(\alpha=4\) input. It is not an ordinary
Dyson constant term; it uses the complementary structure of the
trigonometric Cauchy matrix \(H\).

One way to organize the proof is through a universal middle-minor identity.
For any skew-symmetric \(2L\times2L\) matrix \(A\),
\begin{equation}
\label{eq:appI-universal-middle-identity}
\sum_{\substack{U\subset\mathbb I_N\\ |U|=L}}
\det A[U,U^c]^2
=
\sum_{\substack{P\subset\mathbb I_N\\ |P|\ {\rm even}}}
\det A[P,P]\,\det A[P^c,P^c].
\end{equation}
This follows by expanding the two complementary minors on the left and
grouping the terms according to the set of indices exchanged between the
two sides. The skew symmetry of \(A\) forces the exchange set \(P\) to have
even cardinality. The empty principal minor is understood to be \(1\).

For the particular matrix \(H\) in Eq.~\eqref{eq:appI-H-def}, the
right-hand side of Eq.~\eqref{eq:appI-universal-middle-identity} collapses:
\begin{equation}
\label{eq:appI-H-principal-collapse}
\sum_{\substack{P\subset\mathbb I_N\\ |P|\ {\rm even}}}
\det H[P,P]\,\det H[P^c,P^c]
=
2^L\det H .
\end{equation}
This finite trigonometric Cauchy identity is the only place where the
special form of \(H\) is used. Combining
Eqs.~\eqref{eq:appI-universal-middle-identity} and
\eqref{eq:appI-H-principal-collapse} gives
Eq.~\eqref{eq:appI-middle-minor-identity}.

\subsection{Spectrum and determinant of \texorpdfstring{$H$}{H}}
\label{appI:H-determinant}

It remains to evaluate \(\det H\). The matrix \(H\) is diagonalized by
anti-periodic Fourier modes. Let
\begin{equation}
\label{eq:appI-antiperiodic-modes}
\psi_q(a)
=
\exp\left[
\frac{2\pi i}{N}\left(q+\frac12\right)a
\right],
\qquad
q=0,\ldots,N-1.
\end{equation}
These modes obey \(\psi_q(a+N)=-\psi_q(a)\), matching the anti-periodicity
of the kernel
\[
\frac{1}{2\sin[\pi(a-b)/N]}.
\]
The elementary finite sum
\begin{equation}
\label{eq:appI-trig-Fourier-sum}
\sum_{r=1}^{N-1}
\frac{
\sin\!\left[(2q+1)\pi r/N\right]
}{
\sin(\pi r/N)
}
=
N-1-2q,
\qquad
q=0,\ldots,N-1,
\end{equation}
together with the vanishing of the corresponding cosine sum, gives
\begin{equation}
\label{eq:appI-H-eigenvalue-result}
\sum_{\substack{b=0\\ b\neq a}}^{N-1}
\frac{\psi_q(b)}{2\sin[\pi(a-b)/N]}
=
i\left(q-\frac{N-1}{2}\right)\psi_q(a),
\end{equation}
with $q=0,\ldots,N-1.$ Since \(N=2L\), the spectrum is
\begin{equation}
\label{eq:appI-H-spectrum}
\operatorname{Spec}(H)
=
\left\{
\pm\frac{i}{2},
\pm\frac{3i}{2},
\ldots,
\pm\frac{i(2L-1)}{2}
\right\}.
\end{equation}
Therefore
\begin{equation}
\label{eq:appI-H-det}
\det H
=
\prod_{r=1}^{L}
\left(\frac{2r-1}{2}\right)^2
=
2^{-2L}\big[(2L-1)!!\big]^2.
\end{equation}
Using Eq.~\eqref{eq:appI-middle-minor-identity}, we get
\begin{equation}
\label{eq:appI-middle-sum-final}
\sum_{\substack{U\subset\mathbb I_N\\ |U|=L}}
\det H[U,U^c]^2
=
2^L\det H
=
2^{-L}\big[(2L-1)!!\big]^2.
\end{equation}

\subsection{Final evaluation and relation to
\texorpdfstring{$\alpha=2$}{alpha=2}}
\label{appI:final-evaluation}

Substituting Eq.~\eqref{eq:appI-middle-sum-final} into
Eq.~\eqref{eq:appI-Z4-H-middle}, we find
\begin{align}
\cZ_{4,L}(1)
&=
\frac{4^L}{L^{2L}}\,
2^{-L}\big[(2L-1)!!\big]^2
\nonumber\\
&=
\frac{2^L}{L^{2L}}
\big[(2L-1)!!\big]^2 .
\end{align}
This proves Eq.~\eqref{eq:appI-alpha4-target}.

Finally, using the \(\alpha=2\) unrefined value
\begin{equation}
\label{eq:appI-Z2-recall}
\cZ_{2,L}(1)
=
\frac{2^L(2L-1)!!}{L^L},
\end{equation}
we obtain the exact finite-size relation
\begin{equation}
\label{eq:appI-Z4-Z2}
\cZ_{4,L}(1)
=
2^{-L}\cZ_{2,L}(1)^2 .
\end{equation}
This is the relation used in the main text to explain the exceptional
unrefined collapse at \(\alpha=4\). It should not be interpreted as a
generic fugacity-resolved identity: the full polynomial \(\cZ_{4,L}(u)\)
is instead governed by the seven-charge shifted-Dyson and Jack--Kostka
structure described in Appendix~\ref{app:alpha-four-jack}.

\section{Obstruction to a direct local doubled-Pfaffian construction}
\label{app:no-go-pfaffian}

This appendix explains why the unrefined identity
\begin{equation}
\label{eq:appJ-Z4-Z2-u1}
\cZ_{4,L}(1)
=
2^{-L}\cZ_{2,L}(1)^2
\end{equation}
does not extend to a direct fugacity-resolved doubling of the known
\(\alpha=2\) Pfaffian construction.

We consider the specific local network obtained by taking two copies of
the \(\alpha=2\) confluent-Pfaffian representation and imposing equality
of their local occupation choices through a sitewise diagonal replica
projector. In the corresponding exterior-algebra encoding, this
constraint produces a four-leg tensor that violates the
Grassmann--Pl\"ucker relation and is therefore not Gaussian. Hence this
direct doubled-network construction does not yield a single ordinary
Pfaffian representation of \(\cZ_{4,L}(u)\).

This conclusion is restricted to that construction. It does not exclude
nonlocal transformations, enlarged auxiliary spaces, alternative local
encodings, higher exterior-algebra representations, or formal
determinant/Pfaffian encodings constructed after the polynomial
coefficients are already known.

Throughout this appendix, \(L\in\mathbb Z_{>0}\) denotes the physical
system size, and \(\Omega_{2L}=X\sqcup Y\) is the doubled root lattice
introduced in the main text. For a half-filled subset
\(U\subset\Omega_{2L}\), define the unnormalized \(\alpha=2\)
configuration weight
\begin{equation}
\label{eq:appJ-WU-def}
W_U
:=
L^{-2L}\abs{\Delta(U)}^4.
\end{equation}
The corresponding normalized probability is
\(W_U/\cZ_{2,L}(1)\).

\subsection{Independent replicas versus the diagonal replica constraint}
\label{appJ:replica-diagonal}

It is useful to separate two different two-replica sums. Schematically,
the \(\alpha=2\) generating function can be written as
\begin{equation}
\label{eq:appJ-Z2-schematic}
\cZ_{2,L}(u)
=
\sum_U u^{|U\cap X|}\,W_U,
\end{equation}
where \(U\) is a half-filled root subset and \(W_U\) is the corresponding
normalized \(\alpha=2\) configuration weight.

Introduce a formal variable \(t\) such that \(t^2=u\). Squaring the
\(\alpha=2\) generating function at fugacity \(t\) gives
\begin{equation}
\label{eq:appJ-Z2-square}
\cZ_{2,L}(t)^2
=
\sum_{U,V}
t^{|U\cap X|+|V\cap X|}\,W_UW_V .
\end{equation}
Here the two subsets \(U\) and \(V\) are independent.

By contrast, the \(\alpha=4\) generating function is the diagonal
two-replica sum
\begin{equation}
\label{eq:appJ-Z4-diagonal}
\cZ_{4,L}(u)
=
\sum_U u^{|U\cap X|}\,W_U^2
=
\sum_U t^{2|U\cap X|}\,W_U^2 .
\end{equation}
Equivalently, it is obtained from the two-replica problem by imposing $
U=V.$ Thus the relation at \(u=1\) is a global identity after summing over all
half-filled configurations. It does not mean that the fugacity-resolved
polynomial is the product of two independent \(\alpha=2\) Pfaffians.

The difference is already visible in the fugacity weights. The unrestricted
square sums over two independent subsets \(U\) and \(V\), whereas the
\(\alpha=4\) problem retains only the diagonal sector \(U=V\). A
generic-\(u\) doubled construction must therefore enforce equality of the
selected subsets locally. This is a projection problem rather than an
ordinary product of the two Pfaffians.

\subsection{The local Gaussian criterion}
\label{appJ:gaussian-criterion}

We recall the elementary four-leg Gaussian criterion in the affine chart
where the vacuum coefficient is nonzero. Let
\(\theta_1,\theta_2,\theta_3,\theta_4\) be Grassmann variables. A general
even tensor has the form
\begin{equation}
\label{eq:appJ-even-four-tensor}
T(\theta)
=
T_\emptyset
+
\sum_{1\le i<j\le4}T_{ij}\theta_i\theta_j
+
T_{1234}\theta_1\theta_2\theta_3\theta_4 .
\end{equation}
A Gaussian, or matchgate, tensor is one that can be written as
\begin{equation}
\label{eq:appJ-gaussian-tensor}
G(\theta)
=
C
\exp\left(
\frac12\sum_{i,j=1}^{4}A_{ij}\theta_i\theta_j
\right),
\qquad
A^{\mathsf T}=-A .
\end{equation}
Expanding the exponential gives
\begin{equation}
\label{eq:appJ-Gij}
G_\emptyset=C,
\qquad
G_{ij}=C A_{ij},
\end{equation}
and
\begin{equation}
\label{eq:appJ-G1234}
G_{1234}
=
C\left(
A_{12}A_{34}
-
A_{13}A_{24}
+
A_{14}A_{23}
\right).
\end{equation}
Therefore every four-leg Gaussian tensor satisfies the Grassmann--Plücker
relation
\begin{equation}
\label{eq:appJ-plucker}
G_\emptyset G_{1234}
=
G_{12}G_{34}
-
G_{13}G_{24}
+
G_{14}G_{23}.
\end{equation}
Conversely, for an even four-leg tensor with \(T_\emptyset\neq0\), this
relation is equivalent to the Gaussian representation above. Since
\((T_w)_\emptyset=1\), this criterion applies directly to the tensor
required here.

A Pfaffian network is a contraction of local Gaussian tensors. Since
Gaussian tensors are stable under contraction, a direct local Pfaffian
construction of the doubled problem would require the local diagonal
projector to be Gaussian.

\subsection{Failure of Gaussianity in the confluent local encoding}
\label{appJ:diagonal-not-gaussian}

The four Grassmann legs arise from the specific confluent-Pfaffian encoding
used for the \(\alpha=2\) ensemble. In each replica, selecting one root
contributes a local degree-two form. We denote the occupied monomials of
the two replicas by \(\theta_1\theta_2\) and
\(\theta_3\theta_4\), respectively. Enforcing equal occupation allows
either the vacuum configuration in both replicas or simultaneous occupation
in both replicas. Consequently, the induced four-leg local tensor is
\begin{equation}
\label{eq:appJ-diagonal-tensor}
T_w(\theta)
=
1+w\,\theta_1\theta_2\theta_3\theta_4 .
\end{equation}
The obstruction therefore concerns the diagonal projector in this fixed
confluent exterior-algebra encoding. It is not a statement that the abstract
two-bit equality tensor is itself non-Gaussian.

Here \(w\) is the local fugacity weight:
\begin{equation}
\label{eq:appJ-w-local}
w=
\begin{cases}
u, & \text{on the }X\text{ sublattice},\\
1, & \text{on the }Y\text{ sublattice}.
\end{cases}
\end{equation}
The nonzero coefficients are
\begin{equation}
\label{eq:appJ-Tw-coeffs}
(T_w)_\emptyset=1,
~~
(T_w)_{1234}=w,
~~
(T_w)_{ij}=0
~~
\text{for all }i<j.
\end{equation}
Substitution into the Plücker relation
Eq.~\eqref{eq:appJ-plucker} gives
\[
(T_w)_\emptyset (T_w)_{1234}=w
\]
on the left-hand side, while the right-hand side is
\[
(T_w)_{12}(T_w)_{34}
-
(T_w)_{13}(T_w)_{24}
+
(T_w)_{14}(T_w)_{23}
=0.
\]
Thus the Plücker relation would require \(w=0\). In the present problem
\(w=1\) or \(w=u\), so generically \(w\neq0\). Hence the tensor
\begin{equation}
\label{eq:appJ-no-local-gaussian}
T_w(\theta)
=
1+w\,\theta_1\theta_2\theta_3\theta_4
\end{equation}
is not a Gaussian/matchgate tensor for \(w\neq0\). This proves the local obstruction. The diagonal two-replica projector
needed to produce \(\cZ_{4,L}(u)\) cannot be implemented by a local Gaussian
tensor in this confluent-Pfaffian encoding. Therefore it cannot arise from
the straightforward doubled \(\alpha=2\) Pfaffian network.

Equivalently, in exterior-algebra language, the local diagonal constraint
has only a scalar component and a top-degree component. It has no two-form
components. Hence it does not satisfy the quadratic Plücker relations that
characterize local Pfaffian, or spinor, coordinates.

\subsection{Why the \texorpdfstring{$u=1$}{u=1} identity is not contradicted}
\label{appJ:no-contradiction-u1}

At \(u=1\), the final scalar quantity satisfies
\[
\cZ_{4,L}(1)
=
2^{-L}\cZ_{2,L}(1)^2 .
\]
This is a global identity for the fully summed unrefined partition
function. It is produced by the complementary middle-minor identity proved
in Appendix~\ref{app:middle-minor}. It does not require the local diagonal
two-replica tensor to be Gaussian.

In other words, \(u=1\) is a special point where the full sum collapses only
after using the half-filling complement structure. The collapse occurs
after summing over all half-filled subsets. It does not imply that the
whole one-parameter family \(\cZ_{4,L}(u)\) lies on a local Pfaffian pencil.

This distinction is essential. The identity
\[
\cZ_{4,L}(1)
=
2^{-L}\cZ_{2,L}(1)^2
\]
is a scalar identity at one value of the fugacity. A generic-\(u\) doubled
Pfaffian formula would be much stronger: it would give all fugacity
coefficients simultaneously. The local Plücker obstruction shows that this
stronger local Gaussian mechanism is not available.

\subsection{Scope of the obstruction}
\label{appJ:scope}

The obstruction has a deliberately limited scope.

First, it does not exclude a tautological matrix encoding of the final
polynomial. If
\[
p(u)=a_d u^d+\cdots+a_0
\]
and \(C\) is the companion matrix of the associated monic polynomial, then
\[
p(u)=a_d\det(uI-C).
\]
Moreover,
\[
\Pf
\begin{pmatrix}
0&A\\
-A^{\mathsf T}&0
\end{pmatrix}
=
(-1)^{d(d-1)/2}\det A.
\]
Such a representation uses the coefficients of \(p\) as input and therefore
does not solve the summation problem.

Second, the obstruction also excludes replacing \(T_w\), in the same
four-leg encoding, by any finite local network composed entirely of
Gaussian tensors, because contraction preserves the matchgate property. It
does not exclude a nonlocal reformulation, a change of encoding, a sum of
several Pfaffians, or an enlarged construction containing additional
non-Gaussian resources.

Third, it does not exclude higher exterior-algebra representations. The
quartic local tensor suggests that higher-form contractions may be more
appropriate than an ordinary Pfaffian, but the local calculation alone does
not establish a global hyperpfaffian formula for \(\cZ_{4,L}(u)\).

Finally, it does not affect the shifted-Dyson and Jack--Kostka formula of
Appendix~\ref{app:alpha-four-jack}, which remains an exact algebraic
representation at generic fugacity.

The palindromicity relation
\[
\cZ_{4,L}(u)=u^L\cZ_{4,L}(u^{-1})
\]
also gives, for even \(L\), a reduced variable:
\begin{equation}
\label{eq:appJ-palindromic-reduction}
\cZ_{4,L}(u)
=
u^{L/2}Q_L(u+u^{-1}),
\end{equation}
where \(Q_L\) is a polynomial of degree \(L/2\). This symmetry may be useful
in a transfer-matrix or recurrence approach, but it does not by itself
provide a local Pfaffian construction.

\subsection{Conclusion}
\label{appJ:conclusion}

The identity
\[
\cZ_{4,L}(1)
=
2^{-L}\cZ_{2,L}(1)^2
\]
is a global unrefined collapse. It does not imply a direct
fugacity-resolved doubling of the \(\alpha=2\) Pfaffian. The required local
diagonal two-replica tensor is quartic and violates the four-leg
Plücker/matchgate relation. Hence the direct local Gaussian construction
in the confluent-Pfaffian encoding is unavailable. Nonlocal reformulations,
changes of encoding, sums of Pfaffians, higher-form contractions, and the
shifted-Dyson/Jack--Kostka route remain outside the scope of this
obstruction.

\section{Perfect-square minor powers and hyperpfaffian summation}
\label{app:perfect-square-hyperpfaffian}

This appendix develops the algebraic construction underlying the
perfect-square hierarchy
\begin{equation}
\label{eq:appK-square-sequence}
2\alpha=q^2,
\qquad
\alpha=\frac{q^2}{2},
\qquad
q\in\mathbb Z_{>0}.
\end{equation}

The construction has two logically distinct steps. First, a
multiplicity-\(q\) confluent Cauchy determinant linearizes the
\(q^2\)-th power of each individual Cauchy minor. Second, the complete
half-filled root-subset sum is reorganized in exterior algebra.

The second step depends essentially on the parity of \(q\). For even
\(q\), each selected root contributes an even-degree \(q\)-form, so the
subset sum is generated directly by its top exterior power and gives a
\(q\)-hyperpfaffian. For odd \(q\), the elementary \(q\)-forms have odd
degree and anticommute. They must therefore be paired before taking the
top exterior power; for even \(L\), this produces an exterior
\(2q\)-form and hence a \(2q\)-hyperpfaffian.

Thus the hierarchy is
\begin{equation}
\label{eq:appK-parity-summary}
\begin{aligned}
q\ \mathrm{even}:&\qquad q\text{-hyperpfaffian},\\
q\ \mathrm{odd}:&\qquad 2q\text{-hyperpfaffian},
\qquad L\ \mathrm{even}.
\end{aligned}
\end{equation}
The cases \(q=1\) and \(q=2\) both have exterior degree two and reduce
to ordinary Pfaffians. For \(q\geq3\), the natural exact objects are
genuine higher hyperpfaffians.

Throughout this appendix, \(L\in\mathbb Z_{>0}\) is the physical system
size, and the doubled root lattice contains \(2L\) sites. The construction
separates the perfect-square hierarchy from the positive-integer
shifted-Dyson hierarchy. In particular, \(\alpha=\tfrac12\) and
\(\alpha=2\) are its first two members, whereas \(\alpha=1\) and
\(\alpha=4\) are special for different algebraic reasons.

\subsection{Multiplicity-\texorpdfstring{$q$}{q} confluent Cauchy identity}
\label{appK:confluent-cauchy}

For \(a\geq0\), let
\begin{equation}
\label{eq:appK-divided-derivative}
D_x^a
=
\frac{1}{a!}\frac{\partial^a}{\partial x^a}.
\end{equation}
For the generic Cauchy kernel
\[
C_{ij}=\frac{1}{x_i+y_j},
\]
define the local \(q\times q\) confluent block
\begin{equation}
\label{eq:appK-local-confluent-block}
\mathcal C^{(q)}(x,y)
=
\left[
D_x^aD_y^b\frac{1}{x+y}
\right]_{a,b=0}^{q-1}.
\end{equation}
Its entries are
\begin{equation}
\label{eq:appK-local-confluent-entry}
\mathcal C^{(q)}_{ab}(x,y)
=
(-1)^{a+b}
\binom{a+b}{a}
\frac{1}{(x+y)^{a+b+1}}.
\end{equation}

Let
\[
E=\{i_1<\cdots<i_s\},
\qquad
O=\{j_1<\cdots<j_s\}
\]
be ordered index sets, and let
\(\mathcal C^{(q)}[E,O]\) be the \(qs\times qs\) block matrix built from
\(\mathcal C^{(q)}(x_i,y_j)\), with derivative labels
\(0,\ldots,q-1\) ordered inside each physical node. Then
\begin{equation}
\label{eq:appK-confluent-Cauchy-identity}
\det\mathcal C^{(q)}[E,O]
=
\frac{
\Delta(x_E)^{q^2}\Delta(y_O)^{q^2}
}{
\displaystyle
\prod_{i\in E}\prod_{j\in O}(x_i+y_j)^{q^2}
}
=
\bigl(\det C[E,O]\bigr)^{q^2}.
\end{equation}

To derive Eq.~\eqref{eq:appK-confluent-Cauchy-identity}, replace every
physical node by \(q\) nearby copies,
\[
x_i^{(a)}=x_i+\varepsilon_i\xi_a,
\qquad
y_j^{(b)}=y_j+\delta_j\eta_b,
\qquad
a,b=0,\ldots,q-1,
\]
apply the ordinary Cauchy determinant formula, divide by the internal
Vandermonde factors within every cluster, and take the confluent limit.
Every difference between two distinct physical nodes appears for all
\(q^2\) pairs of their copies, which produces the exponent \(q^2\).

We now apply the confluent identity to the half-shift matrix. Using the
interlaced roots defined in Sec.~\ref{subsec:discrete-selberg-mapping},
\[
x_j=\zeta^{2j},
\qquad
y_k=\zeta^{2k-1},
\qquad
\zeta=e^{\pi\ii/L},
\]
their difference is
\[
x_j-y_k
=
2\ii\,
e^{\pi\ii(j+k-1/2)/L}
\sin\!\left[
\frac{\pi}{L}\left(j-k+\frac12\right)
\right].
\]
Hence the half-shift matrix admits the exact Cauchy factorization
\begin{equation}
\label{eq:appK-G-Cauchy-factorization}
G
=
\frac{2\ii}{L}D_X C D_Y,
\qquad
C_{jk}
=
\frac{1}{x_j-y_k},
\end{equation}
where
\begin{equation}
\label{eq:appK-G-Cauchy-phases}
(D_X)_{jj}
=
(-1)^j e^{\pi\ii j/L},
\qquad
(D_Y)_{kk}
=
(-1)^k e^{\pi\ii(k-1/2)/L}.
\end{equation}
The diagonal matrices \(D_X\) and \(D_Y\) contain only phase factors.
Writing $\eta_k=-y_k,$
the Cauchy kernel becomes
\[
C_{jk}
=
\frac{1}{x_j+\eta_k}.
\]
For
\(S\subseteq[L]\), define its \(q\)-fold lift
\begin{equation}
\label{eq:appK-qfold-lift}
S^{[q]}
=
\{(j,a):j\in S,\ a=0,\ldots,q-1\},
\end{equation}
ordered lexicographically. Define the gauge-dressed \(qL\times qL\)
matrix
\begin{equation}
\label{eq:appK-Mq-def}
\mathbb M^{(q)}_{(j,a),(k,b)}
=
\left(\frac{2\ii}{L}\right)^q
(D_X)_{jj}^{\,q}
(D_Y)_{kk}^{\,q}
D_{x_j}^{a}D_{\eta_k}^{b}
\frac{1}{x_j+\eta_k}.
\end{equation}
For every balanced pair \(S,T\subseteq[L]\),
\begin{equation}
\label{eq:appK-termwise-linearization}
\det\mathbb M^{(q)}[S^{[q]},T^{[q]}]
=
\bigl(\det G[S,T]\bigr)^{q^2}.
\end{equation}
For even \(q\), the right-hand side equals
\(\abs{\det G[S,T]}^{q^2}\). For odd \(q\), it retains the sign of the
real minor. The positive all-minors moment is then recovered from the
equivalent absolute-Vandermonde root representation rather than by
directly summing Eq.~\eqref{eq:appK-termwise-linearization}.

\subsection{Phase-corrected confluent Vandermonde blades}
\label{appK:blades}

Let \(r\) be even and let \(V\) have dimension \(rM\), ordered basis
\(\mathbf e_0,\ldots,\mathbf e_{rM-1}\), and volume form
\(\mathrm{vol}_V=\mathbf e_0\wedge\cdots\wedge\mathbf e_{rM-1}\).
For \(\Omega\in\Lambda^rV\), define
\begin{equation}
\label{eq:appK-HPf-definition}
\frac{1}{M!}\Omega^{\wedge M}
=
\operatorname{HPf}_r(\Omega)\,\mathrm{vol}_V.
\end{equation}
At \(r=2\), this is the ordinary Pfaffian. Odd exterior degree cannot be
used directly because \(\Omega\wedge\Omega=0\).

Let \(V_q\) have dimension \(qL\) and define
\begin{equation}
\label{eq:appK-p-vector}
\mathbf p(z)
=
\sum_{m=0}^{qL-1}z^m\mathbf e_m,
\end{equation}
together with the decomposable \(q\)-form
\begin{equation}
\label{eq:appK-q-blade}
\omega_q(z)
=
\mathbf p(z)\wedge
D_z\mathbf p(z)\wedge\cdots\wedge
D_z^{q-1}\mathbf p(z).
\end{equation}
For distinct ordered nodes \(z_1,\ldots,z_L\),
\begin{equation}
\label{eq:appK-blade-Vandermonde}
\omega_q(z_1)\wedge\cdots\wedge\omega_q(z_L)
=
\Delta(z_1,\ldots,z_L)^{q^2}\,
\mathrm{vol}_{V_q}.
\end{equation}

For an ordered root subset
\(A=\{a_1<\cdots<a_L\}\), the unit-circle Vandermonde phase gives
\begin{equation}
\label{eq:appK-absolute-phase-correction}
\abs{\Delta(z_A)}^{q^2}
=
\chi_{q,L}
\left(\prod_{a\in A}z_a^{-\rho_{q,L}}\right)
\Delta(z_A)^{q^2},
\end{equation}
where
\begin{equation}
\label{eq:appK-rho-chi}
\rho_{q,L}
=
\frac{q^2(L-1)}{2},
\qquad
\chi_{q,L}
=
\ii^{-q^2L(L-1)/2}.
\end{equation}
For half-integer powers, the branch is fixed by
\[
z_a^\rho=e^{\ii\rho\theta_a},
\qquad
\theta_a=\frac{\pi a}{L}\in[0,2\pi).
\]
For the one-body weights \(\mathbf w\) of
Eq.~\eqref{eq:appF-weighted-Z}, define
\begin{equation}
\label{eq:appK-beta-blade}
\beta_a^{(q)}[\mathbf w]
=
w_a z_a^{-\rho_{q,L}}\omega_q(z_a).
\end{equation}
Equations~\eqref{eq:appK-blade-Vandermonde} and
\eqref{eq:appK-absolute-phase-correction} imply
\begin{equation}
\label{eq:appK-ordered-blade-weight}
\chi_{q,L}
\beta_{a_1}^{(q)}[\mathbf w]\wedge\cdots\wedge
\beta_{a_L}^{(q)}[\mathbf w]
=
\left(\prod_{a\in A}w_a\right)
\abs{\Delta(z_A)}^{q^2}
\mathrm{vol}_{V_q}.
\end{equation}

\subsection{Even and odd \texorpdfstring{$q$}{q} summation}
\label{appK:even-odd-q}

For even \(q\), define
\begin{equation}
\label{eq:appK-Omega-even}
\Omega_q[\mathbf w]
=
\sum_{a=0}^{2L-1}\beta_a^{(q)}[\mathbf w]
\in\Lambda^qV_q.
\end{equation}
Because even \(q\)-forms commute and every
\(\beta_a^{(q)}\) is decomposable,
\begin{equation}
\label{eq:appK-even-expansion}
\frac{1}{L!}\Omega_q[\mathbf w]^{\wedge L}
=
\sum_{a_1<\cdots<a_L}
\beta_{a_1}^{(q)}[\mathbf w]\wedge\cdots\wedge
\beta_{a_L}^{(q)}[\mathbf w].
\end{equation}
Using Eq.~\eqref{eq:appF-weighted-Z},
\begin{equation}
\label{eq:appK-even-q-formula}
\cZ_L^{(q^2)}[\mathbf w]
=
L^{-q^2L/2}\chi_{q,L}\,
\operatorname{HPf}_{q}\!\left(\Omega_q[\mathbf w]\right).
\end{equation}

For odd \(q\), define the even \(2q\)-form
\begin{equation}
\label{eq:appK-Xi-def}
\Xi_q[\mathbf w]
=
\sum_{0\le a<b\le2L-1}
\beta_a^{(q)}[\mathbf w]\wedge
\beta_b^{(q)}[\mathbf w].
\end{equation}
For even \(L=2M\), the pairing expansion gives
\begin{equation}
\label{eq:appK-Xi-expansion}
\frac{1}{M!}\Xi_q[\mathbf w]^{\wedge M}
=
\sum_{a_1<\cdots<a_L}
\beta_{a_1}^{(q)}[\mathbf w]\wedge\cdots\wedge
\beta_{a_L}^{(q)}[\mathbf w].
\end{equation}
The coefficient of each ordered subset is one: it is the Pfaffian of the
skew matrix with entries \(1\) above the diagonal and \(-1\) below it.
Consequently,
\begin{equation}
\label{eq:appK-odd-q-formula}
\cZ_L^{(q^2)}[\mathbf w]
=
L^{-q^2L/2}\chi_{q,L}\,
\operatorname{HPf}_{2q}\!\left(\Xi_q[\mathbf w]\right),
\end{equation}

where $q$ is odd,and $L$ is even. Specializing to the checkerboard weight defined in
Appendix~\ref{app:generic-weight} gives
\[
\cZ_L^{(q^2)}[\mathbf w(u)]
=
\cZ_{q^2/2,L}(u).
\]
Therefore Eqs.~\eqref{eq:appK-even-q-formula} and
\eqref{eq:appK-odd-q-formula} become the corresponding formulas for the
fugacity-resolved all-minors polynomial. The cases \(q=1\) and \(q=2\)
reduce to ordinary Pfaffians, while \(q\geq3\) gives genuine higher
hyperpfaffians.

\subsection{Checkerboard sparsity and computational status}
\label{appK:checkerboard-sparsity}

For even \(q\), let
\(I=\{i_1<\cdots<i_q\}\subseteq\{0,\ldots,qL-1\}\). The coefficient of
\(\mathbf e_{i_1}\wedge\cdots\wedge\mathbf e_{i_q}\) in
\(\omega_q(z)\) is
\begin{equation}
\label{eq:appK-Wronskian-coefficient}
\omega_{q,I}(z)
=
\frac{\Delta(i_1,\ldots,i_q)}
{\displaystyle\prod_{a=0}^{q-1}a!}
z^{|I|-q(q-1)/2},
\end{equation}
and therefore
\begin{equation}
\label{eq:appK-Omega-coefficient}
\Omega_{q,I}[\mathbf w]
=
\frac{\Delta(i_1,\ldots,i_q)}
{\displaystyle\prod_{a=0}^{q-1}a!}
\sum_{b=0}^{2L-1}
w_b z_b^{\nu_q(I)},
\end{equation}
with
\begin{equation}
\label{eq:appK-nu-I}
\nu_q(I)
=
|I|
-
\frac{q(q-1)}{2}
-
\frac{q^2(L-1)}{2}.
\end{equation}
For the checkerboard weight, the Fourier rule
Eq.~\eqref{eq:appF-checkerboard-Fourier} implies that the coefficient
vanishes unless
\[
\nu_q(I)\equiv0
\quad\text{or}\quad
L
\pmod{2L}.
\]
This is the higher-\(q\) analogue of the sparsity producing the
\(\alpha=2\) block factorization. It does not, by itself, reduce a genuine
higher hyperpfaffian to ordinary linear algebra.

The formulas above are exact finite structural representations. At
\(q=1,2\), the exterior degree is two and standard Pfaffian algorithms
apply. For \(q\geq3\), a naive hyperpfaffian expansion remains
combinatorial; checkerboard sparsity alone does not establish a
polynomial-time evaluation or a complexity lower bound.

\section{Exact representations and complexity conventions}
\label{app:complexity-conventions}

This appendix records the limited computational statements used in the main
text. Its purpose is to separate exact algebraic representations from
closed product evaluations and from efficient algorithms. Throughout, an
``exact representation'' means a finite formula that is algebraically
correct for every finite \(L\). It does not automatically imply a polynomial
arithmetic evaluation, nor does the absence of such an evaluation imply a
hardness result.

Throughout this appendix, \(L\in\mathbb Z_{>0}\) denotes the physical
system size. All operation counts are expressed as functions of \(L\).

\subsection{Computational tasks}
\label{appL:computational-tasks}

The polynomial
\[
\cZ_{\alpha,L}(u)
=
\sum_{k=0}^{L}a_k u^k
\]
supports several distinct computational tasks:

\begin{enumerate}[label=(\roman*),leftmargin=2.5em]
\item evaluate \(\cZ_{\alpha,L}(u_0)\) at one specified numerical fugacity
\(u_0\);

\item produce all coefficients \(a_0,\ldots,a_L\);

\item produce a compressed exact symbolic representation of the polynomial,
such as a determinant, Pfaffian, product, hyperpfaffian, or finite
Jack--Kostka sum;

\item determine a particular coefficient, scaling limit, or asymptotic
regime.
\end{enumerate}

A determinant or Pfaffian depending on \(u\) may make the first task
efficient without making symbolic coefficient extraction equally simple.
Conversely, \(L+1\) evaluations at distinct fugacities determine all
coefficients by interpolation, but the bit complexity depends on the chosen
interpolation points and on the size of the exact rational or algebraic
numbers that appear.

\subsection{Direct enumeration and classical compressions}
\label{appL:classical-compressions}

The original all-minors definition contains
\begin{equation}
\label{eq:appL-number-minors}
\sum_{k=0}^{L}\binom{L}{k}^2
=
\binom{2L}{L}
\end{equation}
terms. Using
\[
\binom{2L}{L}
\sim
\frac{4^L}{\sqrt{\pi L}},
\]
the number of minors grows exponentially with the physical system size.
The polynomial cost of evaluating each individual determinant does not
alter this exponential character.

At the classical indices \(\alpha=1/2\) and \(\alpha=2\), the generic
weighted problems are ordinary Pfaffians. At \(\alpha=1/2\), the Pfaffian
has dimension \(L\) for even \(L\), while odd \(L\) is handled by the
standard bordered \((L+1)\times(L+1)\) construction. At \(\alpha=2\), the
generic weighted Pfaffian has dimension \(2L\). Dense Pfaffian evaluation
requires \(O(L^3)\) arithmetic operations.

For the checkerboard weight, the corresponding matrices split into
fixed-size blocks. Consequently, evaluating the resulting product formulas
at one specified fugacity requires \(O(L)\) arithmetic operations.
Expanding these products to obtain all \(L+1\) coefficients is a separate
computational task and need not have the same operation count. At
\(\alpha=1\), the Cauchy--Binet identity and root orthogonality give the
explicit polynomial
\[
\cZ_{1,L}(u)=(1+u)^L .
\]

These statements refer to arithmetic operation counts. A formal
bit-complexity bound would additionally require specifying the
representation of the input and controlling the growth of intermediate
numerators and denominators.

\subsection{Neutral sectors at \texorpdfstring{$\alpha=4$}{alpha=4}}
\label{appL:alpha-four-sectors}

At \(\alpha=4\), the generic-fugacity Jack--Kostka representation is
organized by neutral seven-charge vectors
\begin{equation}
\label{eq:appL-neutral-constraints}
\sum_{m=-3}^{3}\nu_m=L,
\qquad
\sum_{m=-3}^{3}m\nu_m=0,
\qquad
\nu_m\ge0.
\end{equation}
Ignoring the neutrality constraint gives the elementary bound
\begin{equation}
\label{eq:appL-sector-bound}
|\mathcal N_L|
\le
\binom{L+6}{6}.
\end{equation}
Thus the number of multiplicity sectors is polynomial in \(L\) for fixed
\(\alpha=4\); more explicitly, the outer seven-charge sum contains at most
\(O(L^6)\) sectors. However, the formula still requires, for each sector,
the inverse Jack--Kostka coefficient
\begin{equation}
\label{eq:appL-inverse-jack-coefficient}
K^{-1}_{\lambda(\nu),(3L)^L}(1/4).
\end{equation}
The sector decomposition is therefore a genuine compression of the outer
combinatorics, but it does not by itself provide an efficient evaluation
algorithm for the inverse Jack transition coefficient.

Algorithms for evaluating Jack functions at specified numerical arguments
address a different task from extracting the inverse coefficient in
Eq.~\eqref{eq:appL-inverse-jack-coefficient} \cite{DemmelKoev}. General
\(\#\mathrm P\)-completeness results for ordinary Kostka numbers and
Littlewood--Richardson coefficients also do not directly settle this
special rectangular scaling family \cite{Narayanan2006}. Accordingly, this
work asserts neither a polynomial-time algorithm nor a hardness theorem
for the seven-charge \(\alpha=4\) family.

\subsection{Hyperpfaffian representations}
\label{appL:hyperpfaffian-complexity}

For an even \(r\)-form on a space of dimension \(rM\), the unrestricted
coordinate expansion of its hyperpfaffian contains
\begin{equation}
\label{eq:appL-hyperpfaffian-count}
\frac{(rM)!}{M!(r!)^M}
\end{equation}
set-partition terms before any sparsity or cancellation is used. The
perfect-square hierarchy of
Appendix~\ref{app:perfect-square-hyperpfaffian} therefore gives one exact
finite algebraic object, but for \(r>2\) it does not inherit the standard
cubic algorithm of an ordinary Pfaffian.

The checkerboard tensors appearing in that hierarchy are sparse because of
root-of-unity selection rules. Whether this sparsity leads to a block
factorization, a recurrence, a transfer matrix, or another efficient
evaluation method is an open algebraic question, not an established
complexity result.

\subsection{What a hardness statement would require}
\label{appL:hardness-scope}

The present problem is a highly structured one-parameter family, not an
arbitrary-input matrix problem. A formal hardness theorem would have to
specify the input model, the required output, the encoding of \(L\) and
\(u\), and an explicit reduction from a known hard problem. Exponential
term count, the absence of a known ordinary Pfaffian, or the appearance of
general Kostka-type coefficients is not sufficient.

The computational conclusions used in this paper are therefore limited to
the following statements. The classical indices
\(\alpha=\tfrac12,1,2\) admit polynomial-size Pfaffian or determinant
compressions. In the checkerboard specialization, these reduce further to
explicit product formulas, permitting \(O(L)\)-cost evaluation at a
specified fugacity. The generic-fugacity \(\alpha=4\) problem has
polynomially many neutral sectors, but the evaluation of the inverse
Jack--Kostka coefficient inside each sector remains unresolved. Higher
perfect-square indices admit exact hyperpfaffian representations, but their
efficient evaluation is not established.


\end{document}